\documentclass[prb,aps,twocolumn,amsmath,amssymb,showpacs]{revtex4}
\usepackage{graphicx}
\usepackage{psfrag}
\usepackage{color}
\usepackage{subfigure}
\usepackage{amsmath}
\definecolor{dred}{rgb}{0.7,0.0,0.0}
\allowdisplaybreaks 
\usepackage{array}

\begin{document}

%
%

\title{Three Orbital Model for the iron-based superconductors}

\author{Maria Daghofer}
\email{M.Daghofer@ifw-dresden.de} 
\thanks{Present address: IFW Dresden, P.O. Box 27 01 16, D-01171 Dresden, Germany}

\author{Andrew Nicholson}
\author{Adriana Moreo}
\author{Elbio Dagotto}
\affiliation{Department of Physics and Astronomy, The University of
  Tennessee, Knoxville, TN 37996} 
\affiliation{Materials Science and Technology Division, Oak Ridge
  National Laboratory, Oak Ridge, TN 32831}

\date{\today}

\begin{abstract}

The theoretical need to study the properties of the Fe-based high-$T_c$ 
superconductors 
using reliable many-body techniques has highlighted the importance of 
determining what is
the minimum number of orbital degrees of freedom that will capture the 
physics of these materials. While the shape of the Fermi surface (FS)
obtained with the local density approximation (LDA) can be reproduced 
by a two-orbital model, it has been argued that the bands that cross the 
chemical potential result from the strong hybridization of three of the Fe 
$3d$ orbitals. For this reason, a three-orbital Hamiltonian for LaOFeAs 
obtained with the  
Slater-Koster formalism by considering the hybridization of the As $p$ 
orbitals with the
Fe $d_{xz}$, $d_{yz}$, and $d_{xy}$ orbitals is discussed here. This model 
reproduces qualitatively the FS shape and 
orbital composition obtained by LDA calculations for undoped LaOFeAs when
four electrons per Fe are considered. Within a mean-field approximation, 
its magnetic and orbital properties in the undoped case are here described for
intermediate values of $J/U$. Increasing the Coulomb repulsion $U$ at zero 
temperature, four different regimes are obtained: {\it (1)} 
paramagnetic, {\it (2)} magnetic  $(\pi,0)$ spin order, 
{\it (3)} the same $(\pi,0)$ spin order but now including orbital order, and 
finally a {\it (4)} magnetic and orbital ordered insulator. The spin-singlet 
pairing operators allowed by the lattice and orbital symmetries are also 
constructed. It is found that for pairs of electrons involving up  
to diagonal nearest-neighbors sites, the only fully gapped and purely 
intraband spin-singlet pairing operator is given 
by $\Delta({\bf k})=f({\bf k})\sum_{\alpha}d_{{\bf k},\alpha,\uparrow}
d_{{\bf -k},\alpha,\downarrow}$ with $f({\bf k})=1$ or $\cos k_x \cos k_y$ 
which would arise only if the electrons in all different orbitals couple with 
equal strength to the source of pairing. 

\end{abstract}
 
 
\maketitle

\section{Introduction}

The discovery of high-$T_c$ superconductivity in a family of iron-based 
compounds\cite{Fe-SC,chen1,chen2,wen,chen3,ren1,55,ren2} is 
offering a new conceptual framework to study the non-standard pairing 
mechanism that seems to induce these exotic superconducting states.\cite{phonon0} 
The magnetism present in several of the parent 
compounds\cite{sdw,neutrons1,neutrons2,neutrons3,neutrons4} and the
high critical temperatures are reminiscent of 
properties observed in the cuprates.\cite{review} But there are
also clear differences, since the parent compounds are
metallic\cite{sdw,neutrons1,neutrons2,neutrons3,neutrons4} and the
Fermi surface (FS) is determined by more than one 
orbital.\cite{first,singh,xu,cao,fang2} The multiorbital nature of 
the problem poses a challenge to the design of minimal models that can be 
studied with powerful techniques, such as numerical
methods. Ab-initio
calculations making use of the local density approximation (LDA) indicate
that the five $3d$ orbitals of Fe strongly hybridize to form the bands that 
are close to the chemical potential.\cite{first,singh,xu,cao,fang2}
However, the FS appears to be determined by bands that have mostly $d_{xz}$ 
and $d_{yz}$ character, and this observation 
has been confirmed by polarized angle resolved
photoemission spectroscopy (ARPES) 
experiments.\cite{FeAs_orb_FS} This supports the notion that the minimum number
of orbitals to be considered to study the pnictides could be two. In fact, a
two-orbital minimal model has been proposed\cite{scalapino,Daghofer:2009p1970} and it has 
been studied with numerical techniques on small clusters,\cite{Daghofer:2009p1970,moreo} as well as
with several other
approximations.\cite{scalapino,yu,lu,chen,calderon,kubo,ku} 
Both numerical\cite{Daghofer:2009p1970,moreo} and mean-field\cite{yu} calculations indicate
that the magnetic metallic regime observed experimentally in the undoped
compounds\cite{sdw,neutrons1,neutrons2,neutrons3,neutrons4} is stabilized for intermediate values of the Coulomb 
repulsion $U$ and the numerical calculations suggest that, upon doping in this 
regime, the most favored pairing operator is interorbital and has symmetry
$B_{2g}$.\cite{Daghofer:2009p1970,moreo} 

On the other hand, several authors have claimed that a 
two-orbital model may miss important features of the real 
system.\cite{plee,first,xu,cao,fang2} 
It has been argued that a minimal model for the pnictides should contain at 
least three orbitals for mainly two reasons: {\it (i)} 
A relatively small portion of the electron-pocket FS of LaOFeAs is
determined by a band of mostly $d_{ xy}$ character and {\it (ii)} the bands that
produce the two hole pockets should be degenerate at the center of the 
Brillouin zone (BZ), which is not the case when only two orbitals are 
considered. The important question is how much these shortcomings of the model 
impact the most relevant properties of the pnictides. The aim of this paper is to construct
a three-orbital model that addresses these concerns and compare its properties 
with the two-orbital case. This is important because in other areas of 
condensed matter physics, such as the manganites, we have learned
that a simple single-orbital model is often sufficient to capture qualitatively 
the phenomenon of colossal 
magnetoresistance,\cite{manga} while clearly a two-orbital model is still
necessary to properly describe additional properties such as 
the magnetic and orbital order observed in  
these materials.\cite{manga} Similarly the minimal two-orbital model
for the pnictides appears to reproduce the experimentally observed
magnetic order, and it is interesting to investigate to what extent the inclusion
of the third orbital $xy$ modifies these results. 

While the pnictides exhibit a structural phase transition at similar
or slightly higher temperature as the onset of antiferromagnetic (AF)
order, experiments have not yet addressed the issue of 
orbital order and also have not provided consensus regarding the symmetry of the 
pairing operator.\cite{nodal1,nodal2,nodal3,Ahilan,nakai,Grafe,Y.Wang,matano,mukuda,millo,wang-nodes,hashimoto,arpes,arpes2,arpes3,C.Martin2,T.Chen,parker,mu} An 
investigation of the magnetic and orbital orders, as well as pairing symmetries,
that are allowed in a three-orbital model compared to those that are possible in the two-orbital case
will shed light on the importance of the role that the additional
$d_{xy}$ orbital should play in theoretical discussions.

This paper is organized as follows. In Sec.~\ref{sec:model}, the three-orbital model is 
introduced.  Section~\ref{sec:results} 
contains a mean-field analysis of the magnetic and orbital properties of the 
undoped system.
Section~\ref{sec:pair} is devoted to the classification and analysis of the spin-singlet pairing
operators allowed by the orbital and lattice symmetries. A summary and
conclusions are presented in Sec.~\ref{sec:conclusions}.

\section{Three Orbital Model for the Pnictides}\label{sec:model}

As explained in the Introduction, it has been suggested that at least
three orbitals may be needed to describe the superconducting pnictides because the 
bands that determine the Fermi surface of LaOFeAs are mostly composed by orbitals 
$d_{xz}$, $d_{yz}$, and $d_{xy}$.\cite{plee,first,xu,cao,fang2,yu:064517}
The notation $xz$, $yz$, and $xy$ will be used for these orbitals, respectively, 
for better readability.
The need to include at least three orbitals in a realistic model was first 
pointed out in Ref.~\onlinecite{plee} where a three-orbital
model was constructed using the symmetry properties of the Fe-As planes and
LDA results. A shortcoming of that proposed model was that it contained an
spurious hole-pocket FS around the $M$ point in the extended BZ notation. It was
argued\cite{plee} that a fourth orbital should be added to remove
the extra pocket. However, it will be shown  in Sec.~\ref{sec:mom} that this spurious pocket
can actually be removed entirely within a three-orbital model formalism, i.e. without 
adding a fourth orbital.

One important issue that needs to be addressed is the electronic 
filling to be used in a three-orbital model. 
Band calculations have determined that the 
undoped pnictides contain six electrons distributed among the five $3d$ 
orbitals of 
each Fe atom. One procedure to determine the filling for a model with a reduced 
number of orbitals is to start from  the crystal field
splitting and fill the levels accordingly from the lowest energy up. 
In the two-orbital model
that considers only the $xz$ and $yz$ orbitals, such a consideration
would suggest that half-filling is the correct 
electronic density,\cite{scalapino,Daghofer:2009p1970,moreo}
because the $x^2-y^2$ and $3z^2-r^2$ orbitals are
assumed to be fully occupied with four of the six electrons and the
$xy$ orbital is assumed empty, leaving two electrons to 
populate the $xz$ and $yz$ orbitals. In addition, this filling 
assignment is the only one that allows to reproduce the LDA calculated FS.   
Applying the crystal-field splitting rationale to the
three-orbital model with $xz$, $yz$, and $xy$ orbitals,
this argument leads to a filling of one third (i.e. two electrons in three orbitals).\cite{zaanen} 
However, for such 
a filling we have not been able to reproduce the LDA shape of the FS. 
Thus, the
filling must be adjusted to approximately reproduce the FS and the orbital
occupation numbers obtained with LDA. In fact, band structure calculations suggest that the 
three-orbital system
should be $more$ than half-filled and actually have a filling of roughly two 
thirds (i.e. four electrons in the three orbitals).\cite{phonon0,haule2} 
Our analysis
shows that a FS with
approximately a similar size for the hole and electron pockets can 
be obtained both at fillings around one and two thirds 
(i.e. two and four electrons in the three orbitals), but the two almost
degenerate hole-pockets around $\Gamma$ demand a filling larger
than half-filling. Thus, the focus of our effort will be on 
a filling of $2/3$, as in Ref.~\onlinecite{yu:064517}. As it will be discussed 
below, non-half-filled 
orbitals allow the $orbital$ degree of freedom to
be active and actually it has been argued that orbital ordering phenomena 
may play a role in these materials.\cite{zaanen} This orbital order is unlikely
in the half-filled case which is the natural filling for the
two\cite{scalapino,Daghofer:2009p1970} and four\cite{yu} orbital
models for the pnictides. Once again, note that some authors have considered half-filling in 
the three-orbital case,\cite{plee} but this leads to the ``unwanted'' hole pocket 
around $M$.

\subsection{Real Space}\label{sec:real}
To construct the tight-binding portion of the 
three-orbital Hamiltonian for the pnictides,
the Slater-Koster procedure described in Ref.~\onlinecite{moreo} will be followed. 
Nearest-neighbor and diagonal next-nearest-neighbor hoppings will be considered for all the 
orbitals. It is clear that the hopping terms for the $xz$ and $yz$ orbitals are the same as in
the previously discussed two-orbital model,
\begin{equation}\begin{split}
H^{xz,yz}&=
-t_1\sum_{{\bf i},\sigma}(d^{\dagger}_{{\bf i},xz,\sigma}
d^{\phantom{\dagger}}_{{\bf i}+\hat y,xz,\sigma}+d^{\dagger}_{{\bf i},yz,\sigma}
d^{\phantom{\dagger}}_{{\bf i}+\hat x,yz,\sigma}+h.c.) \\
&-t_2\sum_{{\bf i},\sigma}(d^{\dagger}_{{\bf i},xz,\sigma}
d^{\phantom{\dagger}}_{{\bf i}+\hat x,xz,\sigma}+d^{\dagger}_{{\bf i},yz,\sigma}
d^{\phantom{\dagger}}_{{\bf i}+\hat y,yz,\sigma}+h.c.) \\
&-t_3\sum_{{\bf i},\hat\mu,\hat\nu,\sigma}(d^{\dagger}_{{\bf i},xz,\sigma}
d^{\phantom{\dagger}}_{{\bf i}+\hat\mu+\hat\nu,xz,\sigma}+d^{\dagger}_{{\bf i},yz,\sigma}
d^{\phantom{\dagger}}_{{\bf i}+\hat\mu+\hat\nu,yz,\sigma}+h.c.) \\
&+t_4\sum_{{\bf i},\sigma}(d^{\dagger}_{{\bf i},xz,\sigma}
d^{\phantom{\dagger}}_{{\bf i}+\hat x+\hat y,yz,\sigma}+d^{\dagger}_{{\bf i},yz,\sigma}
d^{\phantom{\dagger}}_{{\bf i}+\hat x+\hat y,xz,\sigma}+h.c.) \\
&-t_4\sum_{{\bf i},\sigma}(d^{\dagger}_{{\bf i},xz,\sigma}
d^{\phantom{\dagger}}_{{\bf i}+\hat x-\hat y,yz,\sigma}+d^{\dagger}_{{\bf i},yz,\sigma}
d^{\phantom{\dagger}}_{{\bf i}+\hat x-\hat y,xz,\sigma}+h.c.)\\
&-\mu\sum_{\bf i}(n_{{\bf i},xz}+n_{{\bf i},yz}),
\label{ham12}
\end{split}\end{equation}
while the intra-orbital hoppings for the $xy$ orbital are given by
\begin{equation}\begin{split}
H^{xy}=&\ t_5\sum_{{\bf i},{\hat \mu},\sigma}(d^{\dagger}_{{\bf i},xy,\sigma}
d^{\phantom{\dagger}}_{{\bf i}+\hat \mu,xy,\sigma}+h.c.)\\
&-t_6\sum_{{\bf i},\hat\mu,\hat\nu,\sigma}(d^{\dagger}_{{\bf i},xy,\sigma}
d^{\phantom{\dagger}}_{{\bf i}+\hat\mu+\hat\nu,xy,\sigma}+h.c.)\\
&+\Delta_{xy}\sum_{\bf i}n_{{\bf i},xy}-\mu\sum_{\bf i}n_{{\bf i},xy},\;
\label{ham3}
\end{split}\end{equation}
where $\Delta_{xy}$ is the energy difference between the $xy$ and
the degenerate $xz$/$yz$ orbitals. The hybridization
between the $xz$/$yz$ and the $xy$ orbitals is given by
\begin{align}
H^{xz,yz;xy}=&-t_7\sum_{{\bf i},\sigma}[(-1)^{|i|}d^{\dagger}_{{\bf i},xz,\sigma}
d^{\phantom{\dagger}}_{{\bf i}+\hat x,xy,\sigma}+h.c.]\nonumber\\
&-t_7\sum_{{\bf i},\sigma}[(-1)^{|i|}d^{\dagger}_{{\bf i},xy,\sigma}
d^{\phantom{\dagger}}_{{\bf i}+\hat x,xz,\sigma}+h.c.]\nonumber\\
&-t_7\sum_{{\bf i},\sigma}[(-1)^{|i|}d^{\dagger}_{{\bf i},yz,\sigma}
d^{\phantom{\dagger}}_{{\bf i}+\hat y,xy,\sigma}+h.c.] \nonumber\\
&-t_7\sum_{{\bf i},\sigma}[(-1)^{|i|}d^{\dagger}_{{\bf i},xy,\sigma}
d^{\phantom{\dagger}}_{{\bf i}+\hat y,yz,\sigma}+h.c.] \nonumber\\
&-t_8\sum_{{\bf i},\sigma}[(-1)^{|i|}d^{\dagger}_{{\bf i},xz,\sigma}
d^{\phantom{\dagger}}_{{\bf i}+\hat x+\hat y,xy,\sigma}+h.c.] \nonumber\\
&+t_8\sum_{{\bf i},\sigma}[(-1)^{|i|}d^{\dagger}_{{\bf i},xy,\sigma}
d^{\phantom{\dagger}}_{{\bf i}+\hat x+\hat y,xz,\sigma}+h.c.] \nonumber\\
&-t_8\sum_{{\bf i},\sigma}[(-1)^{|i|}d^{\dagger}_{{\bf i},xz,\sigma}
d^{\phantom{\dagger}}_{{\bf i}+\hat x-\hat y,xy,\sigma}+h.c.] \nonumber\\
&+t_8\sum_{{\bf i},\sigma}[(-1)^{|i|}d^{\dagger}_{{\bf i},xy,\sigma}
d^{\phantom{\dagger}}_{{\bf i}+\hat x-\hat y,xz,y\sigma}+h.c.] \nonumber\\
&-t_8\sum_{{\bf i},\sigma}[(-1)^{|i|}d^{\dagger}_{{\bf i},yz,\sigma}
d^{\phantom{\dagger}}_{{\bf i}+\hat x+\hat y,xy,\sigma}+h.c.] \nonumber\\
&+t_8\sum_{{\bf i},\sigma}[(-1)^{|i|}d^{\dagger}_{{\bf i},xy,\sigma}
d^{\phantom{\dagger}}_{{\bf i}+\hat x+\hat y,yz,\sigma}+h.c.] \nonumber\\
&+t_8\sum_{{\bf i},\sigma}[(-1)^{|i|}d^{\dagger}_{{\bf i},yz,\sigma}
d^{\phantom{\dagger}}_{{\bf i}+\hat x-\hat y,xy,\sigma}+h.c.] \nonumber\\
&-t_8\sum_{{\bf i},\sigma}[(-1)^{|i|}d^{\dagger}_{{\bf i},xy,\sigma}
d^{\phantom{\dagger}}_{{\bf i}+\hat x-\hat y,yz,\sigma}+h.c.].
\label{ham12_3}
\end{align}
The hopping parameters $t_i$ in Eqs.(\ref{ham12}-\ref{ham12_3}) will be determined by fitting
the band dispersion to band structure calculation results. The chemical
potential $\mu$ is set to a two-thirds filling, as already discussed. The operator 
$d^{\dagger}_{{\bf i},\alpha,\sigma}$ ($d^{\phantom{\dagger}}_{{\bf
    i},\alpha,\sigma}$) creates (annihilates) an electron at site ${\bf i}$,
orbital $\alpha=xz,yz,xy$, and with spin projection $\sigma$. 
$n_{{\bf i},\alpha}=n_{{\bf
    i},\alpha,\uparrow}+n_{{\bf i},\alpha,\downarrow}$ are the corresponding density
operators. Previously proposed three-orbital
models\cite{plee,yu:064517} only contained the nearest-neighbor (NN) 
hybridization $t_7$, but since next-nearest neighbor (NNN) terms are 
included for the intra-orbital component, as well
as for the hybridization between $xz$ and $yz$, they should also be included in
the hybridization with $xy$. 
In Sec.~\ref{sec:mom}, it is shown that these NNN
terms with hopping $t_8$ turn out to be \emph{crucial} to provide the proper
orbital character for the electron pockets when compared with LDA results.
Finally note that the hybridization Eq.~(\ref{ham12_3})
contains factors $(-1)^{|i|}$ that arise from the two-iron unit cell of the 
original FeAs planes.

\subsection{Momentum Space}\label{sec:mom}

The Hamiltonian Eqs.~(\ref{ham12}-\ref{ham12_3}) can be transformed to 
momentum space
using $d^{\dagger}_{{\bf k},\alpha,\sigma}={\frac{1}{\sqrt{N}}}\sum_{\bf i}
e^{-i{\bf k.i}}d^{\dagger}_{{\bf i},\alpha,\sigma}$, where ${\bf k}$ is the 
wavevector and $N$ the number of sites. 
Note that the Fourier transformed Hamiltonian is defined in the extended or 
unfolded BZ.\cite{mazin,kuroki} As pointed out in 
Ref.~\onlinecite{plee}, the 
real-space Hamiltonian presented in Eqs.~(\ref{ham12}-\ref{ham12_3}) is 
invariant under a translation along the $x$ or $y$ directions followed by a 
reflexion about the $x$-$z$ plane. When the eigenstates of the Hamiltonian are 
labeled in terms of the eigenvalues of these symmetry operations, then the 
momentum-space Hamiltonian can be expressed in terms of a pseudocrystal 
momentum that will be called ${\bf k}$ that expands the unfolded BZ that 
corresponds to a single Fe unit-cell in real space.
Note that folding the three
bands used here into the reduced diamond-like cell defined in the first 
quadrant by $k_x+k_y\leq\pi$, and 
symmetric points in the other three quadrants,
doubles the number of bands to six as expected. 

Equations~(\ref{ham12}-\ref{ham12_3}) become in momentum space
\begin{eqnarray}\label{E.H0k}
H_{\rm TB}(\mathbf{ k}) &=& \sum_{\mathbf{ k},\sigma,\mu,\nu} T^{\mu,\nu}
(\mathbf{ k})
d^\dagger_{\mathbf{ k},\mu,\sigma} d^{\phantom{\dagger}}_{\mathbf{ k},\nu,\sigma},
\end{eqnarray}
with
\begin{eqnarray}
T^{11} &=& 2t_2\cos  k_x +2t_1\cos  k_y +4t_3 \cos  k_x 
\cos  k_y-\mu, \label{eq:t11}\\
T^{22} &=& 2t_1\cos  k_x +2t_2\cos  k_y +4t_3 \cos  k_x 
\cos  k_y-\mu, \label{eq:t22}\\
T^{33} &=& 2t_5(\cos  k_x+\cos  k_y) \nonumber\\
       & & +4t_6\cos  k_x\cos  k_y -\mu +\Delta_{xy},
\label{eq:t33} \\
T^{12} &=& T^{21} =4t_4\sin  k_x \sin  k_y, \label{eq:t12}\\
T^{13} &=& \bar{T}^{31} = 2it_7\sin  k_x + 4it_8\sin  k_x \cos  k_y,
\label{eq:t13} \\
T^{23} &=& \bar{T}^{32} 2it_7\sin  k_y + 4it_8\sin  k_y \cos  k_x\;,
\label{eq:t23}
\end{eqnarray}
where a bar on top of a matrix element denotes the complex conjugate.
Since the
Hamiltonian for a one-iron unit cell has been considered, then ${\bf k}$ runs 
within the corresponding extended BZ $-\pi<k_x,k_y\le\pi$. 

\begin{table}
\caption{Parameters for the tight-binding portion of the three-orbital model
  Eqs.(\ref{eq:t11}) to (\ref{eq:t23}). The overall energy unit is
  electron volts.\label{tab:hopp3}}
 \begin{tabular}{|ccccccccc|}\hline
$t_1$ & $t_2$ & $t_3$ & $t_4$ & $t_5$ & $t_6$ & $t_7$ & $t_8$ &
   $\Delta_{xy}$\\
\hline
  0.02   &0.06    &0.03   &$-0.01$&$0.2$ & 0.3 & $-0.2$ & $-t_7/2$&
  0.4\\ \hline
 \end{tabular}
\end{table}

\begin{figure}
\subfigure{\includegraphics[width = 0.27\textwidth]{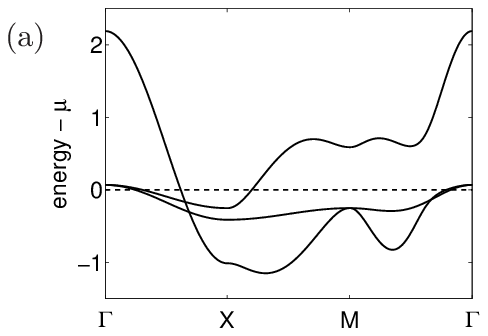}\label{fig:bands_U0}}\\
 \begin{minipage}{0.22\textwidth}
 \vspace*{-18em}
\subfigure[]{\includegraphics[width = \textwidth]{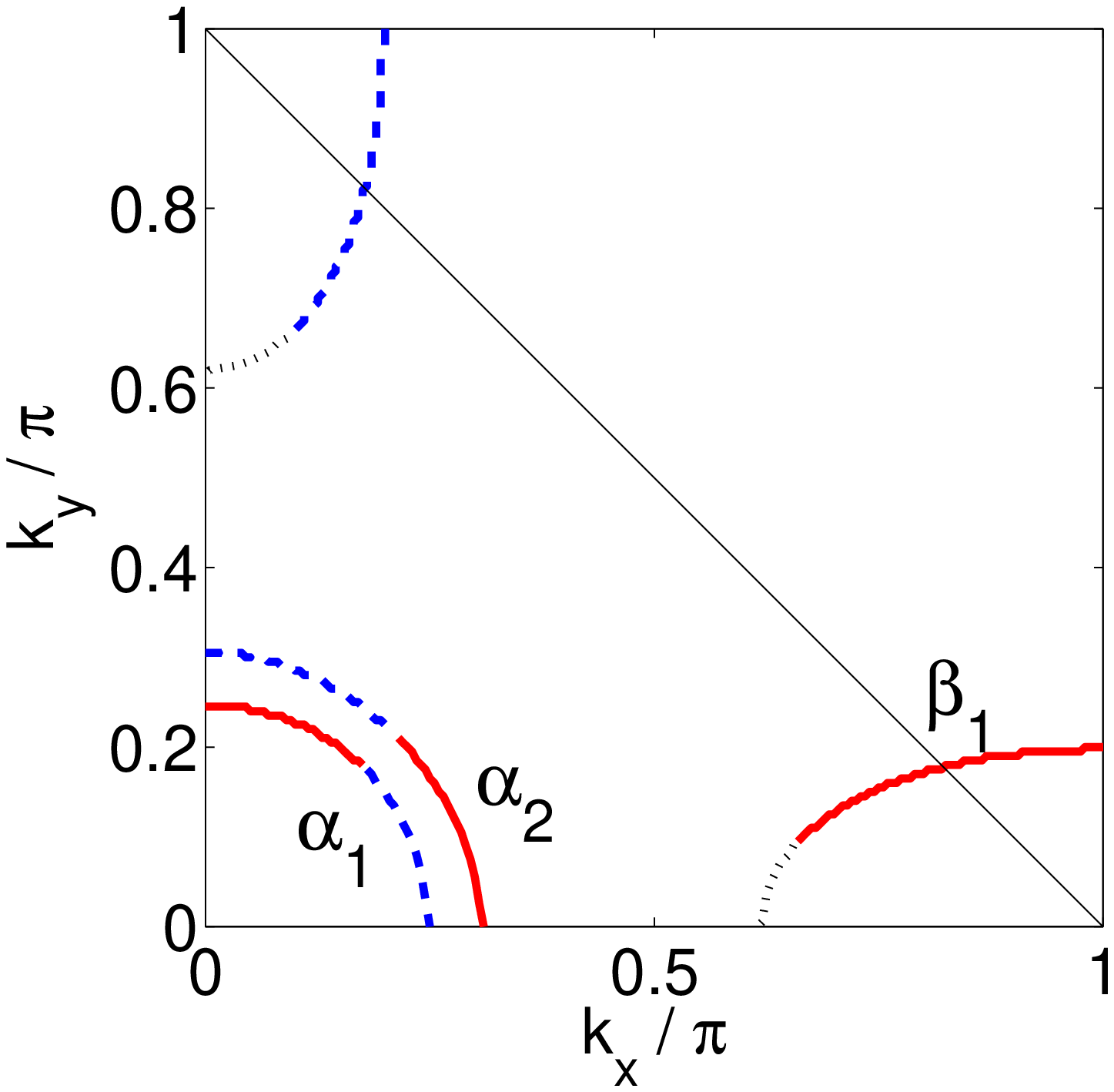}\label{fig:fs_U0}}
 \end{minipage}\hspace{-0.5em}
\subfigure{
\includegraphics[width =
0.25\textwidth]{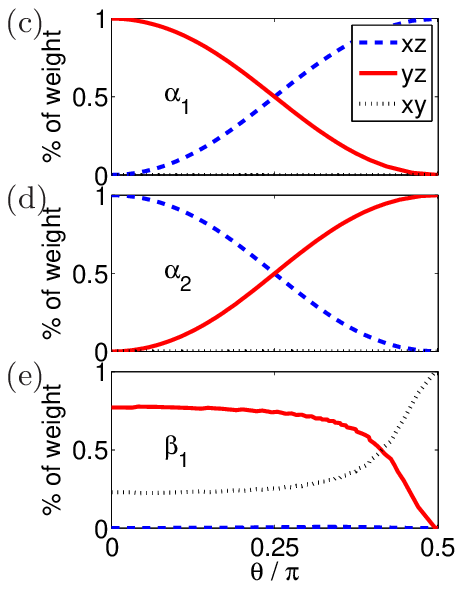}\label{fig:fs_U0_h1}}
\subfigure{\label{fig:fs_U0_h2}}
\subfigure{\label{fig:fs_U0_e}}
\caption{(Color online) \subref{fig:bands_U0} 
Band structure and \subref{fig:fs_U0} Fermi 
surface of the tight-binding (i.e. non-interacting) 
three-orbital model, with parameters from 
Tab.~\ref{tab:hopp3} and in the unfolded BZ. The diagonal thin solid line 
in \subref{fig:fs_U0} indicates the boundary of the 
folded BZ.
  In panels (c-e), the orbital contributions to the two hole and one of the 
electron pockets are given. The winding angle $\theta$
  is measured with respect to the $k_y$-axis. The second electron
  pocket is analogous to the one shown simply replacing $xz$ by
  $yz$. In all panels, the dashed lines refer to the $xz$ orbital,
  the solid to $yz$, and the dotted to $xy$.\label{fig:bands_fs_orb_U0}}
\end{figure}

The hopping parameters $t_i$ and $\Delta_{xy}$ were chosen to
reproduce the shape of the LaOFeAs FS obtained using LDA calculations. 
These parameters are given in Tab.~\ref{tab:hopp3}. The
chemical potential $\mu=0.212$ was fixed to this value to ensure a filling of two thirds,
as discussed in Sec.~\ref{sec:model}. The resulting  
band dispersion is presented in Fig.~\ref{fig:bands_U0} along high symmetry
directions in the extended BZ and the corresponding FS is shown in
Fig.~\ref{fig:fs_U0}. The two hole-pocket FSs, labeled $\alpha_1$ and
$\alpha_2$, are formed by two bands that are degenerate at $\Gamma$ in
agreement with LDA, and both of them are found around the
$\Gamma$-point instead of $M$ in the extended zone.
Thus, one of the shortcomings of the two-orbital model has been
corrected. It can also be observed that there is no hole-pocket FS around $M$ which was a 
problem encountered in Refs.~\onlinecite{yu:064517,plee}.

The orbital composition of the hole pockets is displayed in
Figs.~\ref{fig:bands_fs_orb_U0} (c) and (d). They are given by a linear combination of the $xz$ and 
$yz$ orbitals, in agreement with LDA\cite{phonon0,kuroki} and ARPES 
results.\cite{FeAs_orb_FS} The electron pockets at $X$ ($Y$) arise from a  
linear combination of the $yz$ ($xz$) and $xy$ orbitals. The orbital composition for the electron pocket 
around $X$, labeled $\beta_1$ in Fig.~\ref{fig:fs_U0}, is displayed
in Fig.~\ref{fig:fs_U0_e}: here it can be observed that the orbital character
changes from purely $xy$ along $\Gamma$-$X$ to predominantly $yz$ along
$X$-$M$, as predicted by LDA\cite{phonon0,kuroki} and also found with  
ARPES techniques.\cite{FeAs_orb_FS} As can be deduced from
Eqs.~(\ref{eq:t13}) and (\ref{eq:t23}), setting $t_8 = - t_7/2$
ensures that the electron pockets have pure $xy$ character along the
$\Gamma$-$X$ and $\Gamma$-$Y$ directions.\cite{note:signt8} A large $yz$
($xz$) contribution  along $X$-$M$ ($Y$-$M$) requires the hybridizations
$t_7$ and $t_8$ to be quite robust.\cite{note_t7}

In summary, a tight-binding Hamiltonian has been constructed  that 
captures the generic shape and orbital composition of the FS for undoped 
LaOFeAs by considering only the three $xz$, $yz$, and $xy$ orbitals, 
and assuming that they 
share four of the six electrons that populate the five Fe $3d$ levels. 

The Coulombic interacting portion of the Hamiltonian is given by:
\begin{equation}\begin{split}  \label{eq:Hcoul}
  H_{\rm int}& =
  U\sum_{{\bf i},\alpha}n_{{\bf i},\alpha,\uparrow}n_{{\bf i},
    \alpha,\downarrow}
  +(U'-J/2)\sum_{{\bf i},
    \alpha < \beta}n_{{\bf i},\alpha}n_{{\bf i},\beta}\\
  &\quad -2J\sum_{{\bf i},\alpha < \beta}{\bf S}_{\bf{i},\alpha}\cdot{\bf S}_{\bf{i},\beta}\\
  &\quad +J\sum_{{\bf i},\alpha < \beta}(d^{\dagger}_{{\bf i},\alpha,\uparrow}
  d^{\dagger}_{{\bf i},\alpha,\downarrow}d^{\phantom{\dagger}}_{{\bf i},\beta,\downarrow}
  d^{\phantom{\dagger}}_{{\bf i},\beta,\uparrow}+h.c.),
\end{split}\end{equation}
where $\alpha,\beta=xz,yz,xy$ denote the orbital, ${\bf S}_{{\bf i},\alpha}$
($n_{{\bf i},\alpha}$) is the spin (electronic density) in orbital $\alpha$ at site
${\bf i}$, and the relation $U'=U-2J$ between these Kanamori parameters 
has been used (for a discussion in the manganite context 
see Ref.~\onlinecite{manga} and references therein).
The first line terms give
the energy cost of two electrons located in  the same orbital or in
different orbitals on the same site, respectively. The second line contains the Hund's
rule coupling that favors the ferromagnetic (FM) alignment of the spins in
different orbitals at the same lattice site. The ``pair-hopping'' term is in the third line and its 
coupling is equal to $J$ by symmetry. 
The values used for $U$ and $J$ can be substantially smaller
than the atomic ones, because the interactions may be screened by
bands not included in our Hamiltonian. These Coulombic interaction terms have
been used and discussed in several previous publications\cite{Daghofer:2009p1970,moreo,yu}
where more details can be found by the readers. All the
energies in this paper are given in electron volts.

\section{Numerical results}\label{sec:results}

\subsection{Mean-field Approximation and Ordered Phases}\label{sec:competing}

\begin{table}
\caption{Magnetic and orbital ordering wavevectors for the possible ordered
  phases discussed in the text. The third column indicates the panel of
  Figs.~\ref{fig:poss_order_af} and~\ref{fig:poss_order_fm}, where a schematic sketch for the corresponding 
  ordered phase can be found.\label{tab:phases}}
 \begin{tabular}{|c|c|c|}\hline
spin: $q_1$ & orbital: $q_2$ & panel in Figs.~\ref{fig:poss_order_af}
or~\ref{fig:poss_order_fm}\\
\hline
$(\pi,0)$ & $(0,0)$&\ref{fig:x_pi0}, \ref{fig:xy_pi0}\\
$(\pi,\pi)$ & $(0,0)$&\ref{fig:xy_pipi}\\
\hline
$(\pi,0)$ & $(\pi,0)$&\ref{fig:xy_0pi_0pi}, \ref{fig:mp_0pi_0pi}\\
$(\pi,\pi)$ & $(\pi,\pi)$&\ref{fig:xy_pipi_pipi}\\
\hline
$(\pi,0)$ & $(0,\pi)$&\ref{fig:xy_0pi_pi0}\\
$(\pi,0)$ & $(\pi,\pi)$&\ref{fig:xy_0pi_pipi}\\
$(\pi,\pi)$ & $(\pi,0)$&\ref{fig:xy_pipi_0pi}\\
\hline
$(0,0)$ & $(0,0)$&\ref{fig:xy_fm_fo}\\
$(0,0)$ & $(\pi,0)$&\ref{fig:xy_fm_pi0}\\
$(0,0)$ & $(\pi,\pi)$&\ref{fig:xy_fm_pipi}\\
 \hline\end{tabular}
\end{table}

\begin{figure}
\subfigure[]{\includegraphics[width=0.155\textwidth,trim=110 60 110 60, clip]
{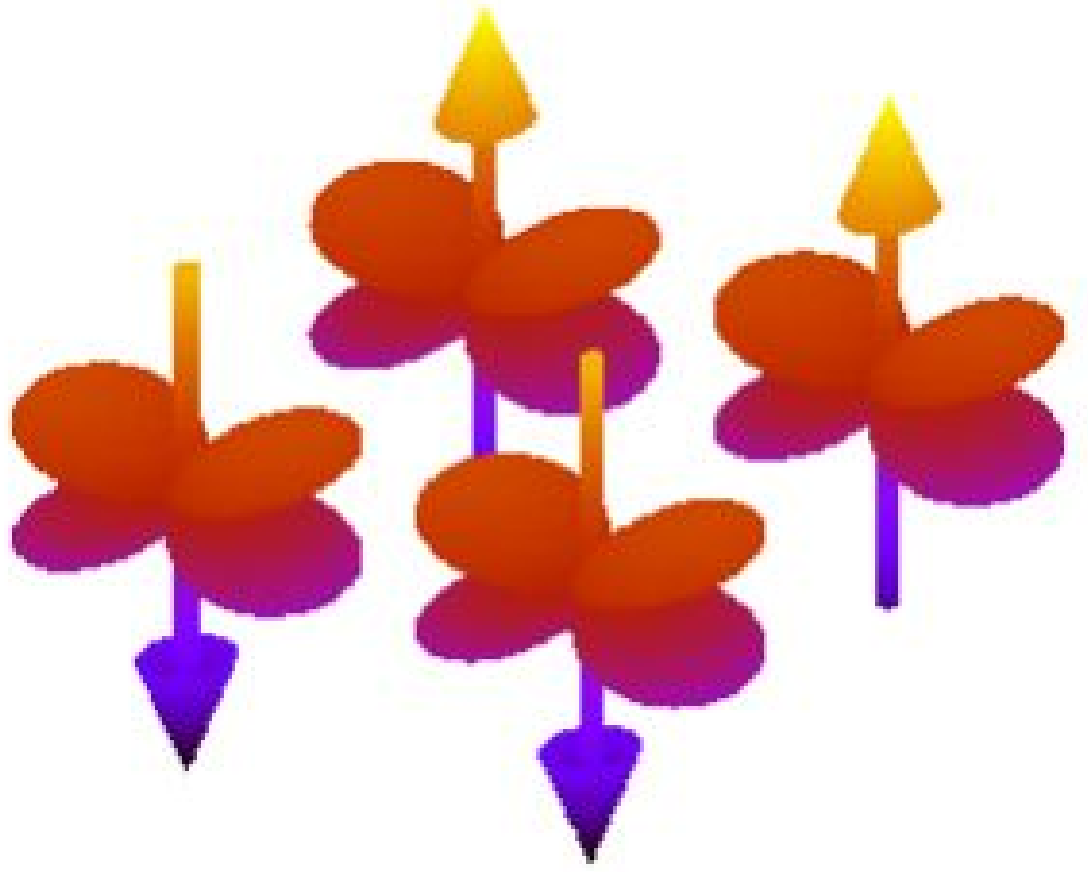}\label{fig:x_pi0}}
\subfigure[]{\includegraphics[width=0.155\textwidth,trim=110 60 110 60, clip]
{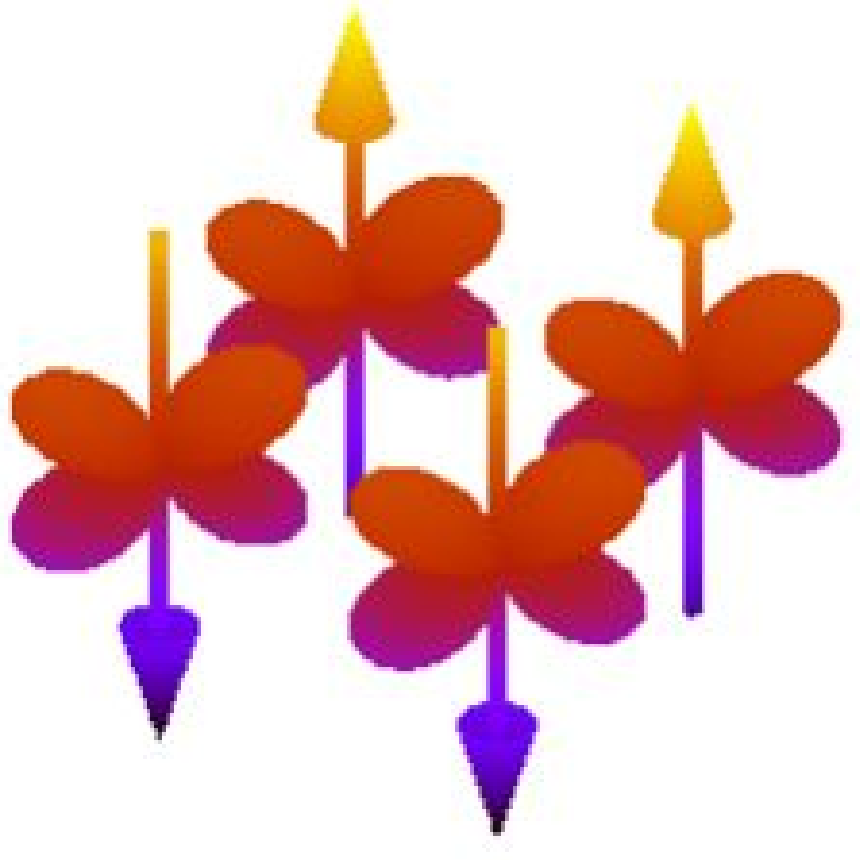}\label{fig:xy_pi0}}
\subfigure[]{\includegraphics[width=0.155\textwidth,trim=110 60 110 60, clip]
{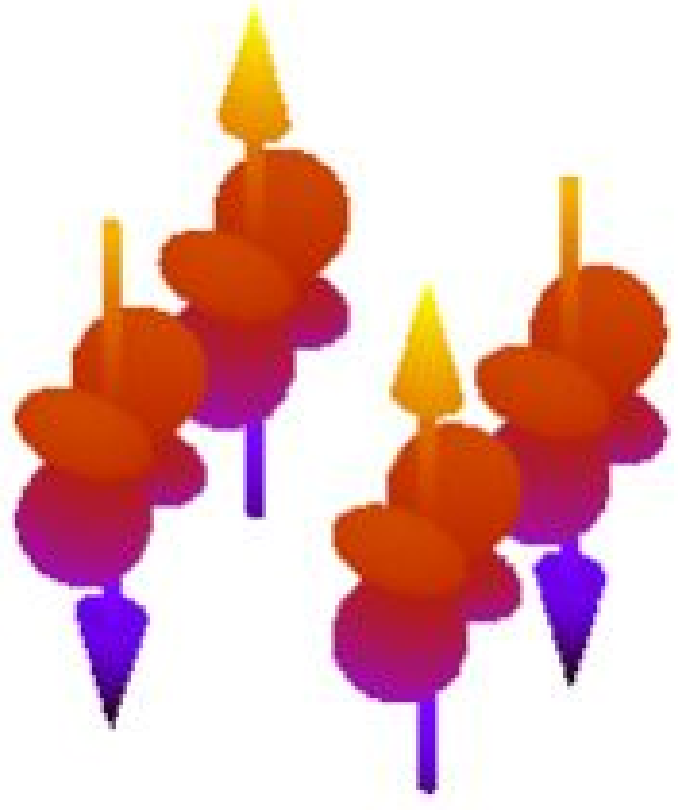}\label{fig:xy_pipi}}\\
\subfigure[]{\includegraphics[width=0.155\textwidth,trim=110 60 110 60, clip]
{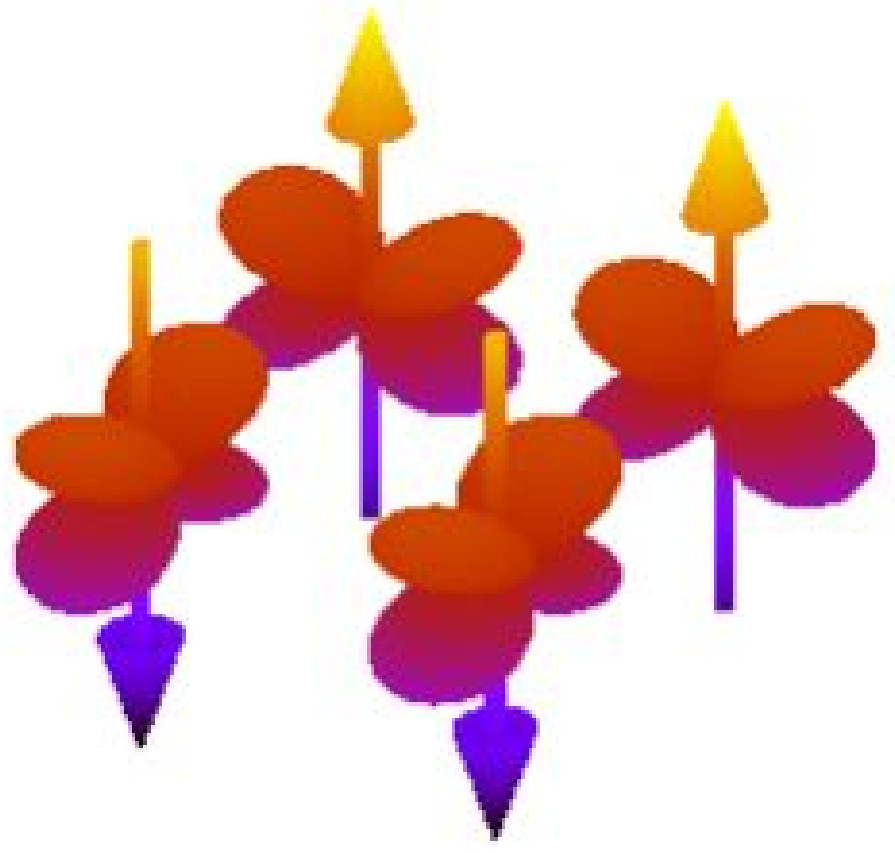}\label{fig:xy_0pi_0pi}}
\subfigure[]{\includegraphics[width=0.155\textwidth,trim=110 60 110 60, clip]
{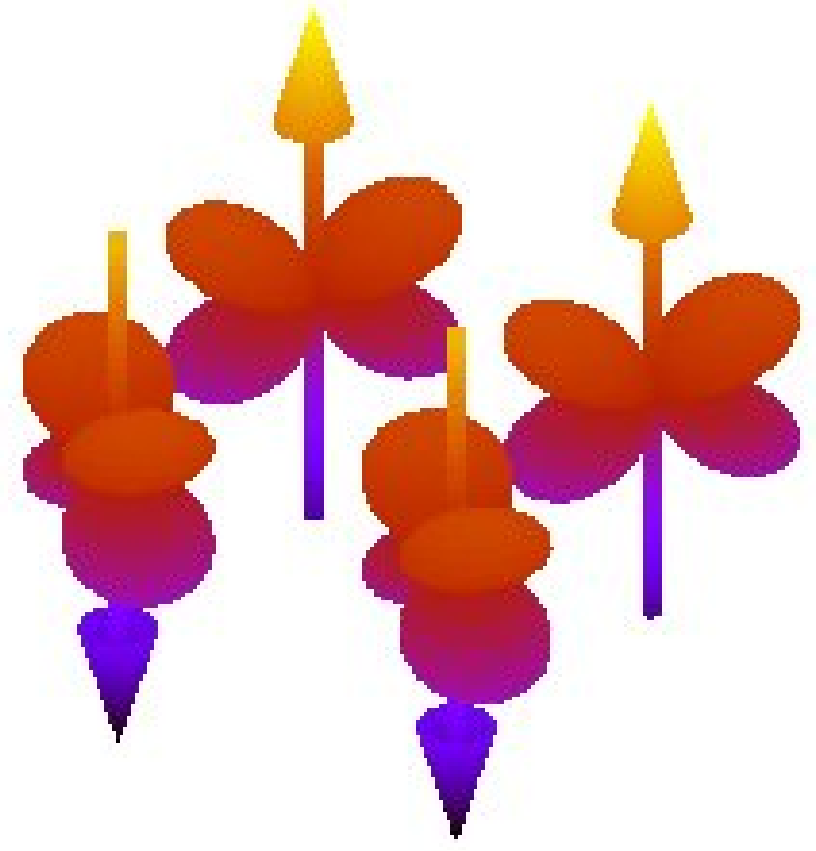}\label{fig:mp_0pi_0pi}}
\subfigure[]{\includegraphics[width=0.155\textwidth,trim=110 60 110 60, clip]
{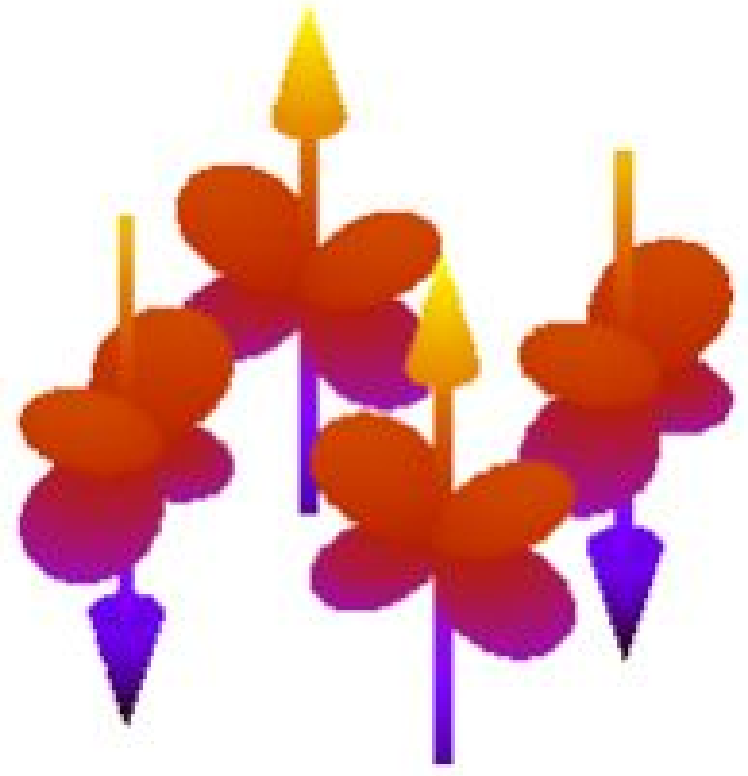}\label{fig:xy_pipi_pipi}}\\
\subfigure[]{\includegraphics[width=0.155\textwidth,trim=110 60 110 60, clip]
{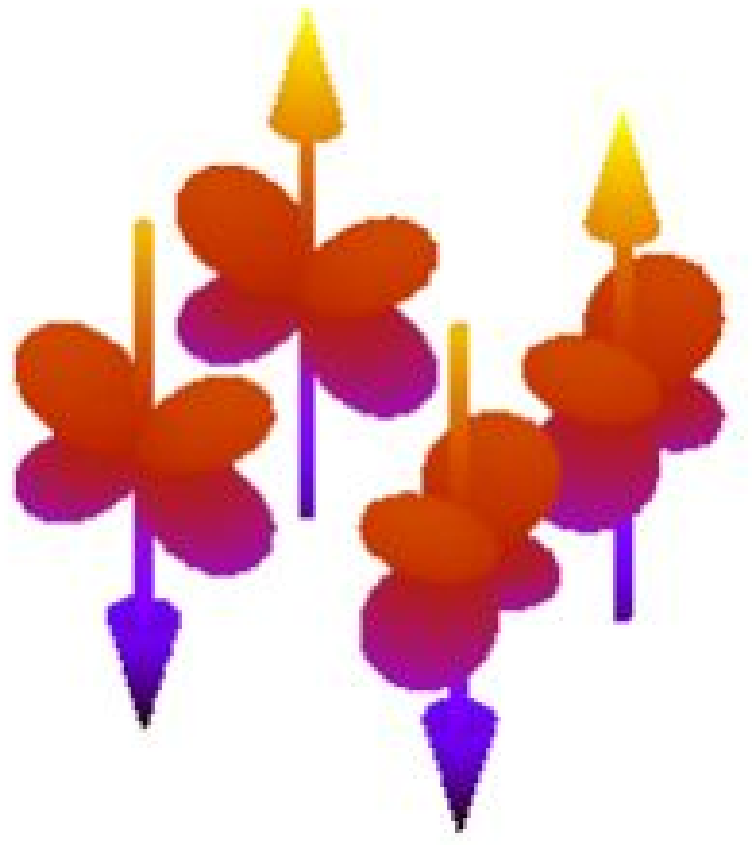}\label{fig:xy_0pi_pi0}}
\subfigure[]{\includegraphics[width=0.155\textwidth,trim=110 60 110 60, clip]
{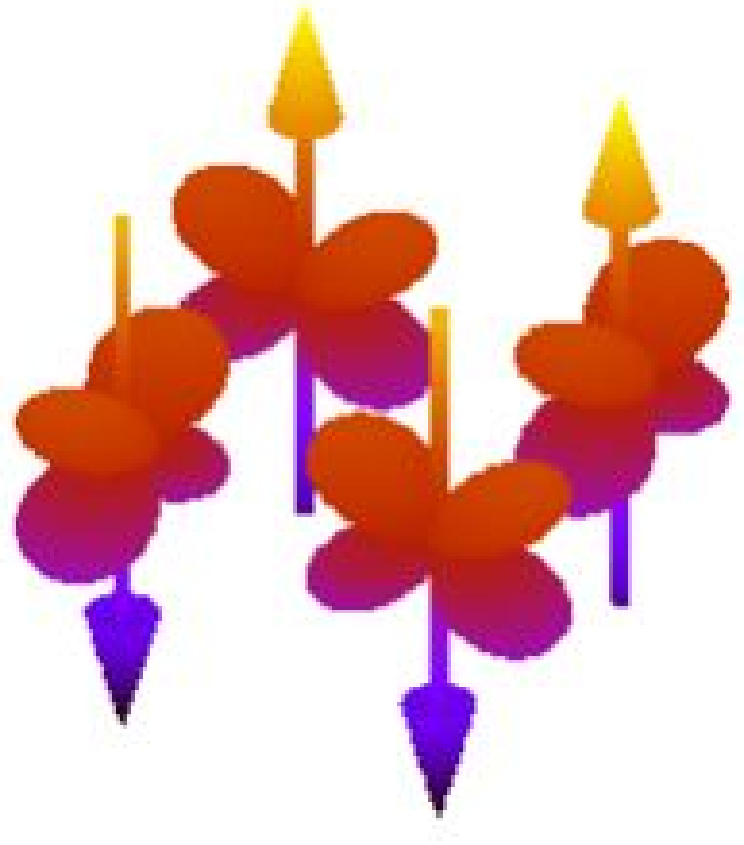}\label{fig:xy_0pi_pipi}}
\subfigure[]{\includegraphics[width=0.155\textwidth,trim=110 60 110 60, clip]
{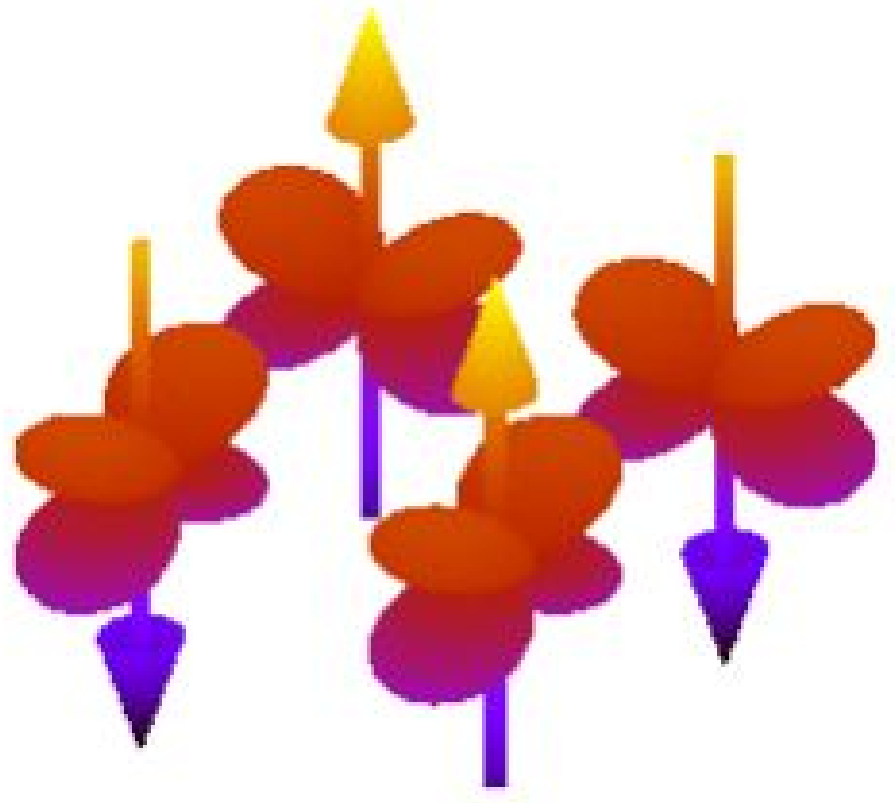}\label{fig:xy_pipi_0pi}}\\
\caption{(Color online) Cartoons representing the spin- and orbital-order 
  configurations considered in this effort. Since four electrons occupy
  three orbitals, perfectly ordered states have one doubly occupied
  orbital and the unpaired electrons are located in the other two orbitals, each 
  singly occupied, 
  forming a net spin $S=1$. In this cartoon, the orbital drawn
  indicates the doubly occupied one and the arrows indicate the orientation 
  of the total spin at that site. \subref{fig:x_pi0} $(\pi,0)$-spin
  and ferro-orbital (FO) order favoring the $yz$ orbital,
  \subref{fig:xy_pi0} $(\pi,0)$-spin and FO order favoring $(xz+yz)/\sqrt{2}$, 
  \subref{fig:xy_pipi} $(\pi,\pi)$-spin and FO order.  
  \subref{fig:xy_0pi_0pi}-\subref{fig:xy_pipi_pipi}: states with the same
  ordering vector for the spins and the orbitals:
  $(\pi,0)$ for \subref{fig:xy_0pi_0pi} and \subref{fig:mp_0pi_0pi},
  $(\pi,\pi)$ in \subref{fig:xy_pipi_pipi}.
  \subref{fig:xy_0pi_pi0} $(\pi,0)$ for spins and $(0,\pi)$ for orbitals; 
  \subref{fig:xy_0pi_pipi} $(\pi,0)$ for spins and $(\pi,\pi)$ for orbitals; 
  \subref{fig:xy_pipi_0pi} $(\pi,\pi)$ for spins and $(\pi,0)$ for orbitals; 
  \subref{fig:xy_pi0} and
  \subref{fig:mp_0pi_0pi} illustrate phases where the orbital order
  does not feature alternating $xz$ and $yz$ orbitals, but the combinations
  $(xz+yz)/\sqrt{2}$ and $(xz-yz)/\sqrt{2}$ do.\label{fig:poss_order_af}}
\end{figure}

\begin{figure}
\subfigure[]{\includegraphics[width=0.155\textwidth,trim=110 60 110 60, clip]
{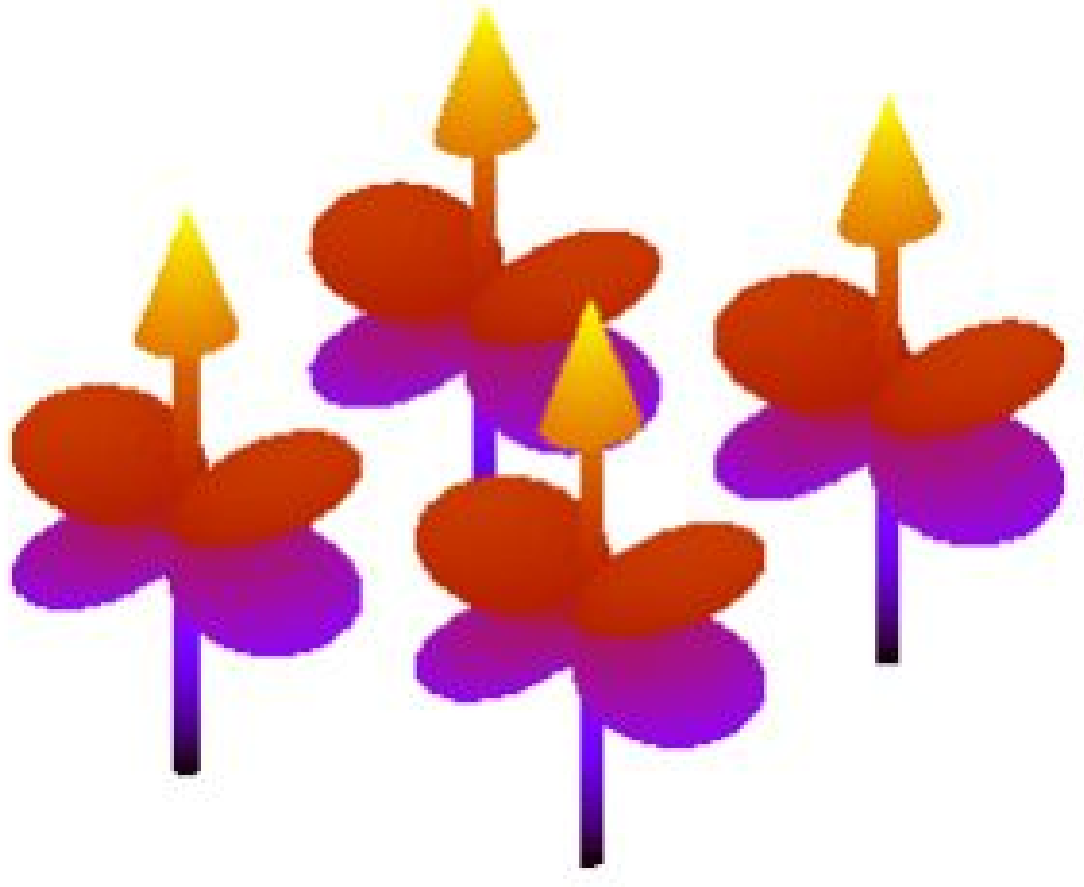}\label{fig:xy_fm_fo}}
\subfigure[]{\includegraphics[width=0.155\textwidth,trim=110 60 110 60, clip]
{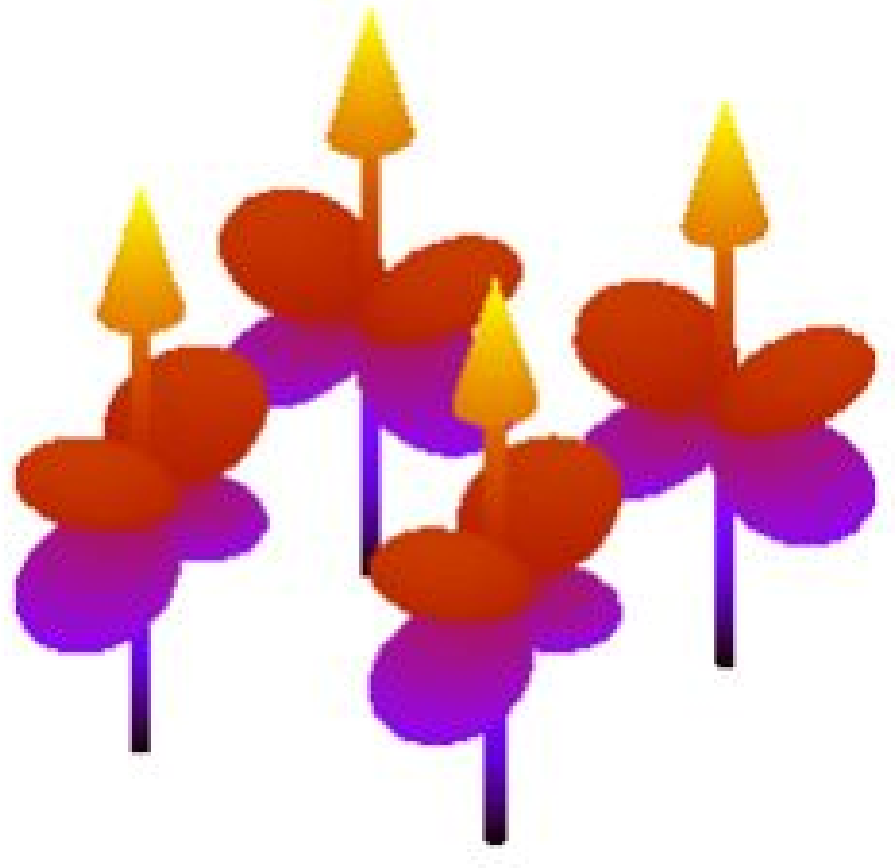}\label{fig:xy_fm_pi0}}
\subfigure[]{\includegraphics[width=0.155\textwidth,trim=110 60 110 60, clip]
{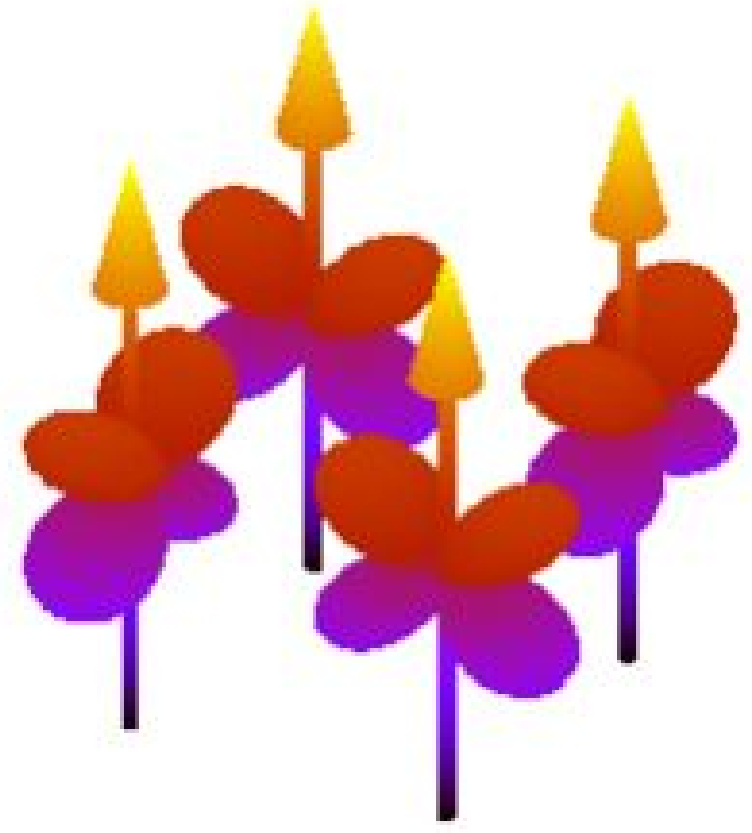}\label{fig:xy_fm_pipi}}\\
\caption{Color online) Cartoons for 
  the spin ferromagnetic phases taken into consideration in this effort. Results are depicted with the convention of
  Fig.~\ref{fig:poss_order_af}, but for ferromagnetic spins combined with \subref{fig:xy_fm_fo}
  FO, \subref{fig:xy_fm_pi0} $(\pi,0)$, and \subref{fig:xy_fm_pipi}
  $(\pi,\pi)$ orbital order. 
\label{fig:poss_order_fm}}
\end{figure}

To investigate the magnetic and orbital order properties of the three-orbital model
in the presence of the onsite Coulomb repulsion, here a mean-field approximation will be used. 
As it was carried out for the two- and
four-orbital models in Ref.~\onlinecite{yu}, here only the 
mean-field values for diagonal operators will be
considered~\cite{Nomura:2000p647} (for details see App.~\ref{app:mf}).
In the present
case, the degenerate $xz$ and $yz$ orbitals make up most of the weight
close to the chemical potential. In previous investigations of  the two-orbital $xz$-$yz$ model, very weak
orbital order has been found within a mean-field approximation at intermediate 
Coulomb repulsion in the half-filled case, and
none in the large-$U$ limit.~\cite{kubo} Numerical simulations did not 
provide indications of orbital order at half-filling either.~\cite{Daghofer:2009p1970} 
However, the $xz$ and $yz$ orbitals are not expected to be half filled in 
the present case: In the noninteracting $U=0$ limit, the $xy$ orbital
is approximately half filled, so that the subsystem consisting of the two degenerate
$xz$ and $yz$ orbitals turns out to be approximately $3/4$ filled. Consequently, orbital 
order within the $xz$/$yz$-subspace could now occur,
analogously to the case of quarter filling in the $xz$/$yz$
subsystem.~\cite{zaanen} 

Three possible orbital-order patterns will be considered: {\it (i)} Ferro-orbital (FO) order
which corresponds to the orbitals $xz$ and $yz$ having different electronic densities,  {\it (ii)} alternating orbital (AO) order, and {\it (iii)} stripe orbital (SO) order.
Combined with the magnetic spin order, these orbital orders lead to a 
large variety of possible combinations of polarized or alternating spin and 
orbital order.~\cite{note_GK} Here, phases that can be expressed using
(at most) two ordering vectors have been considered, i.e. ${\bf q}_1$ for magnetic order and
${\bf q}_2$ for orbital order. The expectation values for the mean-field proposed states
can then be expressed as
\begin{align}
\label{eq:mf_xy}
\langle n_{{\bf r},xy,\sigma}\rangle &= n_{xy} + \frac{\sigma}{2}
\textrm{e}^{i{\bf q}_1.{\bf r}} m_{xy}\\
\langle n_{{\bf r},\alpha,\sigma}\rangle &= n + \frac{\sigma}{2}
\textrm{e}^{i{\bf q}_1.{\bf r}} m+
\frac{\alpha}{2} \textrm{e}^{i{\bf q}_2.{\bf r}} p + \frac{\sigma\alpha}{2}
\textrm{e}^{i({\bf q}_1+{\bf q}_2.){\bf r}} q\;,
\label{eq:mf_xz_yz}
\end{align}
where the first equation with the mean-field
parameters $n_{xy}, m_{xy}$ describes the $xy$ orbital and the second
equation with parameters $n, m, p$, and $q$ applies to the $xz$/$yz$ 
subsystem, with $\alpha=\pm 1$ indicating the $xz$/$yz$ orbitals.

The fact that the $xz$/$yz$ space is not $SU(2)$ symmetric introduces
another degree of freedom in addition to the ordering vector:
Ferro-orbital order (i.e., site-independent orbital densities) can
favor either the $xz$ or $yz$ orbitals [as shown in Fig.~\ref{fig:x_pi0}], 
or any linear
combination $|\phi\rangle = \cos\phi |xz\rangle + \sin\phi
|yz\rangle$. As an example, Fig.~\ref{fig:xy_pi0} illustrates a state
with $\phi = \pi/4$ corresponding to a symmetric combination
$(|xz\rangle+|yz\rangle)/\sqrt{2}$. The same holds for
alternating orbital order: alternating order corresponding to $\phi =
\pi/4$ can be seen in Fig.~\ref{fig:mp_0pi_0pi}. Consequently,
mean-field calculations were performed for several values of $\phi$
for each phase.
The $xy$ orbital might be similarly involved in such a
linear combination, because the crystal-field splitting separating it
from the $xz$ and $yz$ orbitals is not very large. However, Exact Diagonalization
of $2\times 2$ clusters did not give any indication for such
behavior: the dominant states in the low-energy eigenstates that we analyzed involved
almost exclusively singly occupied or unoccupied $xy$ orbitals. Moreover, the $xy$ orbital is almost singly occupied at $U=0$ and has
low weight at the Fermi energy, so that ordering phenomena in the most
relevant intermediate-$U$ regime may be expected to involve mostly
$xz$ and $yz$ orbitals. The states considered are shown in 
Figs.~\ref{fig:poss_order_af} and \ref{fig:poss_order_fm} and the corresponding
values of ${\bf q}_i$ are given in Tab.~\ref{tab:phases}. Which
spin-orbital pattern is stabilized depends on the interaction
parameters $U$ and $J$, as described in Sec.~\ref{sec:phases}. The phase with
the largest stability range, and which is stable for the most
realistic parameter choices, is the $(\pi,0)$-AF orbital disordered (OD) phase, 
which will be discussed in more detail in Sec.~\ref{sec:mf}. 

\subsection{Magnetic and Orbital Orders in the Undoped Regime}
\label{sec:phases}

\begin{figure}
\includegraphics[width=0.45\textwidth]{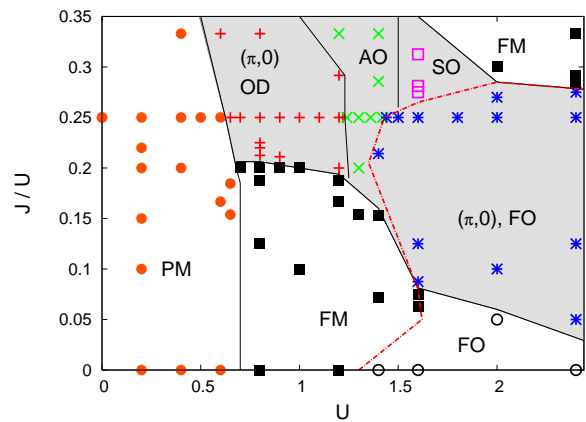}
\caption{(Color online) Qualitative phase diagram in the $J/U$ vs. $U$
  plane. The shaded area denotes the stability region of the realistic $(\pi,0)$
  magnetic ordering. The lines are guides to the eye, the dashed line
  approximately indicates  the metal-insulator transition. The data
  points were obtained by comparing the energies of the various phases
  within the mean-field approximation described in
  App.~\ref{app:mf}. The meaning of the symbols is the following: 
$+$: $(\pi,0)$-AF state without pronounced
  orbital order (orbital disorder, OD); $\times$: $(\pi,0)$-AF state with $(\pi,\pi)$
  alternating orbital (AO) order with $\phi=0$; empty squares:
  $(\pi,0)$-AF state with $(0,\pi)$ orbital stripes (SO) ($\phi=0$); $\ast$:
  $(\pi,0)$-AF state with ferro-orbital (FO) order ($\phi=0$); filled
  squares: FM with FO ordering tendencies, $\phi=\pi/4$. This FO order
  is weaker for small $U$ and larger $J\approx 0.2$. The filled circles denote parameters
  that do not support magnetically ordered states. For small $J$, some
  FO order with $\phi=\pi/4$ is found, similar to the FM phase. The
  empty circles at large $U$ and small $J$ denote similar states
  without magnetic ordering, but with extreme orbital order, where the
  $xy$ orbital is (almost) empty, while $xz$ and $yz$ are (almost) filled.
\label{cartoon}}
\end{figure}

Magnetic order with wavevector ${\bf q}_1=(\pi,0)$ (or $(0,\pi)$) and OD was 
found to be 
stable in a broad - and
especially the most realistic - range of interaction parameters. However, the
phase diagram in the $J/U$ vs. $U$ plane turns out to contain a large
variety of metallic disordered phases, metallic phases with different kinds of magnetic
and/or orbital order, and insulating magnetically and orbitally ordered 
regions. Figure~\ref{cartoon} shows a qualitative rendition of the
resulting phase diagram. A more accurate quantitative determination of the boundaries
as well as a  detailed description of all phases will be 
presented in a future publication. For a realistic Hund's rule
coupling $J=U/4$, we discuss the properties of the ground
state in the different regimes encountered varying $U$ in
Sec.~\ref{sec:mf} below.

It has been found that at large interaction strengths $U$, after the magnetic 
order with ${\bf q}_1=(\pi,0)$ is established, alternating- 
[${\bf q}_2=(\pi,\pi)$] and ferro-orbital [${\bf q}_2=(0,0)$] order develops
for $1/5\le J/U\le 1/3$. For very small values of $J/U$, ferromagnetic order
becomes stable instead of $(\pi,0)$-antiferromagnetism, and the strongest 
competitor in the limit $J\to 0$
is ${\bf q}_1=(\pi,\pi)$ antiferromagnetism. At $J=0$ and for small
$3/4\lesssim U\lesssim 1.5$, both the FM and the $(\pi,\pi)$-AF state have pronounced orbital order 
corresponding to $\phi=\pi/4$: One of the orbitals given by the linear 
combinations $(|xz\rangle \pm |yz\rangle)/\sqrt{2}$ is almost filled, 
the other contains $\approx 1.5$ electrons, and $xy$ the remaining
$0.5$. In the the FM state, $(|xz\rangle + |yz\rangle)/\sqrt{2}$ is
the almost filled orbital, while $(|xz\rangle - |yz\rangle)/\sqrt{2}$
has higher occupancy in the $(\pi,\pi)$-AF state with only slightly
higher energy. At larger $U > 1.5$, both $xz$ and $yz$ are almost
filled for $J\to0$, $xy$ is nearly empty, and there are hardly any
unpaired spins that could support magnetically ordered phases. Notice that 
such small ratios of $J/U$ are not expected to be realistic for the pnictides, 
where the onsite Coulomb repulsion $U$ is strongly 
screened.\cite{jeroen_small_U} 
With growing $J/U$, the magnetic ordering vector
switches to $(\pi,0)$ at $J/U\approx 1/5$ for small $U$, and even
smaller $J/U$ for larger $U$.   

For $U\lesssim 2$, the $(\pi,0)$ magnetic order remains stable
for all values of $J/U\gtrsim 1/5$ up to $J/U=1/3$. At this ratio, the Hund's  
coupling is such that the inter-orbital repulsion felt by two electrons in different
orbitals, but on the same site, vanishes, and we therefore did not
consider $J/U>1/3$. 


\subsection{Evolution of the ground state as a function of $U$ for 
$J= U/4$}
\label{sec:mf}

\begin{figure}
\subfigure{\includegraphics[width=0.4\textwidth]{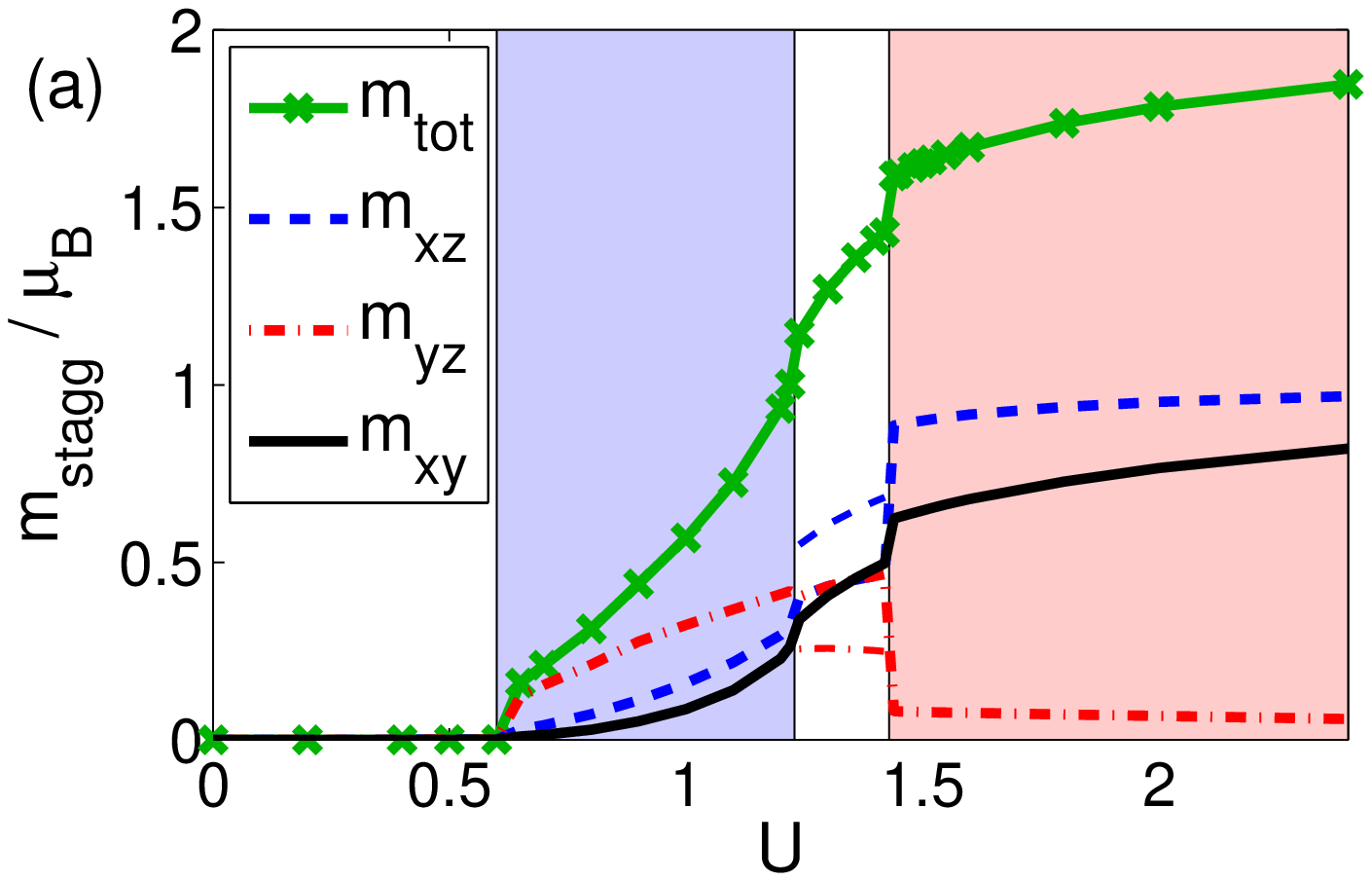}\label{fig:U_mag}}\\[-1.5em]
\subfigure{\includegraphics[width=0.4\textwidth]{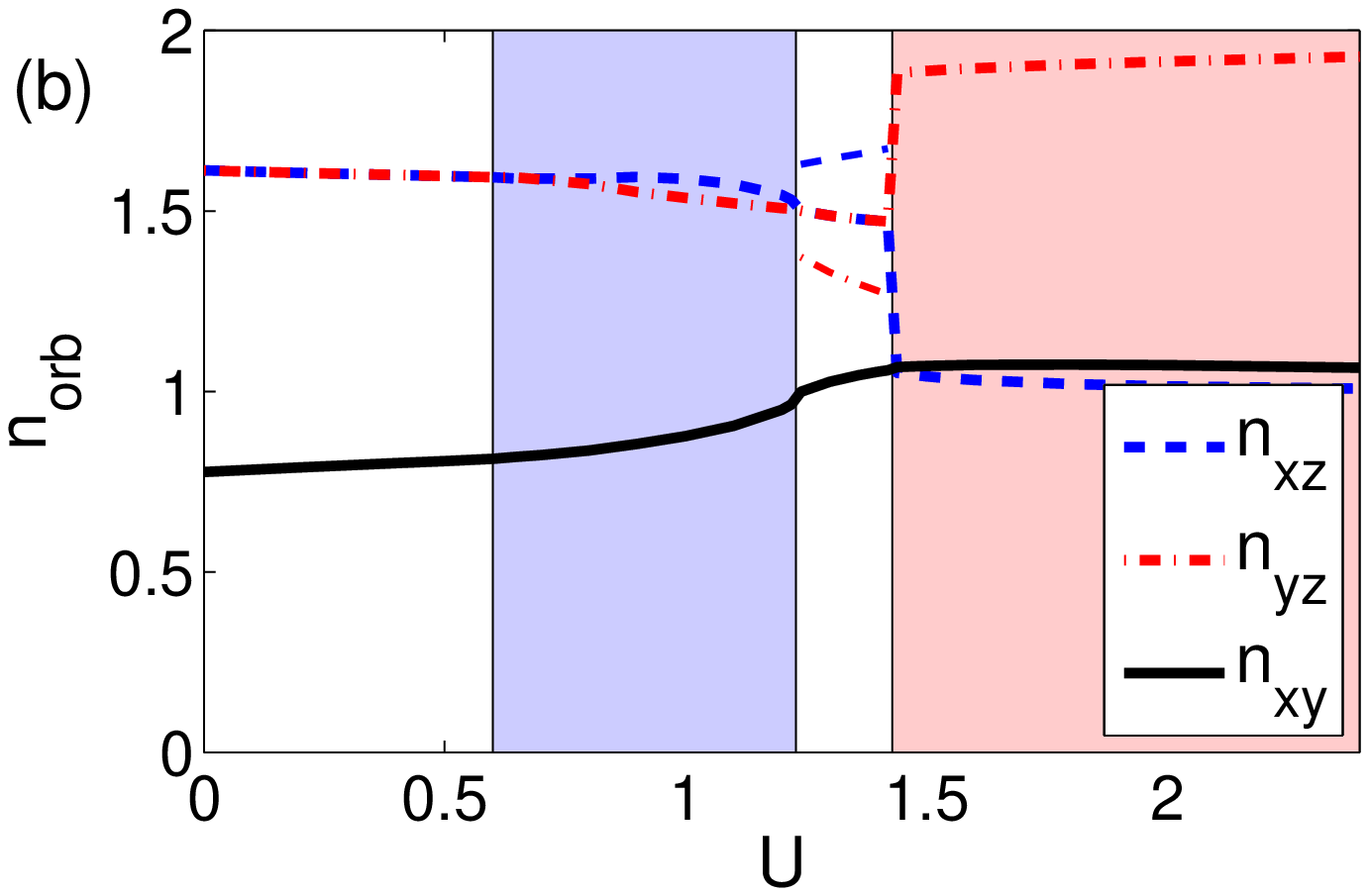}\label{fig:U_n}}
\caption{(Color online) (a) Orbital magnetization and (b) occupation number 
  as a function of the Coulomb repulsion strength $U$, obtained with a 
  mean-field
  approximation. The colors indicate the different phases (for increasing $U$):
  uncorrelated metal, itinerant $(\pi,0)$ antiferromagnet without orbital order,
  itinerant $(\pi,0)$ antiferromagnet with alternating orbital order [small
  white window, spin-orbital order as in Fig.~\ref{fig:xy_0pi_pipi}],
  and a ferro-orbitally-ordered $(\pi,0)$ antiferromagnetic insulator
  [spin-orbital order as in Fig.~\ref{fig:x_pi0}].  
  Hopping parameters are from Tab.~\ref{tab:hopp3}, and $J=U/4$. For
  the phase with alternating orbital order, the thin lines show (a)
  $m\pm q$ and (b) $2n\pm p$.
\label{fig:U_n_mag}}
\end{figure}

\begin{figure}
\subfigure{\includegraphics[width=0.47\textwidth,trim = 15 20 20 45,clip]
 {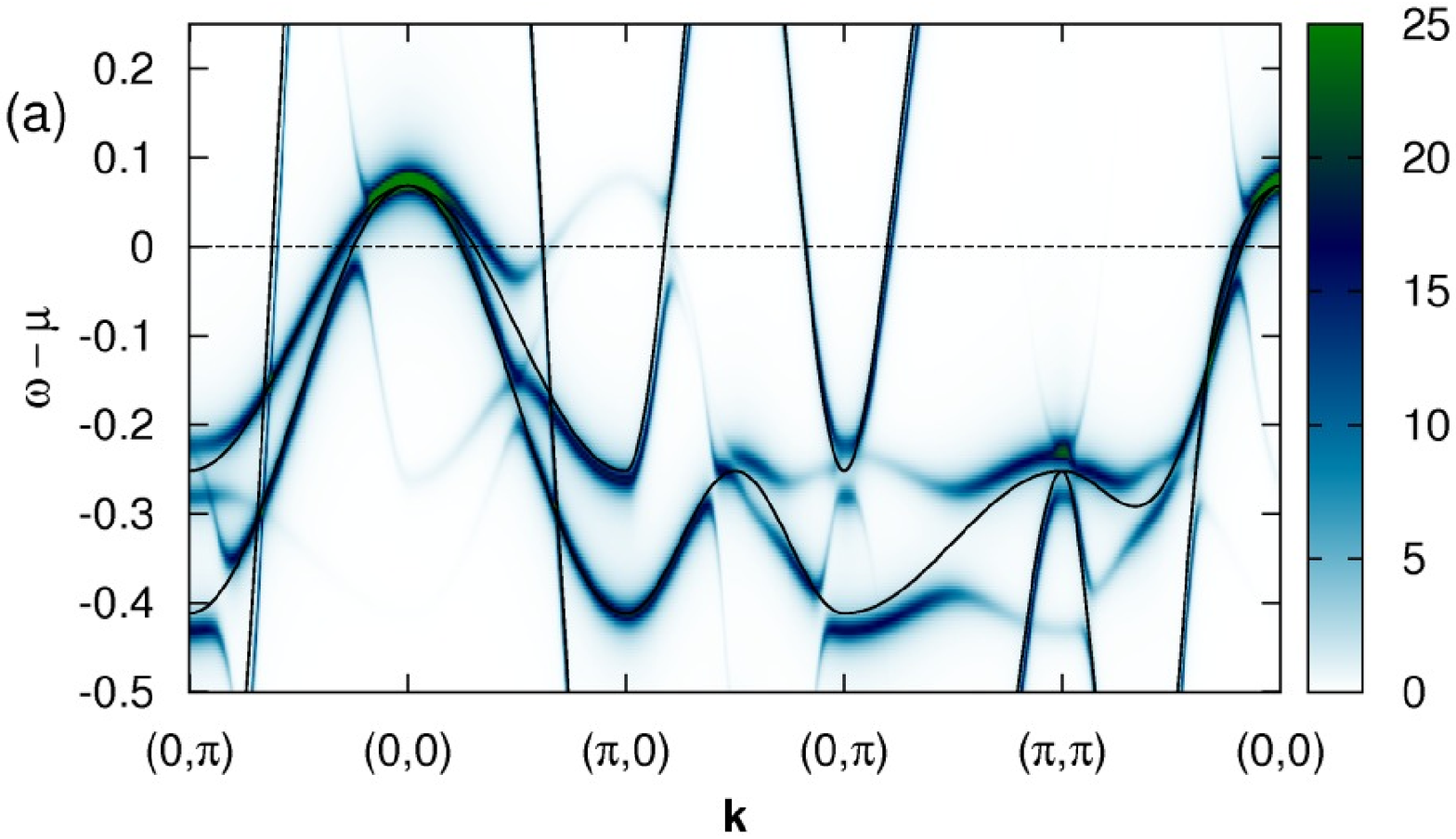}\label{fig:Ak_U35}}\\[-0.5em]
\subfigure{\includegraphics[width=0.47\textwidth,trim = 15 20 20 45,clip]
 {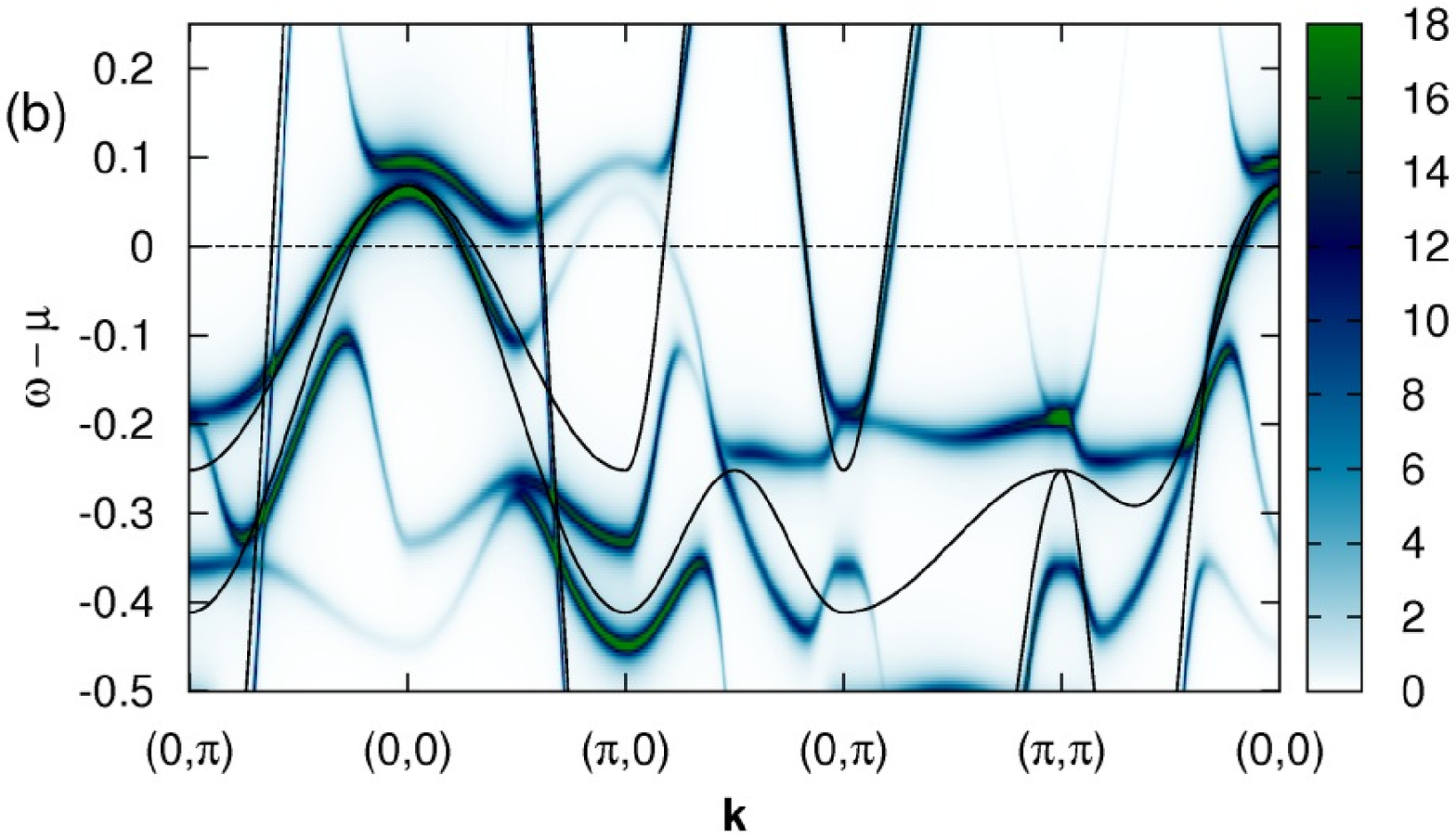}\label{fig:Ak_U45}}\\[-0.5em]
\subfigure{\includegraphics[width=0.47\textwidth,trim = 15 20 20 45,clip]
 {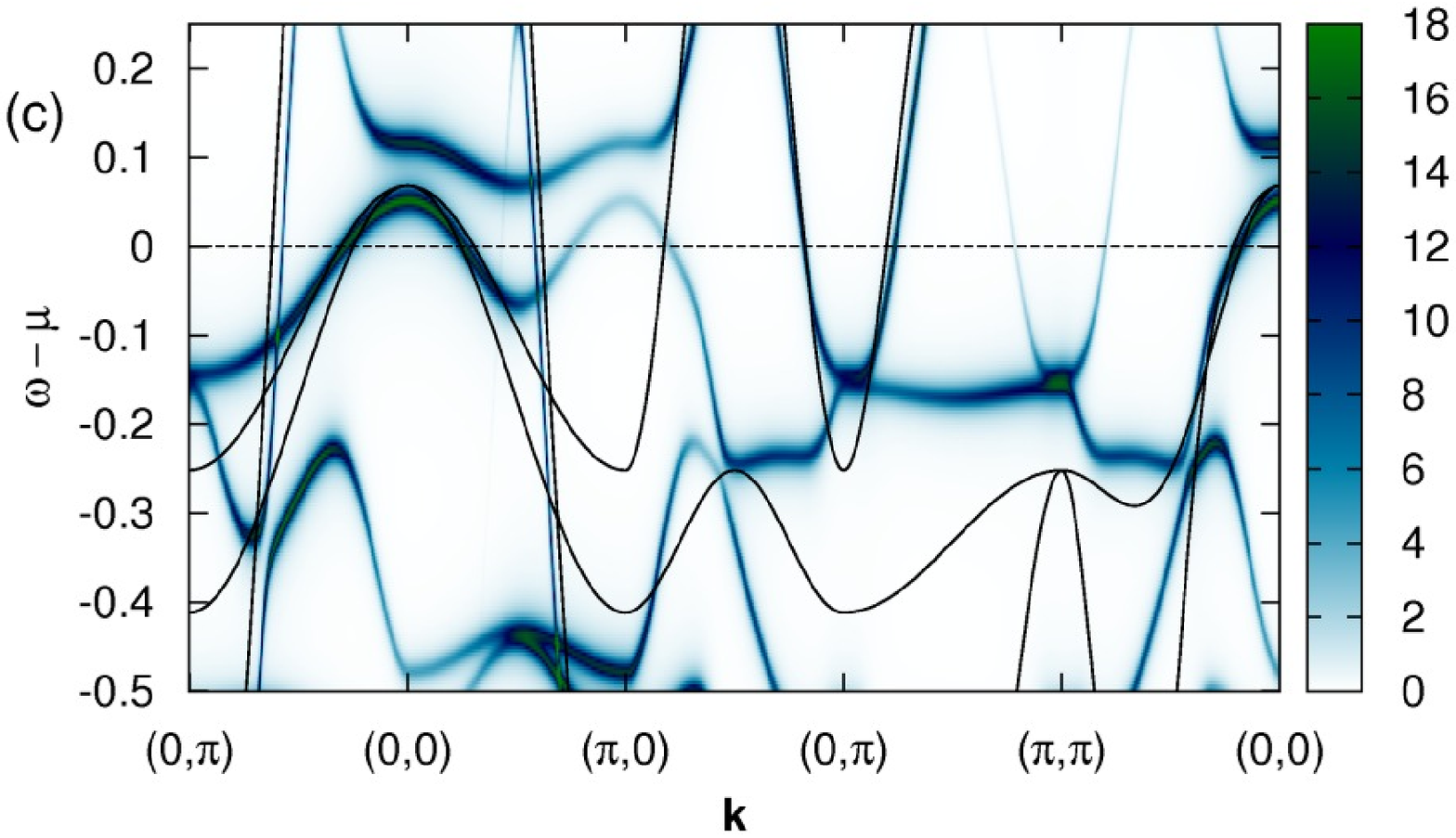}\label{fig:Ak_U55}}\\[-0.5em]
\caption{(Color online) Spectral density $A({\bf k},\omega)$ for the 
antiferromagnet orbital-disordered metallic phase 
at (a) $U=0.7$, (b) $U=0.9$, and (c) $U=1.1$. The BZ is for the one-Fe unit
  cell, and $J=U/4$ was used. The uncorrelated $A({\bf k},\omega)$ for $U=0$ is included as
  solid lines for comparison.\label{fig:Ak_AF}}
\end{figure}

In this section,  the ground state properties at a fixed
ratio  $J/U=1/4$ will be discussed. Around this value of $J/U$ is where we 
have found the realistic AF order with ordering momentum ${\bf q}_1=(\pi,0)$ for all values of 
$U>U_{c_1}$. Figure~\ref{fig:U_mag} shows
how the staggered magnetization with ordering momentum $(\pi,0)$
increases with the Coulomb repulsion $U$. As previously found for two-
and four-orbital models,~\cite{yu} intermediate $U$ leads to an
antiferromagnetic metal. The system remains nonmagnetic for small $U$ up 
to $U_{c_1}\approx
0.6$. For $U>U_{c_1}$, the spin $(\pi,0)$-ordered magnetic moment starts to
grow, see Fig.~\ref{fig:U_mag}, but the band structure remains metallic, as 
can be deduced from the spectral functions presented in
Fig.~\ref{fig:Ak_AF}, calculated for several values of the Coulomb repulsion 
in this regime. 
Note that the spin-$(\pi,0)$ AF order triggered by $U$ introduces gaps and 
magnetically-induced ``shadow''
bands;\cite{haas} for comparison, the
uncorrelated $A({\bf k},\omega)$ is 
included in Fig.~\ref{fig:Ak_AF}.
Since several
bands are involved, the gaps are not necessarily located at the
chemical potential, as it has been discussed for the two- and four-orbital 
models in
Ref.~\onlinecite{yu}, and the system remains metallic. As it can be seen
in Fig.~\ref{fig:Ak_AF}, the overall features of the spectral density
in this regime remain similar to those of the noninteracting limit, with the
bandwidth being slightly reduced with increasing $U$. However, the onset of AF
order does affect some details of $A({\bf k},\omega)$, especially
low-energy features at the chemical potential, where one of the
hole pockets disappears and additional pockets arise.

\begin{figure}
\subfigure{\includegraphics[width=0.23\textwidth, trim = 70 10 110.5 50,
  clip]{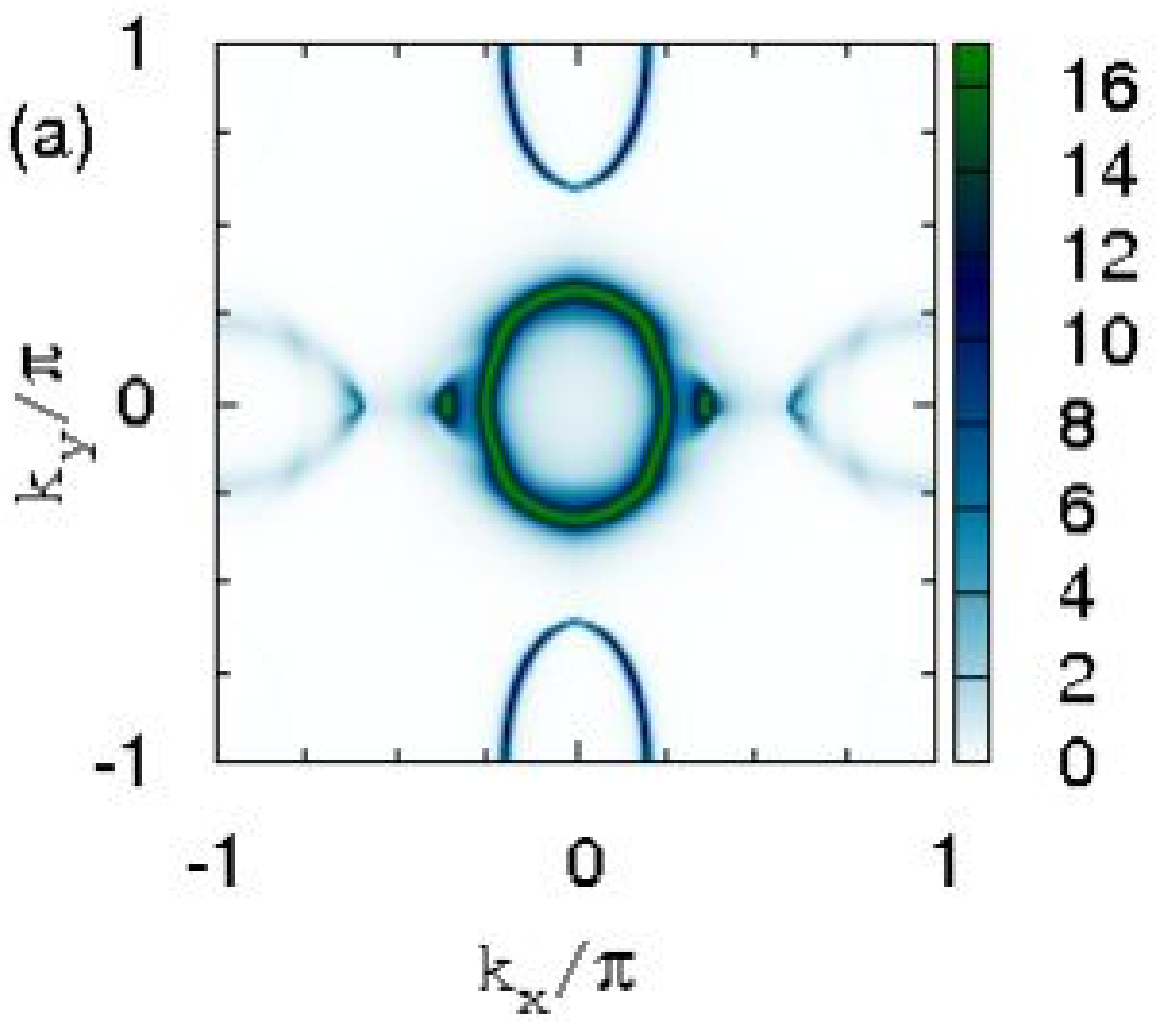}\label{fig:fs_U35_u}}
\subfigure{\includegraphics[width=0.23\textwidth, trim = 70 10 110.5 50,
  clip]{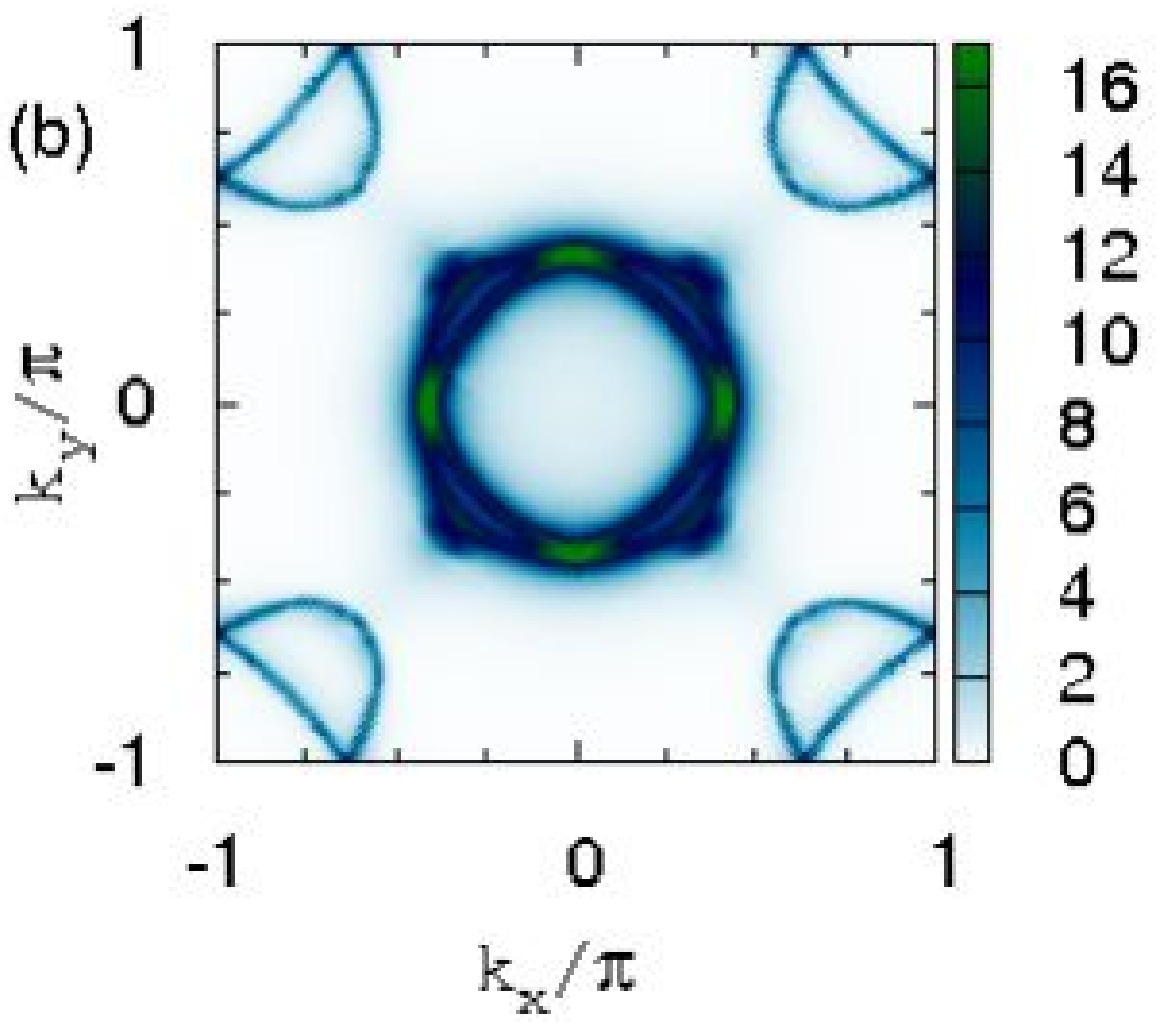}\label{fig:fs_U35fr}}
\subfigure{\includegraphics[width=0.23\textwidth, trim = 70 10 110.5 50,
   clip]{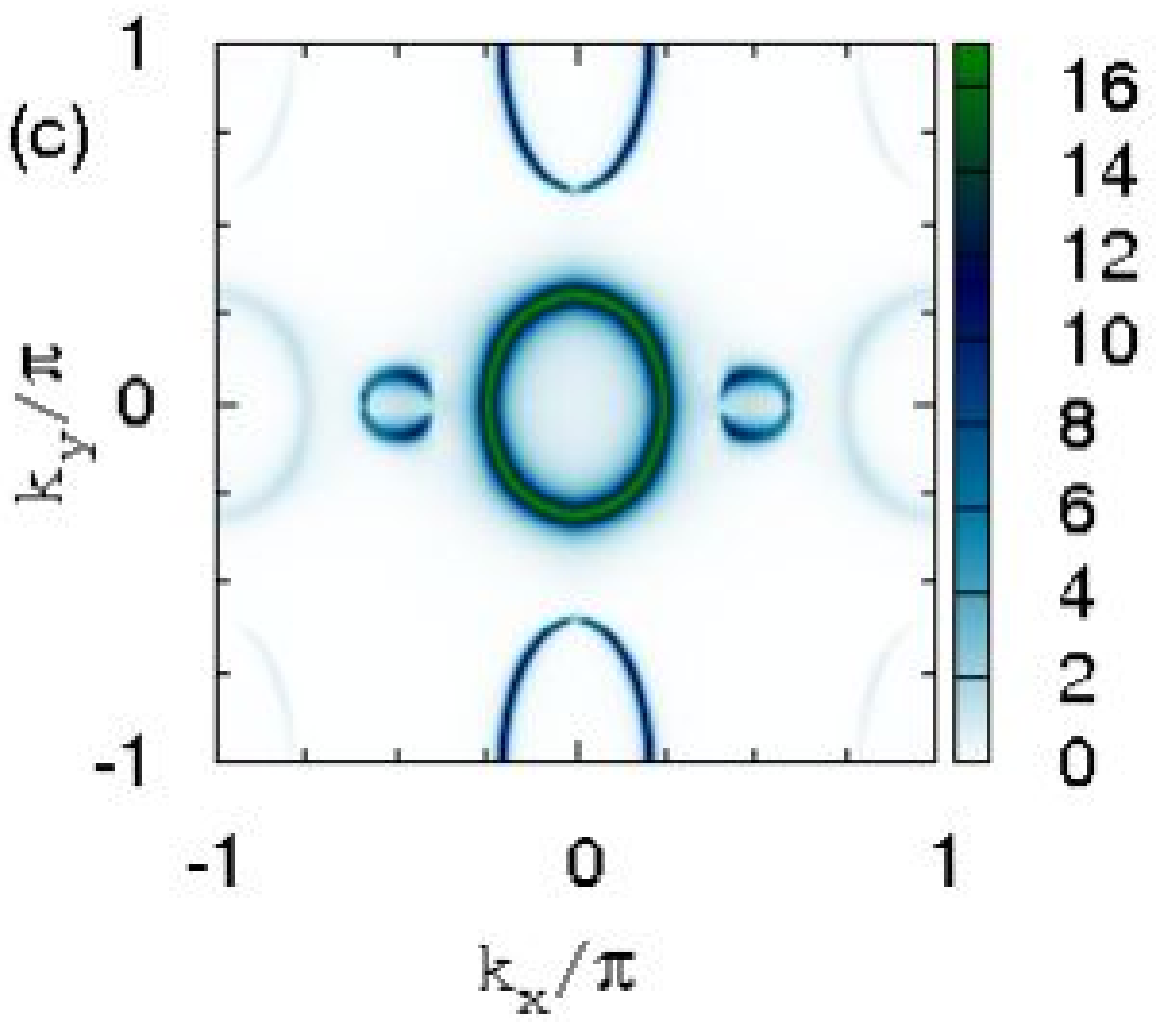}\label{fig:fs_U45_u}}
\subfigure{\includegraphics[width=0.23\textwidth, trim = 70 10 110.5 50,
   clip]{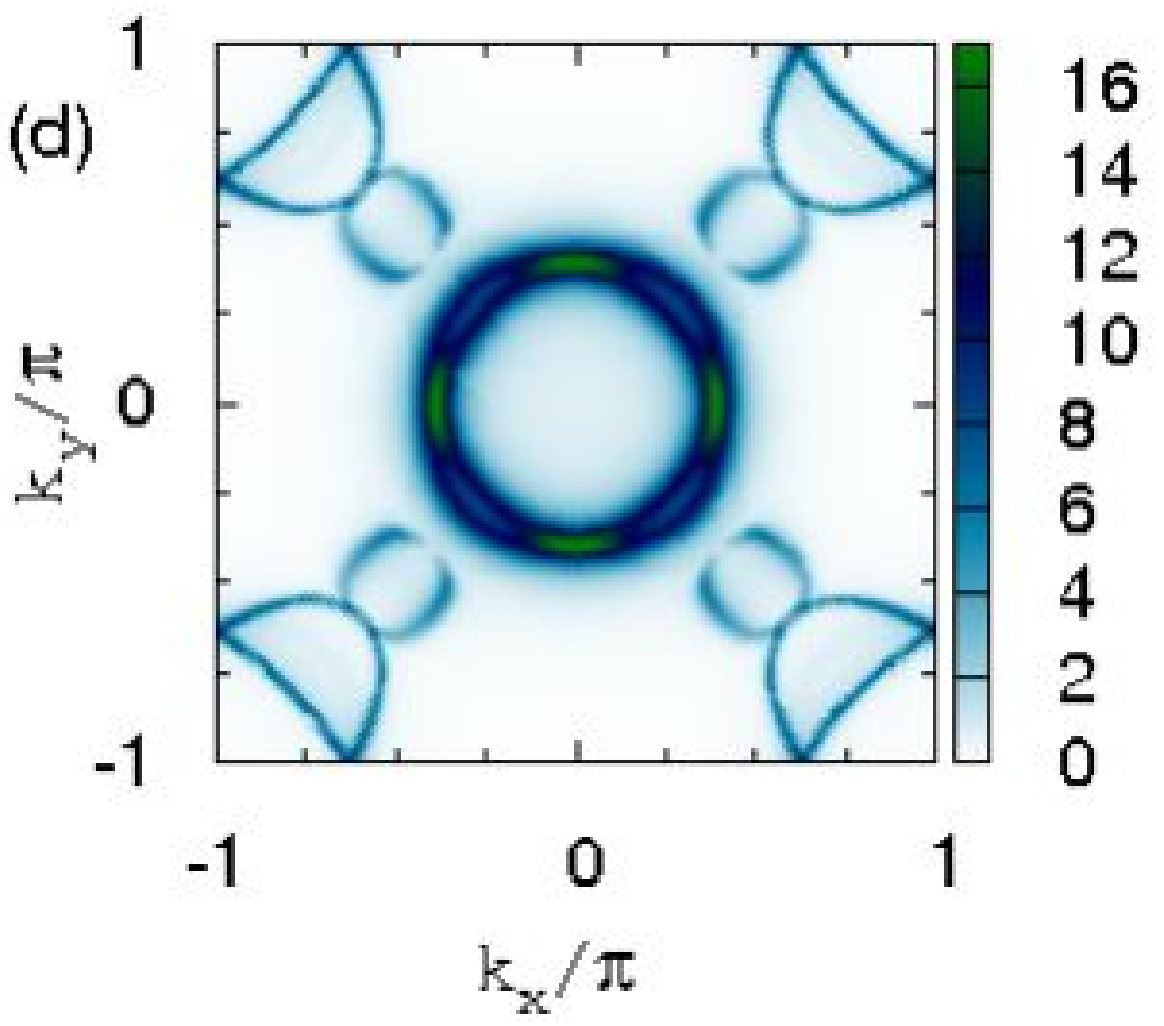}\label{fig:fs_U45fr}}
\subfigure{\includegraphics[width=0.23\textwidth, trim = 70 10 110.5 50,
   clip]{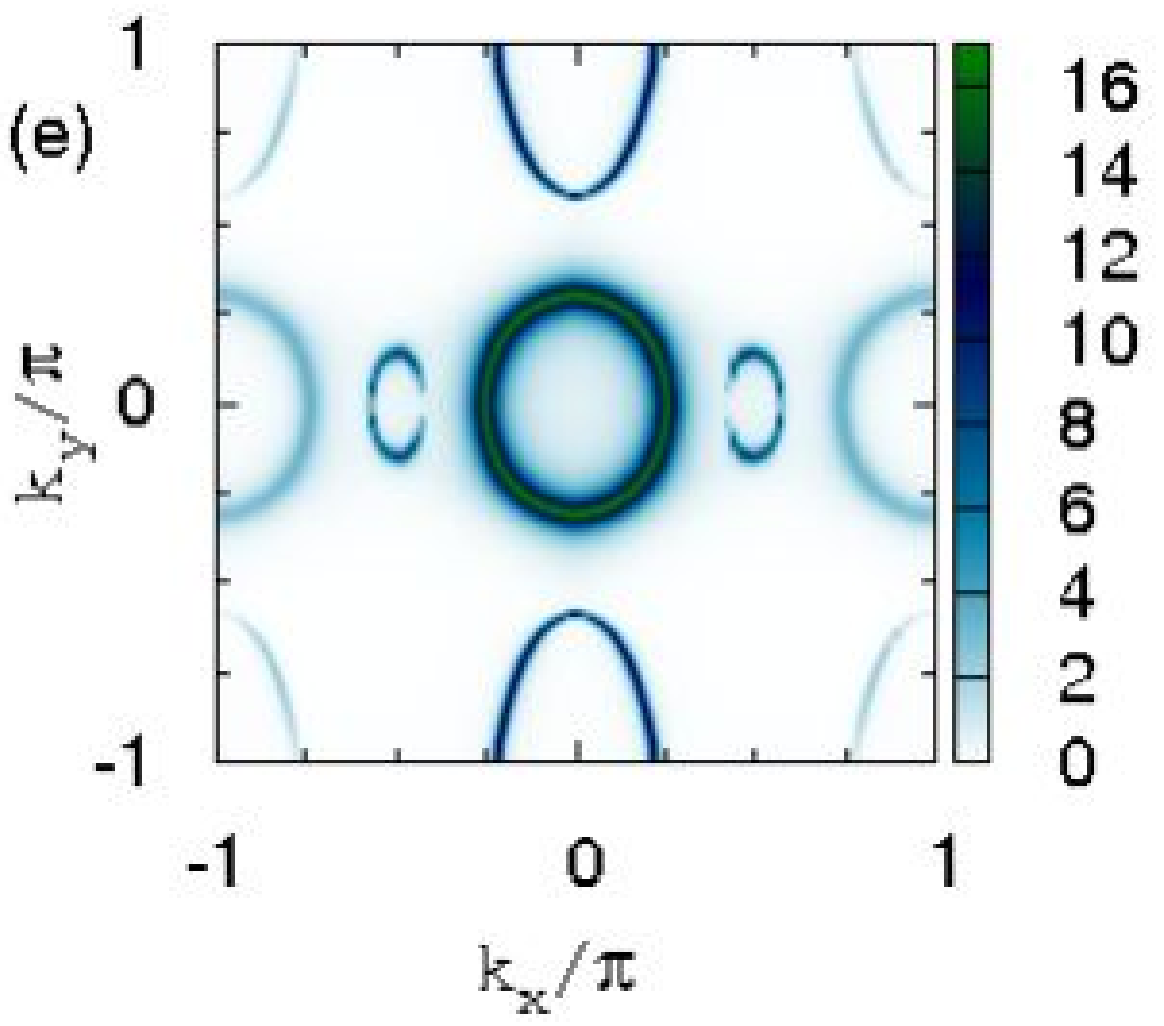}\label{fig:fs_U55_u}}
\subfigure{\includegraphics[width=0.23\textwidth, trim = 70 10 110.5 50,
   clip]{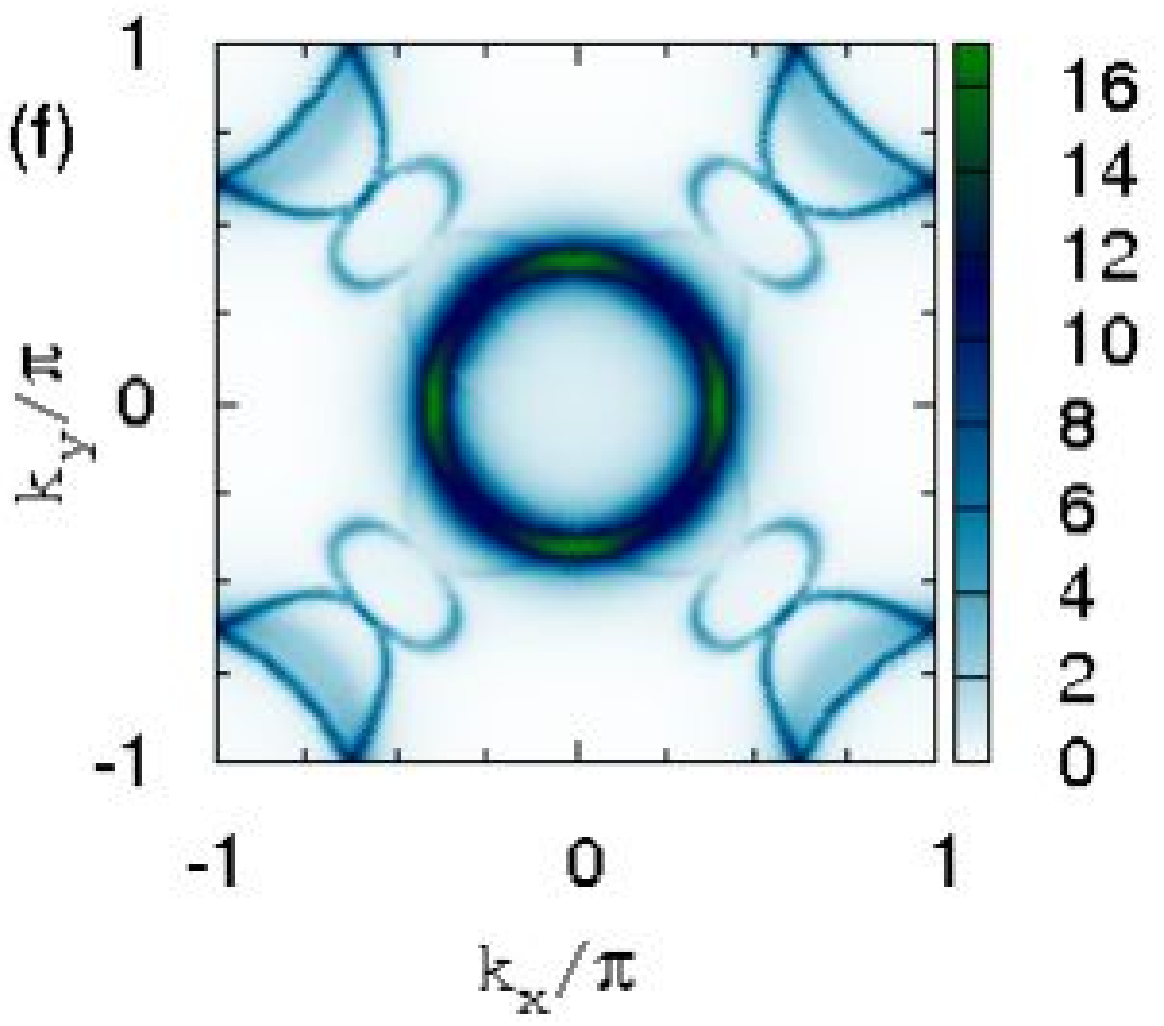}\label{fig:fs_U55fr}}
\caption{(Color online) Fermi surface in the orbital-disordered 
spin-antiferromagnetic metallic phase
  with (a,b) $U=0.7$, (c,d) $U=0.9$, and (e,f) $U=1.1$. (a,c,e) show the
  unfolded BZ containing one Fe, for the antiferromagnetic ordering
  vector $q=(\pi,0)$.  
  (b,d,f) depict the superposition of the FSs for  ${\bf q}=(\pi,0)$
  and ${\bf q}=(0,\pi)$ in the (rotated) folded BZ corresponding to
  two Fe atoms. The ratio 
$J=U/4$ was used. The color scale is the same as in Fig.~\ref{fig:Ak_AF}.
\label{fig:fs_U3545}} 
\end{figure}

The Fermi surface for $U=0.7$, where the Coulomb repulsion is just barely 
strong enough to induce $(\pi,0)$
antiferromagnetism, is shown in
Figs.~\ref{fig:fs_U3545}(a,b). More specifically, Fig.~\ref{fig:fs_U35_u} shows the Fermi surface in the
extended BZ for spin stripes running along the $y$-direction, i.e., for
the ordering vector $(\pi,0)$. While the electron pocket at $(0,\pi)$ is
hardly affected, the pocket at $(\pi,0)$ has almost disappeared. Of the
two hole pockets, the inner one has also disappeared for momenta
$(0,k_y)$, because a gap has here developed at the chemical
potential $\mu$, see Fig.~\ref{fig:Ak_U35}. For momenta $(k_x,0)$, in
contrast, the gap in the \emph{outer} pocket lies below $\mu$, and the
band consequently forms a very small electron
pocket. This result is in qualitative agreement with the unconventional 
electronic reconstruction observed with ARPES in 
${\rm (Ba,Sr)Fe_2As_2}$.\cite{ARPESFS}
Figure~\ref{fig:fs_U35fr} shows the superposition of the Fermi
surfaces obtained for the two equivalent ordering vectors $(\pi,0)$
and $(0,\pi)$ in the reduced BZ corresponding to the two-Fe unit cell.
If $U$ is increased to $U=0.9$, the gap in the outer hole pocket along
$(k_x,0)$ increases and pushes the outer band above the chemical
potential; the small electron pockets seen for $U=0.7$ in
Fig.~\ref{fig:fs_U35_u} consequently disappear, and only
one hole pocket remains around $\Gamma$, see Fig.~\ref{fig:fs_U45_u}. The $(0,\pi)$ electron
pocket remains unaffected, but at $(\pi,0)$, a hole-like shadow
pocket with very low spectral weight has replaced the original
electron pocket. The band that formed
the vanished electron pocket at $U=0$ has been deformed strongly
enough to create a small \emph{hole}-like pocket at $\approx
(\pi/2,0)$. As it can be seen in Fig.~\ref{fig:fs_U45fr}, this hole pocket
touches the $(\pi,0)$ electron pocket once the results for ordering
vectors $(\pi,0)$ and $(0,\pi)$ are combined. As $U$ continues to increase within the 
magnetic metallic phase no further qualitative changes are observed, as it can be 
seen in Fig.~\ref{fig:Ak_U55} and Fig.~\ref{fig:fs_U55_u} 
and \ref{fig:fs_U55fr} where the spectral functions and the FS are shown 
for $U=1.1$.

The average electronic
occupation numbers for the three orbitals,
shown in Fig.~\ref{fig:U_n}, are not significantly affected by the
onset of antiferromagnetism. We believe that the
small difference in electronic population observed is driven by the
different orbital magnetization (see Fig.~\ref{fig:U_mag}) and is due
to the orbital anisotropy relative 
to the direction of the magnetic $(\pi,0)$ stripes. Note that the difference between
$ m_{xz}$  and $ m_{yz}$  in Fig.~\ref{fig:U_mag} is larger than the 
difference between $ n_{xz}$ and $ n_{yz}$ in Fig.~\ref{fig:U_n} 
indicating that $q$ is more important than $p$ in Eq.~(\ref{eq:mf_xz_yz}).
In addition, the behavior of the spectral functions appears to be dominated
by the magnetization in this phase. 

\begin{figure}
\subfigure{\includegraphics[width=0.47\textwidth,trim = 15 20 20 45,clip] 
{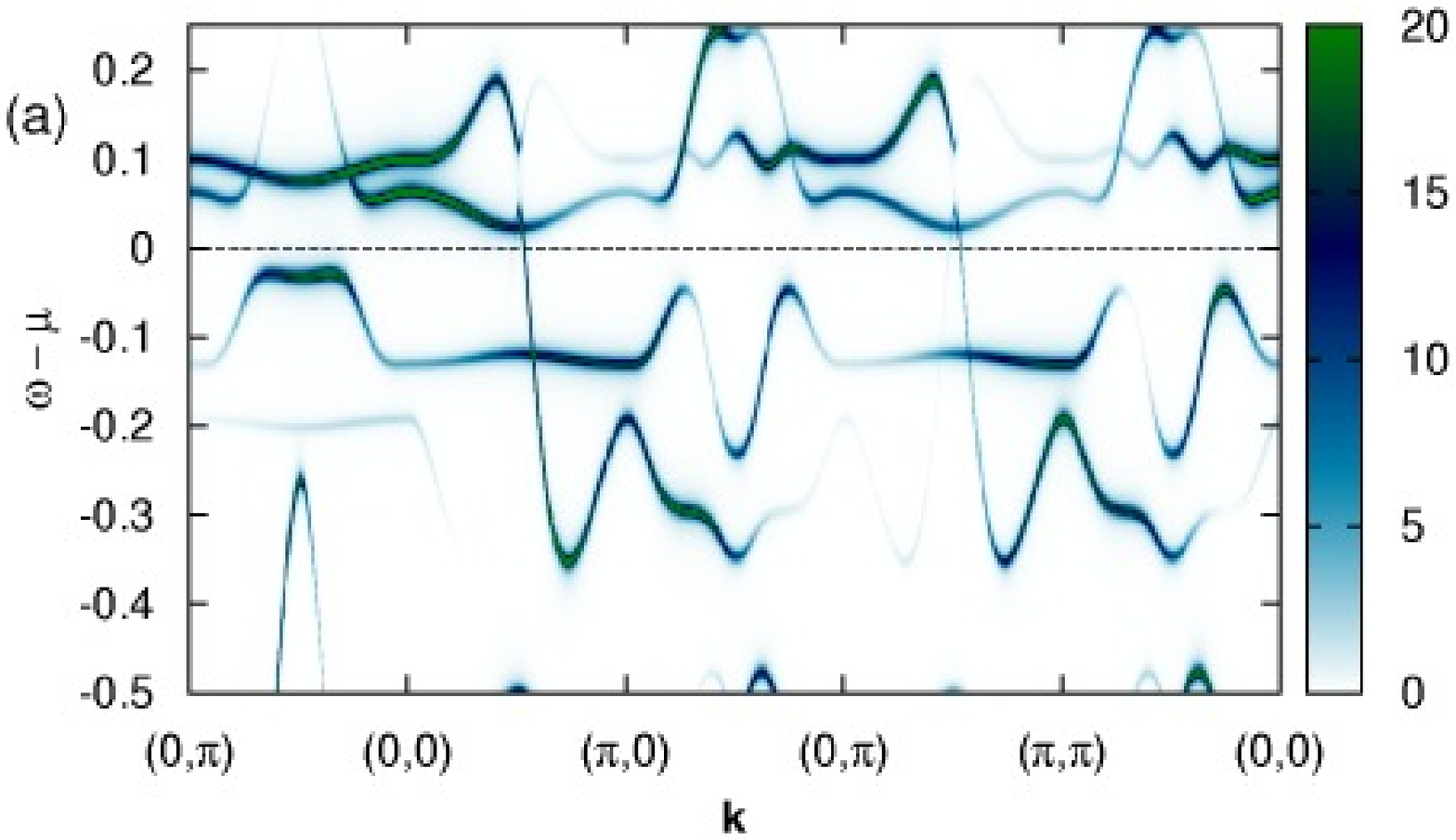}\label{fig:Ak_68}}\\[-0.5em]
\subfigure{\includegraphics[width=0.47\textwidth,trim = 15 20 20 45,clip]
{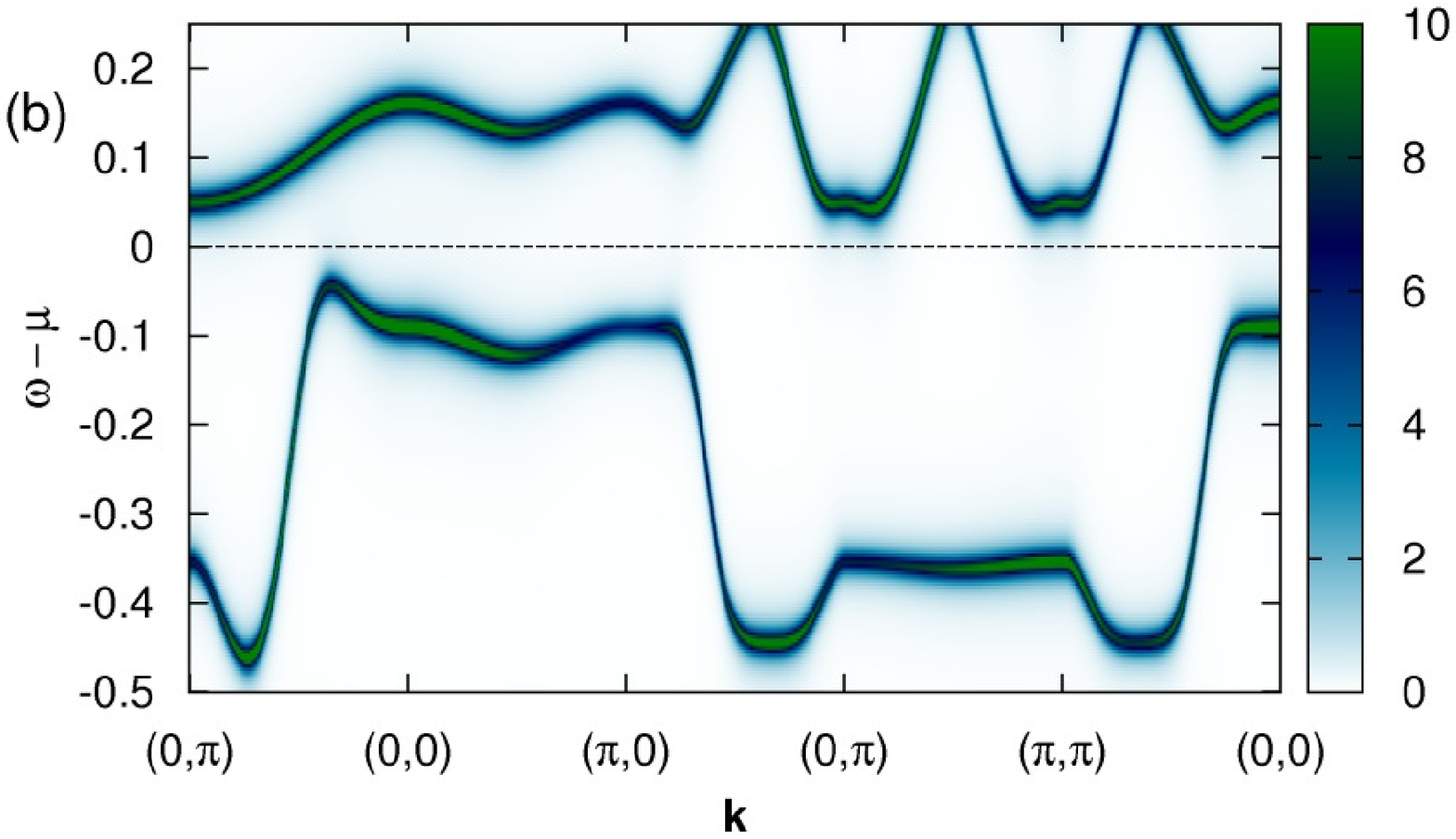}\label{fig:Ak_8}}
\caption{(Color online) Spectral density $A({\bf k},\omega)$ for (a)
  the orbitally-ordered spin-$(\pi,0)$ antiferromagnetic metallic phase at
  $U=1.36$, and (b) the orbitally polarized spin-$(\pi,0)$ antiferromagnetic
  insulator at $U=1.6$. The ratio $J=U/4$ was used, and 
the unfolded BZ is for the one-Fe unit
  cell.\label{fig:Ak_2}}
\end{figure}

When a second critical coupling
$U_{c_2}\approx 1.23$ is reached, the system develops orbital order
with an ordering momentum $(\pi,\pi)$, different from the magnetic
ordering vector $(\pi,0)$; the spin-orbital order realized is the one schematically depicted in
Fig.~\ref{fig:xy_0pi_pipi} (the order occurs between the $xz$ and $yz$
orbitals, i.e., $\phi=0$.) The system remains a metal through this
second transition as well, but the spectral density is profoundly affected,
see Fig.~\ref{fig:Ak_68}. The original hole and electron pockets around 
$\Gamma$ and $M$ completely disappear and only correlation-induced
pockets remain: the hole pocket around ${\bf k}= (\pi/2,0)$ is mirrored at ${\bf k}\approx
(\pi/2,\pi)$, and four more small pockets can be seen in the FS (not shown)
away from the high-symmetry directions plotted in
Fig.~\ref{fig:Ak_68}. Different states with different orbital ordering
patterns have only slightly higher energies in this regime. In
contrast, phases with different \emph{magnetic} ordering have
significantly higher energy, suggesting that the $(\pi,0)$ AF ``stripes''
may be more robust than the alternating orbital order. 
                                                               
If $U$ is further increased, a metal-insulator 
transition finally occurs at a third critical $U_{c_3}\approx
1.43$. At this point, the orbital order changes: as can be concluded
from the orbital densities shown in Fig.~\ref{fig:U_n}, the system
develops ferro-orbital order. The spin-$(\pi,0)$ antiferromagnetism
persists, and the spectral density in Fig.~\ref{fig:Ak_8} has a full
gap. The ferro-orbital spin-$(\pi,0)$ order in this insulator is the one
depicted schematically in Fig.~\ref{fig:x_pi0}. With growing $U$, the
staggered magnetization converges to its maximal possible value
$2\mu_\textrm{Bohr}$, as shown in Fig.~\ref{fig:U_mag}.

\begin{figure}
\includegraphics[width=0.3\textwidth]{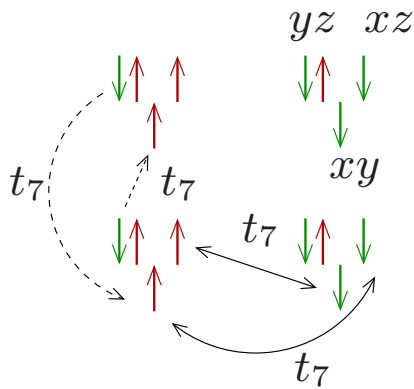}
\caption{(Color online) Magnetic order and orbital occupation at large
  $U$ shown for four sites. The magnetic ordering vector is $(\pi,0)$,
  i.e, the spin stripes run along the $y$ direction. For each site, the
  $xy$, $xz$, and $yz$ orbitals are shown: the $xy$ is the one below
  the other two, and the $yz$ is doubly occupied. Dashed (continuous)
  lines indicate inter-orbital hopping $t_7$ connecting the $xy$
  orbital to $xz$ ($yz$) along the $x$- ($y$-) direction.\label{fig:SE_AF}}
\end{figure}

The direction of the magnetic stripes determines
which of the two degenerate $xz$ and $yz$ orbitals is (almost)
doubly occupied in the FO order realized at large $U$. 
For ordering vector $(\pi,0)$ ($(0,\pi)$) it is the
$yz$ ($xz$) orbital. This can be understood by considering the
interorbital hopping $t_7$ between the $xy$ orbital 
and the $yz$($xz$) orbital along the $y$-($x$-)direction, which has to be 
large in
order to ensure $xz$/$yz$ character for the hole pockets, see
Sec.~\ref{sec:mom}. For spin stripes along the $y$-direction, i.e.
ordering vector $(\pi,0)$, bonds in the $x$-direction are
anti-ferromagnetic and the electrons can better take advantage of the AF
superexchange if the two orbitals connected by $t_7$ are both singly
occupied (see the schematic illustration in Fig.~\ref{fig:SE_AF}). The
remaining $yz$ orbital then has to be doubly occupied. This does not
cost any AF superexchange energy, because its connection to
the $xy$ orbital via $t_7$ lies along the spin-aligned $y$-direction where such
AF superexchange would in any case not occur. In fact, the
additional electron in the $yz$ orbital has the opposite spin from the
majority spin of the stripes and can, thus, gain some kinetic energy by
hopping via $t_7$  to the $xy$ orbital along the FM $y$-direction.
Thus, in this regime of large $U$ the ground state corresponds to the cartoon shown in
Fig.~\ref{fig:x_pi0} if the magnetic order is $(\pi,0)$.

Summarizing, our mean-field calculations indicate the existence of four 
distinct phases that are stabilized with growing Coulomb repulsion $U$: (i) a disordered,
paramagnetic phase for $U<U_{c_1}$, (ii) a metallic phase with $(\pi,0)$
or $(0,\pi)$ magnetic order for $U_{c_1}<U<U_{c_2}$ , (iii) a
metallic magnetic phase for $U_{c_2}<U<U_{c_3}$ 
with alternating orbital order with ordering vector $(\pi,\pi)$,
and (iv) a ferro-orbitally ordered insulator with spin-$(\pi,0)$
magnetic order for $U>U_{c_3}$, where the $yz$ [$xz$]  orbital has
larger electronic occupation for magnetic ordering vector 
$(\pi,0)$ [$(0,\pi)$].

\section{Pairing Operators in a three-orbital model for 
pnictides}\label{sec:pair}

In this section, the spin-singlet
pairing operators that are allowed by the lattice and orbital symmetries in 
the three-orbital model for LaOFeAs will be constructed.
 This classification of operators has 
previously been made for the two-orbital model.\cite{wang,shi,2orbitals,wan}
An approach similar to  
Ref.~\onlinecite{wan} will be followed. To achieve this goal, the 
three-orbital tight-binding portion of the Hamiltonian, $H_{\rm TB}$, presented in 
Eq.~(\ref{E.H0k}) will be rewritten in terms of the $3 \times 3$ matrices $\lambda_i$
which correspond to the eight Gell'mann matrices\cite{schiff} for the cases 
$i=1$ to 8, 
while $\lambda_0$ is the $3\times 3$ identity (see App.~\ref{app:lambda}). Then, $H_{\rm TB}$
becomes
\begin{equation}
H_{\rm TB}( {\bf k})=\sum_{ {\bf k},\sigma}\Phi^{\dagger}_{ {\bf k},\sigma}\xi_{ {\bf k}}
\Phi_{ {\bf k},\sigma},
\label{8}
\end{equation}
where  
$\Phi^{\dagger}_{ {\bf k},\sigma}=
(d^{\dagger}_{xz}({\bf k}),d^{\dagger}_{yz}({\bf k}),
d^{\dagger}_{xy}({\bf k}))_{\sigma}$ and
\begin{equation}
\xi_{ {\bf k}}=\epsilon_{ {\bf k}}\lambda_0+\delta_{ {\bf k}}\lambda_3+\gamma_{ {\bf k}}\lambda_1+
\alpha^{(1)}_{ {\bf k}}\lambda_5+\alpha^{(2)}_{ {\bf k}}\lambda_7+h_{ {\bf k}}\lambda_8,
\label{9}
\end{equation}
with
\begin{align}
\epsilon_{ {\bf k}}&=(T^{11}+T^{22}+T^{33})/3\\
&=\frac{2}{3}(t_1+t_2+t_5)(\cos  k_x+\cos k_y)\nonumber\\
&\quad +\frac{4}{3}(2t_3+t_6)\cos k_x\cos k_y
-\mu+\frac{\Delta_{xy}}{3},\\
\label{10a}
\delta_{ {\bf k}}&=(T^{11}-T^{22})/2=-(t_1-t_2)(\cos k_x-\cos k_y),\\
\gamma_{ {\bf k}}&=T^{12} = 4t_4\sin k_x\sin k_y,\\
\alpha^{(1)}_{\bf k}&=T^{13}/i=-2t_7\sin k_x-4t_8\sin k_x \cos k_y,\\
\alpha^{(2)}_{\bf k}&=T^{23}/i=-2t_7\sin k_y-4t_8\sin k_y \cos k_x,
\end{align}
and
\begin{equation}\begin{split}
h_{ {\bf k}}&=\frac{T^{11}+T^{22}}{2\sqrt{3}} - \frac{T^{33}}{\sqrt{3}}\\
&=\frac{1}{\sqrt{3}}(t_1+t_2-2t_5)(\cos k_x+\cos k_y)\\
&\quad+\frac{4}{\sqrt{3}}
(t_3-t_6)\cos k_x\cos k_y-\frac{\Delta_{xy}}{\sqrt{3}}.
\label{10}
\end{split}\end{equation}

It can be shown
that each element in Eqs.~(\ref{9}-\ref{10}) transforms according 
to one irreducible representation of the $D_{4h}$ group 
corresponding to the Fe
lattice. The classification is given in Tab.~\ref{table5}.
 \begin{table}
 \caption{Symmetry properties of the several terms in the $H_{\rm TB}$ 
Hamiltonian of the 
three-orbital model.}
 \begin{tabular}{|c|c|}\hline
Term & IR \\
\hline
$\epsilon_{ {\bf k}}$ & $A_{1g}$ \\
$\delta_{ {\bf k}}$ & $B_{1g}$ \\
$\gamma_{ {\bf k}}$ & $B_{2g}$ \\
$(\alpha^{(1)}_{ {\bf k}},\alpha^{(2)}_{ {\bf k}})$ & $E_{g}$ \\
$h_{ {\bf k}}$ & $A_{1g}$ \\ \hline
 \end{tabular}
 \label{table5}
 \end{table}

Since the Hamiltonian has to transform according to $A_{1g}$, 
the Gell'mann matrices in the orbital basis here chosen transform as 
indicated in Tab.~\ref{table6}.
 \begin{table}
\caption{Symmetry properties of Gell'mann matrices for the orbital assignment
that defines the proposed three-orbital model.}
 \begin{tabular}{|c|c|}\hline
Matrix & IR \\
\hline
$\lambda_0$ & $A_{1g}$ \\
$\lambda_1$ & $B_{2g}$ \\
$\lambda_2$ & $A_{2g}$ \\
$\lambda_3$ & $B_{1g}$ \\
($\lambda_4,\lambda_6$) & $E_{g}$ \\
($\lambda_5,\lambda_7$) & $E_{g}$ \\
$\lambda_8$ & $A_{1g}$ \\ \hline
 \end{tabular}
 \label{table6}
 \end{table}

In multiorbital systems the general form of a spin-singlet pairing operator is 
given by\cite{ib}
\begin{equation}
\Delta^{\dagger}({\bf k})=f({\bf k)}
(\lambda_i)_{\alpha,\beta}
(d^{\dagger}_{{\bf k},\alpha,\uparrow} d^{\dagger}_{{\bf -k},\beta,\downarrow}
-d^{\dagger}_{{\bf k},\beta,\uparrow} d^{\dagger}_{{\bf -k},\alpha,\downarrow}),
\label{2}
\end{equation}
\noindent where a sum over repeated indices is implied; 
the operators $d^{\dagger}_{{\bf k},\alpha,\sigma}$ have been defined in the 
previous sections 
and $f({\bf k})$ is the form factor
that transforms according to one of the irreducible representations of 
the crystal's symmetry group. Although $f({\bf k})$ may, in general, have a 
very complicated form, a short pair-coherence length requires the two
electrons that form the pair to be very close to each other. Consequently, 
for simplicity we
focus on nearest and diagonal next-nearest neighbors, and form factors
that are allowed in a lattice with $D_{4h}$ symmetry. The momentum 
dependent expression, as well as the irreducible representation according to 
which each form factor transforms, are given in Tab.~\ref{table1}. Note that
if the pairing mechanism is non-BCS and if the Coulomb repulsion is strong 
on-site pairing, then f(k)=1 corresponding to onsite pairing is an
unlikely factor.

\begin{table}
\caption{Form factors $f({\bf k})$ for pairs up to distance (1,1) classified
according to their symmetry under $D_{4h}$ operations.}
 \begin{tabular}{|l|c|r|} \hline
 \# & $f({\bf k})$ & IR \\
\hline
1 & 1 & $A_{1g}$\\
2 & $\cos k_x +\cos k_y $ & $A_{1g}$\\
3 & $\cos k_x \cos k_y $ & $A_{1g}$ \\
4 & $\cos k_x -\cos k_y $ & $B_{1g}$\\
5 & $\sin k_x \sin k_y $ & $B_{2g}$ \\
6 & $(\sin k_x, \sin k_y )$ & $E_{g}$\\
7 & $(\sin k_x \cos k_y ,\sin k_y \cos k_x )$& $E_{g}$\\
\hline
\end{tabular}
\label{table1}
\end{table}

\subsection{Intraorbital Pairing}

The previous discussion shows that the symmetry of the pairing operator will be exclusively 
determined by the 
symmetry of $f({\bf k})$ \emph{only} if $\lambda_i$ transforms according to $A_{1g}$.
Table~\ref{table6} indicates that this is the case for pairing operators
constructed by using $\lambda_0$ or $\lambda_8$ in Eq.~(\ref{2}). These 
two matrices are diagonal, which means that such pairing 
operators define \emph{intraorbital} pairings. 
For intraorbital pairing, with a symmetry fully determined by the spatial 
form factor, the basis functions are then given by:
\#I:$f({\bf k})\lambda_0$ or \#II: $f({\bf k})\lambda_8$.

For \#I, the superconducting order parameter (OP) will be the same for the 
three
orbitals while \#II allows the OP for $xy$ to be different than  for $xz$
and $yz$ which need, by symmetry, to have OPs that can only differ by a
relative sign. Thus, the addition of a third
orbital may allow the possibility of different superconducting gaps in 
the band representation, reminiscent of the two gaps in 
MgB$_2$.\cite{akimitsu,Louie}

 \begin{table}
 \caption{Product table for the irreducible representations of the group
 $D_{4h}$
 relevant to this work.}
 \begin{tabular}{!{\vrule width 1pt}c!{\vrule width 1pt}c|c|c|c|c!{\vrule
 width 1pt}} \noalign{\hrule height 1pt}
 & \textbf{$A_{1g}$} & $A_{2g}$& $B_{1g}$& $B_{2g}$& $E_{g}$ \\
 \noalign{\hrule height 1pt}
 $A_{1g}$ & $A_{1g}$ & $A_{2g}$& $B_{1g}$& $B_{2g}$&
 $E_{g}$ \\
 \hline
 $A_{2g}$ & $A_{2g}$ & $A_{1g}$& $B_{2g}$& $B_{1g}$&
 $E_{g}$ \\
 \hline
 $B_{1g}$ & $B_{1g}$ & $B_{2g}$& $A_{1g}$& $A_{2g}$&
 $E_{g}$ \\
 \hline
 $B_{2g}$ & $B_{2g}$ & $B_{1g}$& $A_{2g}$& $A_{1g}$&
 $E_{g}$ \\
 \hline
 $E_{g}$ & $E_{g}$ & $E_{g}$& $E_{g}$& $E_{g}$& $A_{1g}+A_{2g}+B_{1g}+B_{2g}$ \\
 \noalign{\hrule height 1pt}
 \end{tabular}
 \label{prod}
 \end{table}
                                                                               
When any of the 
remaining seven matrices $\lambda_i$ appear in Eq.~(\ref{2}), the symmetry of
the pairing operator is given by the irreducible representation of 
$D_{4h}$ 
resulting from the product of the symmetry of the form factor and the
symmetry of the orbital
component,\cite{ib} according to the product table given in Tab.~\ref{prod}. 
For $\lambda_i=\lambda_3$, the basis function is given by \#III: 
$f({\bf k})\lambda_3$. The pairing is still intraorbital but
since $\lambda_3$ transforms according to $B_{1g}$ the symmetry of the operator
will be $B_{1g}$ if $f({\bf k})$ transforms according to $A_{1g}$, etc.
Note that this pairing operator does not involve the $xy$ orbital 
and it has already been presented in the context of the two-orbital 
model.\cite{wan} However, since the orbital composition of the bands is not 
the same as for the two-orbital model, it will be important to determine 
whether the gap structure of this pairing operator has changed.

\subsection{Interorbital Pairing}

The remaining six $\lambda_i$ matrices lead to interorbital pairing.
Note that $\lambda_1$ and $\lambda_2$ do not involve the orbital $xy$ and 
the pairing operators that they generate have already been discussed in the 
two-orbital model.\cite{wan} We are interested in the spin-singlet pairing 
operator for orbitals $xz/yz$ that has a basis \#IV:$f({\bf k})\lambda_1$.
This operator, with $f({\bf k})=\cos k_x+\cos k_y$, has been found to be favored
for intermediate values of the Coulomb repulsion $U$ in numerical calculations 
of the two-orbital model for pnictides.\cite{Daghofer:2009p1970,moreo}
The addition of the $xy$ orbital leads to the possibility of new
interorbital pairing operators, i.e., pairing between electrons in the orbitals
$xz$ and $yz$ with electrons in the $xy$ orbital.
Thus, now the focus will be on the 
interorbital spin-singlet pairing operators that result from the addition of
$xy$.

The interorbital case becomes very interesting because we need to combine
($xz$,$yz$) that transform as the two-dimensional representation 
$E_{g}$ with $xy$ that transforms as $B_{2g}$; thus, the product transforms 
as $E_{g}$. The $\lambda_i$ matrices
that can  appear in this intraorbital pairing are ($\lambda_4,\lambda_6$) or
($\lambda_5,\lambda_7$). Since the focus here is 
on pairing operators that are spin singlets, it will be
required that the operator is even under orbital exchange. Thus, only 
($\lambda_4,\lambda_6$) will be considered since the other two 
matrices ($\lambda_5,\lambda_7$) that transform according to $E_{g}$ will 
produce operators odd under orbital exchange. 
Let us further restrict the analysis to the case of 
pairing operators that transform according to one dimensional representations 
of the point group because we assume that the ground state is non degenerate.
Thus, the only spatial form factors $f({\bf k})$ that we 
should consider must transform according to $E_{g}$. 
 This leaves us with
\begin{equation}
f({\bf k})=(\sin k_x,\sin k_y),
\label{11}
\end{equation}
for nearest neighbor pairs and
\begin{equation}
f({\bf k})=(\sin k_x \cos k_y,\cos k_x \sin k_y),
\label{12}
\end{equation}
for diagonal pairs. Since the direct product of two $E_{g}$ 
representations is 
$E_{g}\times E_{g}= A_{1g}+A_{2g}+B_{1g}+B_{2g}$, 
pairing operators transforming according to four
irreducible representations will be obtained. The basis for the 
new pairing operators, labeled V$_i$, are presented in 
Tab.~\ref{table7}.

 \begin{table}
\caption{Properties of pairing operators in the three-orbital 
model. $f$ indicates the symmetry of $f({\bf k})$.}
 \begin{tabular}{|c|c|c|c|}\hline
No. & IR & Basis & Gap \\ 
\hline
I& $f$ & $f({\bf k})\lambda_0$ & Full or Nodal \\
II& $f$ & $f({\bf k})\lambda_8$ & Full or Nodal \\
III& $f$$B_{1g}$ & $f({\bf k})\lambda_3$ &  Nodal \\
IV& $f$$B_{2g}$ & $f({\bf k})\lambda_1$ &  Nodal \\
V$_a$& $A_{1g}$ & $\sin k_x \lambda_4+\sin k_y \lambda_6$ & Nodal \\
V$_b$& $A_{1g}$ &$\lambda_4\sin k_x\cos k_y+\lambda_6\cos k_x\sin k_y$ & Nodal\\
V$_c$& $B_{1g}$ &$\sin k_x \lambda_4-\sin k_y \lambda_6$ & Nodal\\
V$_d$ & $B_{1g}$& $\lambda_4\sin k_x\cos k_y-\lambda_6\cos k_x\sin k_y$ & Nodal\\
V$_e$ & $A_{2g}$& $\sin k_x \lambda_6+\sin k_y \lambda_4$ & Nodal\\
V$_f$ & $A_{2g}$& $\lambda_4\cos k_x\sin k_y+\lambda_6\sin k_x\cos k_y$  
& Nodal\\
V$_g$ & $B_{2g}$ & $\sin k_x \lambda_6-\sin k_y \lambda_4$ & Nodal\\
V$_h$ & $B_{2g}$ &$\lambda_4\cos k_x\sin k_y-\lambda_6\sin k_x\cos k_y$ & Nodal\\ 
\hline
 \end{tabular}
 \label{table7}
 \end{table}

\subsection{Band Representation}

To obtain the gap structure of the pairing operators \#I to \#IV 
(shown in Table~\ref{table7}) 
the Bogoliubov-de Gennes Hamiltonian (BdG) is constructed and it is given by
\begin{equation}
H_{\rm BdG}=\sum_{{\bf k}}\Psi^{\dagger}_{\bf k}H^{\rm MF}_{\bf k}\Psi_{\bf k},
\label{13}
\end{equation}
\noindent with the definitions
\begin{eqnarray}
\Psi^{\dagger}_{\bf k}=(d^{\dagger}_{{\bf k},xz,\uparrow},d^{\dagger}_{{\bf k},yz,\uparrow},d^{\dagger}_{{\bf k},xy,\uparrow},\nonumber\\
d_{-{\bf k},xz,\downarrow},d_{-{\bf k},yz,\downarrow},d_{-{\bf k},xy,\downarrow}),
\label{14}
\end{eqnarray}
\noindent and
\begin{equation}
H^{\rm MF}_{\bf k}=
 \left(\begin{array}{cc}
H_{\rm TB}({\bf k}) & P( {\bf k}) \\
P^{\dagger}({\bf k}) & -H_{\rm TB}({\bf k})
\end{array} \right),
\label{15}
\end{equation}
where each element represents a $3\times 3$ block with
$H_{\rm TB}({\bf k})$ given by Eq.~(\ref{E.H0k})
and 
\begin{equation}
P({\bf k})_{\alpha,\beta}=Vf({\bf k})(\lambda_i)_{\alpha,\beta},
\label{16}
\end{equation}
with $i=0$, 8, 3, and 1 for pairing \#I, \#II, \#III, and \#IV, respectively.
$V$ is the magnitude of the OP given by the product of the
pairing attraction $V_0$ and a mean-field parameter $\Delta$ that should be 
obtained from minimization of the total energy.\cite{ib} For pairing
\#V$_i$ the basis listed in Table~\ref{table7} should be used instead of
$f({\bf k})\lambda_i$.

Up to this point we have worked using the orbital representation because
this basis renders it straightforward to obtain the form of the Hamiltonian, as well as the  
pairing operators allowed by the symmetry of the lattice and 
orbitals.
However, the experimentally observed superconducting gaps occur at the FS 
determined by the bands that result from the hybridization of the orbitals.
For this reason, it is convenient to express Eq.~(\ref{15}) in the band 
representation. $H_{\rm TB}({\bf k})$ can be expressed in the band 
representation via the transformation 
$H_{\rm Band}({\bf k})=U^{\dagger}({\bf k})H_{\rm TB}({\bf k})U({\bf k})$,
where $U({\bf k})$ is the unitary change of basis matrix and 
$U^{\dagger}({\bf k})$ is the 
transpose conjugate of $U({\bf k})$. 
Since $U$ is unitary it is known that for each value of ${\bf k}$,
$\sum_i(U_{i,j})^*U_{i,k}=\sum_i(U_{j,i})^*U_{k,i}=\delta_{j,k}$. 
Then, $H'_{\rm MF}=G^{\dagger}H_{\rm MF}G$ where 
$G$ is the $6\times 6$ unitary matrix composed of two $3\times 3$ blocks 
given by $U$. Then,
\begin{equation}
H'^{\rm MF}_{\bf k}=
 \left(\begin{array}{cc}
H_{\rm Band}({\bf k}) & P_{B}( {\bf k}) \\
P_{B}^{\dagger}({\bf k}) & -H_{\rm Band}({\bf k})
\end{array} \right),
\label{15a}
\end{equation}
with
\begin{equation}
P_{B}({\bf k})=U^{-1}({\bf k})P({\bf k})U({\bf k}).
\label{15b}
\end{equation}

A standard assumption in superconducting multiband systems is that the pairing 
interaction should be purely intraband, meaning that $P_{B}({\bf k})$ is
diagonal. Thus, let us explore what 
kind of 
purely intraband pairing operators are allowed by the symmetry properties of the
three-orbital model for LaOFeAs. In the band representation, the most general 
BdG matrix with purely intraband pairing is given by
\begin{equation}
H'_{\rm MF}=\hspace{-0.5em}
 \left(\hspace{-0.5em}\begin{array}{cccccc}
\epsilon_1({\bf k}) & 0&0&\Delta_1( {\bf k})&0&0 \\
0&\epsilon_2({\bf k}) &0&0&\Delta_2( {\bf k})&0 \\
0 & 0&\epsilon_3({\bf k})&0&0&\Delta_3( {\bf k}) \\
\Delta^*_1( {\bf k})& 0&0&-\epsilon_1({\bf k}) &0&0\\
0&\Delta^*_2( {\bf k}) &0&0&-\epsilon_2({\bf k}) &0 \\
0& 0&\Delta^*_3( {\bf k}) &0&0&-\epsilon_3({\bf k}) 
\end{array} \hspace{-0.5em}\right),
\label{17}
\end{equation}
where $\epsilon_i({\bf k})$ are the eigenvalues of $H_{\rm TB}({\bf k})$
and $\Delta_i({\bf k})$ denotes the band and momentum dependent pairing
interactions.\cite{hop} As it can be deduced from the properties of the unitary 
change of basis matrix $U$, if all three bands have the same pairing interaction, 
i.e. $\Delta_1({\bf k})=\Delta_2({\bf k})=
\Delta_3({\bf k})=\Delta({\bf k})$, then the matrix $P({\bf k})$ in the orbital 
representation will also be diagonal. In this case,
the pairing operator is given by Eq.~(\ref{2}) with an arbitrary
$f({\bf k})$ and $\lambda_i=\lambda_0$, i.e., the pairing operator
is intraorbital and the OP is the same for the three orbitals. This 
corresponds to pairing operator \#I which describes a pairing interaction that
is the {\it same} for each of the three orbitals. However, symmetry only requires that 
the orbitals $xz$ and $yz$ must have the same OP, while $xy$ can have a different 
one. Thus, there does not seem to be a reason to assume that electrons in the 
many bands that determine the FS should be affected by the same pairing 
interactions. In fact, in MgB$_2$ the electron-phonon interaction that provides
the pairing is stronger on the $\sigma$-bands than on the $\pi$-bands giving, 
as a result, two different superconducting gaps. Thus, it can be asked whether the
symmetry of the three-orbital model allows for the possibility of two different 
OPs 
with a pure intraband pairing interaction. If it is assumed
that in Eq.(\ref{17}) $\Delta_1=\Delta_2=\Delta=f({\bf k})C$ and 
$\Delta_3=\Delta'=f({\bf k})C'$, then in the orbital representation
\begin{equation}
P({\bf k})_{\alpha,\beta}=U_{\alpha,3}({\bf k})
U^*_{\beta,3}({\bf k})f({\bf k})(C'-C),
\label{15c}
\end{equation}
for the off-diagonal elements and
\begin{equation}
P({\bf k})_{\alpha,\alpha}=\Delta+|U_{\alpha,3}({\bf k})|^2f({\bf k})(C'-C),
\label{15d}
\end{equation}
for the diagonal ones.

Now let us concentrate on the diagonal part. This has to arise from a linear 
combination of intraorbital pairing operators with compatible symmetries. 
There are two possibilities: 
\begin{equation}
P({\bf k})_{\alpha,\alpha}=f({\bf k})[A(\lambda_0)_{\alpha,\alpha}+
B(\lambda_8)_{\alpha,\alpha}],
\label{15e}
\end{equation}
\noindent or
\begin{equation}
P({\bf k})_{\alpha,\alpha}=Df({\bf k})(\lambda_3)_{\alpha,\alpha},
\label{15f}
\end{equation}
where $A$, $B$, and $D$ are independent of momentum. It can be shown that
Eq.~(\ref{15e}) requires $|U_{13}|^2 =|U_{23}|^2$  while  Eq.~(\ref{15f}) 
requires $|U_{13}|^2 =-|U_{23}|^2$, which are not satisfied by the 
elements of the matrix $U$ determining the change of basis. This means that 
any purely intraband pairing interaction allowed by the symmetry of the 
three-orbital model should be the same for the three bands. On the other hand, if we
had a case in which $|U_{13}|^2 =|U_{23}|^2=0$, which means that one of the
three orbitals does not hybridize with the other two, it would be possible to
have a system with two different gaps. Note that this is the situation for MgB$_2$ 
in which the $z$ orbital that forms the $\pi$ band does not hybridize with the
$x$ and $y$ orbitals that constitute the $\sigma$ band. 

Summarizing, it has been found that independent gaps in different Fermi surfaces cannot 
arise if the hybridization among all the orbitals is strong and the pairing 
interaction is purely intraband. 

\subsection{The $s\pm$ Pairing Operator}

The next issue to be considered  is whether the orbital and lattice symmetries 
allow for the possibility of the often discussed $s\pm$ pairing scenario. ARPES
experiments indicate the existence of two hole-pockets around  
$\Gamma$. The interior pocket, which is almost nested with the electron 
pockets with a nesting vector ${\bf q}=(\pi,0)$ or $(0,\pi)$, develops a 
constant gap $\Delta_h$ which has the same magnitude than the gap on the 
electron pockets $\Delta_e$. In addition, they find a smaller gap 
$\Delta_h'\approx\Delta_h/2$ on the exterior hole pocket.\cite{arpes3} 
The ARPES results can be interpreted in two different ways in the context of 
a three-orbital model:

{\it (i)} Assume that the inner hole pocket observed in ARPES corresponds to two 
almost degenerate FS that cannot be resolved, and assign the external hole 
pocket to a band that arises when extra orbitals are added. This is the same
assumption made in the two-orbital model for which it was 
shown in Ref.~\onlinecite{moreo} that the $s\pm$ pairing state is compatible
with the lattice and orbital symmetries. 

Under this assumption, in the three-orbital model the $s\pm$ pairing state 
corresponds
to our pairing operator \#I with $f({\bf k})=\cos k_x \cos k_y$, which in the 
band representation leads to a purely intraband pairing attraction given by
$\Delta_i({\bf k})=V \cos k_x \cos k_y$ for each of the three-bands. For hole 
pockets almost degenerate with each other, the gap in both bands will be the 
same and there will be a sign difference with the gap at the electron pockets
whose Fermi momentum differs from those of the hole pockets by $(0,\pi)$
or $(\pi,0)$.

{\it (ii)} Assume that the inner and outer hole pockets observed by ARPES are 
described by the two hole pockets in the three-orbital model. This would force us 
to request that, for example,  $\Delta_1({\bf k})=-\Delta_3({\bf k+q})$ 
where ${\bf q}=(\pi,0)$ 
or $(0,\pi)$ and $\Delta_2({\bf k})$ is independent. Then, let us assume 
that $\Delta_1({\bf k})=
\Delta_3({\bf k})=\cos  k_x\cos  k_y\Delta_0$ and 
$\Delta_2({\bf k})=\Delta({\bf k})$. 
Let us concentrate on the $\Gamma$-$X$ direction. Along this direction, 
$\gamma_{{\bf k}}$ and $\alpha_{{\bf k}}^{(i)}$ vanish, meaning that there is no
hybridization among the three orbitals. From Fig.~\ref{fig:bands_fs_orb_U0} 
we observe that each of the two hole FS results from the crossing of $xz$ 
and $yz$, while the electron FS has pure $xy$ character. From the orbital 
symmetry, then, it is deduced that the only reason for having different gaps at the 
two hole-like FS would be a strong momentum dependence of the gap since 
symmetry enforces $|\Delta_{xz}({\bf k})|=|\Delta_{yz}({\bf k})|$; 
in addition, the gap at the electron pockets does not need to be related to 
the gap in the hole pockets, unless the pairing operator contains $\lambda_0$. 
Thus, we observe that the $s\pm$ pairing operator could be supported under this 
assumption if it is given by Eq.~(\ref{2}) with $f({\bf k})=\cos k_x \cos k_y$ 
with $\lambda_i=\lambda_0$ and the additional condition that if 
${\bf k}^{FS_h}$
represents the Fermi momentum of the internal hole pocket and 
${\bf k}^{FS_{h'}}={\bf k}^{FS_h}+\delta$ is the Fermi momentum of the 
external hole 
pocket it is necessary that $f({\bf k}^{FS_h})/f({\bf k}^{FS_{h'}})\approx 2$ 
which would 
require fine tuning of the parameters. It would also be expected that in this 
scenario the ratio $\Delta_{h'}/\Delta_h$ should not be 1/2 for all materials. 

Thus, it is concluded that the $s\pm$ pairing could be supported by a three-orbital model.
It corresponds to pairs of electrons in the same 
orbital at distance one along the diagonals of the square lattice, i.e., on
next-nearest neighbor sites, with the same  
pairing potential for all three orbitals. Then, if experiments show 
that $s\pm$ is indeed the correct pairing operator, it will remain to be understood
why the pairing interaction does not appear to depend on the 
symmetry of each different orbital or, equivalently, why it is the same for 
electrons in different bands. This should be contrasted with the case of 
MgB$_2$ in which the strength of the electron-phonon coupling that leads to 
pairing is stronger on the $\sigma$-band FS than in the
$\pi$-band FS.

Since the pairing mechanism for the pnictides is not known and the $s\pm$ 
pairing state is just one of many proposed states, 
our discussion will continue by analyzing the other new pairing states that 
are allowed by symmetry when
the $xy$ orbital is considered.

\subsection{Properties of the Pairing Operators}

In the previous subsection, it was shown that only pairing \#I can lead to a purely
intraband pairing interaction in the context of the three-orbital model, and that 
the widely proposed $s\pm$ pairing state indeed belongs to the class represented by
pairing \#I. On the other hand, given the complexity of the problem,
only numerical simulations can clarify 
whether the three-orbital model becomes superconducting upon doping and what is 
the symmetry of the dominant pairing state. Since numerical results are not yet available, 
here the properties of the other possible spin singlet 
pairing operators will be discussed. 

Let us start with pairing \#II, i.e. the intraorbital pairing operator 
containing $\lambda_8$, that allows a different pairing strength for the 
$xy$ orbital. This pairing operator is not purely intraband. 
This fact can be easily deduced from the properties of the unitary change of basis
matrices. In this case, 
$P({\bf k})_{\alpha,\beta}=C_{\alpha}\delta_{\alpha\beta}$ with
$C_1=C_2={Vf({\bf k})/{\sqrt{3}}}$ corresponding to orbitals $xz$ and $yz$, 
and $C_3=-{2Vf({\bf k})/{\sqrt{3}}}$ for orbital $xy$. Then, 
\begin{equation}
P_{B}({\bf k})_{\alpha,\beta}=\sum_iC_iU^*_{i,\alpha}({\bf k})
U_{i,\beta}({\bf k}),
\label{15g}
\end{equation}
\noindent which does not vanish for all values of ${\bf k}$ for 
$\alpha\ne\beta$, thus indicating the existence of interband pairing terms. 
Similar calculations for all the pairing operators presented in 
Table~\ref{table7} show nonvanishing interband pairing terms.

Then, it is concluded that starting from the orbital 
representation, the only way to obtain pure intraband pairing in the band 
representation is by considering a pairing interaction that affects equally
all the orbitals involved, producing equal gaps in all the orbitals and/or 
bands with symmetry determined by the spatial form factor. This shows that the
requirement of purely intraband pairing induces a strong constraint regarding
the coupling of the electrons in the different orbitals with the source of 
the pairing attraction. On the other hand, if the requirement is relaxed, 
interband pairing occurs at least in some regions 
of the BZ.\cite{ib} It was verified that this is the case for the remaining 
operators \#II, \#III, \#IV, and \#V$_i$. It has also been observed, by 
monitoring the
eigenvalues of $H_{\rm BdG}$ for operators \#III and \#IV, that there is a 
nodal structure in the superconducting gap for all the values of $f({\bf k})$ 
shown in Tab.~\ref{table1}, while operator \#II becomes nodeless for
$f({\bf k})=\cos k_x \cos k_y$ or $1$ at a finite value of $V$. In addition,
some linear combinations of pairing \#I and \#II with 
$f({\bf k})=\cos k_x \cos k_y$ or $1$ are also nodeless for all finite 
values of $V$ but they lead to interband pairing interactions. 
These nodeless states that we call  $s_{IB}$ will be discussed 
in Sec.\ref{sec:ARPES}.

\subsubsection {Pairing with Pseudocrystal Momentum ${\bf Q}=(\pi,\pi)$}

In Sec.~\ref{sec:mom}, it was explained that although the three-orbital 
Hamiltonian retains the two-iron unit cell of the original FeAs planes, it 
is possible to express it in terms of three orbitals in the space of 
pseudocrystal momentum ${\bf k}$ defined in the extended Brillouin zone 
corresponding to a base with one single Fe atom per unit cell. In terms of the
real momentum, the Hamiltonian consists of two $3\times 3$ blocks 
$H_1({\bf k})$ and $H_2({\bf k})$ with $H_1({\bf k})=H_{\rm TB}({\bf k})=
H_2({\bf k+Q})$; thus, these 
two blocks provide the same eigenvalues in the unfolded 
Brillouin zone, but both blocks need to be considered if 
the actual reduced BZ is used. 
This means that in the reduced BZ the bands arise 
as 
combinations of six orbitals labeled by the orbital index $\alpha=1$, 2, or 3,
and the Hamiltonian block index $i=1$ or 2. 
Then, to consider all the possible 
interorbital pairing operators it is important to include pairs formed by electrons
in orbitals in the two different blocks. In the extended BZ, this is equivalent 
to considering pairs with both pseudocrystal momentum 0 and ${\bf Q}$. Note that  
Cooper pairs with pseudocrystal momentum ${\bf Q}$ still have zero center-of-mass 
momentum. Exact diagonalization studies of the  two-orbital
model\cite{Daghofer:2009p1970,moreo} did not favor such operators, and it is
possible that this kind of pairing does not occur in the three-orbital model either. 
However, since symmetry allows such a possibility, pairing operators with nonzero pseudocrystal momentum
will here be discussed for completeness.

The generalized Bogoliubov-de Gennes matrix $H^{\rm MF}_{\bf k}$ that allows us to 
consider interorbital pairs with pseudocrystal momentum ${\bf Q}$ is given 
by:
\begin{equation}
H^{\rm MF}_{\bf k}\hspace{-0.5em}=\hspace{-0.5em}
 \left(\hspace{-0.5em}\begin{array}{cccc}
H_{\rm TB}({\bf k}) &\hspace{-0.5em}0 &\hspace{-1em}0&\hspace{-0.5em}P( {\bf k}) \\
0& \hspace{-0.5em}-H_{\rm TB}({\bf k})&P({\bf k+Q})  &\hspace{-0.5em}0\\
0&\hspace{-1em}P^{\dagger}({\bf k+Q})&\hspace{-0.5em}H_{\rm TB}({\bf k+Q})&\hspace{-1em}0\\
\hspace{-1em}P^{\dagger}({\bf k})&\hspace{-0.5em}0&\hspace{-0.5em}0& \hspace{-0.5em}-H_{\rm TB}({\bf k+Q})
\end{array} \hspace{-0.5em}\right),
\label{155}
\end{equation}
where $P({\bf k})$ has the form given in Eq.~(\ref{16}).

By finding the eigenvalues of $H^{\rm MF}_{\bf k}$, the structure of the gap 
of the possible pairing operators with pseudocrystal momentum ${\bf Q}$ can be obtained. Our analysis shows that all the pairing operators with pseudocrystal 
momentum ${\bf Q}$ lead to inter and intraband pairing in the band 
representation
and nodes on the FS for small $V$. We have observed that a nodeless gap
for pairs with pseudocrystal momentum ${\bf Q}$ develops at a finite value of 
$V$ for operators \#I and \#II with $f({\bf k})=1$
or $\cos k_x \cos k_y$ in a manner characteristic of systems with interband
pairing.\cite{ib}

\subsubsection{Spectral Functions}\label{sec:ARPES}

It is straightforward to calculate the spectral functions $A({\bf k},\omega)$
for all the pairing operators presented in this manuscript. However, 
due to their
large number, we will concentrate on {\it (i)} pairing operator
\#I with $f({\bf k})=\cos k_x \cos k_y$, i.e., the $s\pm$ pairing 
operator, {\it (ii)} a linear combination of pairing operators \#I and \#II 
with $f({\bf k})=\cos k_x \cos k_y$,
that we will call the $s_{IB}$ pairing operator;
{\it (iii)} pairing operator \#IV with 
$f({\bf k})=\cos k_x +\cos k_y$, i.e., the $B_{2g}$ pairing 
operator, favored by numerical calculations in the magnetic 
metallic regime of the two-orbital model, which will be called $B_{2g}$,
and {\it (iv)} a linear combination of 
pairing operator \#IV with $f({\bf k})=\cos k_x +\cos k_y$ and pairing operator
\#$V_g$ with pseudocrystal momentum ${\bf Q}$, which is the natural 
extension to three orbitals of $B_{2g}$ and will be called 
$B_{2g}^{\rm ext}$. 

\begin{figure}
\includegraphics[width=0.47\textwidth]{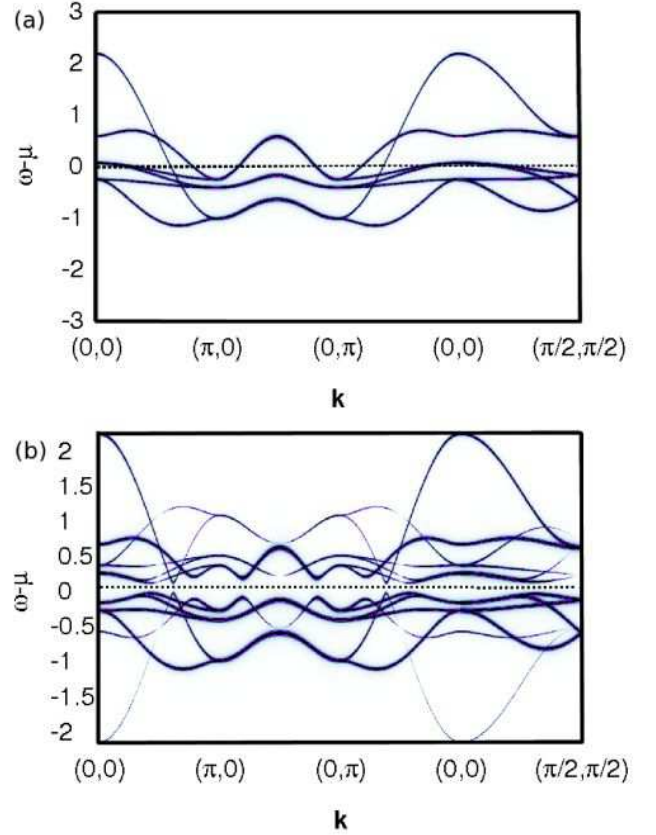}
\caption{(Color online) The intensity of the points represents the values 
of the
spectral function $A({\bf{k},\omega})$ for the three-orbital model with pairing
interaction (a) $V=0$; (b) $V=0.2$, for the $s\pm$ pairing operator given 
in the text.\label{akwv0spm}}
\end{figure}

In Fig.~\ref{akwv0spm}(a) the spectral functions $A({\bf k},\omega)$ along high
symmetry directions in the reduced Brillouin zone are shown for the
three-orbital Hamiltonian with $V=0$, i.e., without pairing, in order to 
illustrate the 
changes induced by the various pairing interactions considered here. 
Note that these results correspond to the system with two Fe atoms 
per unit cell, which leads to the six bands seen in Fig.~\ref{akwv0spm}(a).
The 
results for the $s\pm$ pairing operator with intensity $V=0.2$ are 
presented in 
Fig.~\ref{akwv0spm}(b). It can be observed that a gap opens at the FS and 
shadow Bogoliubov
bands, which should be visible in ARPES experiments, appear. Numerically, 
we have 
verified that no nodes occur anywhere in the BZ. Note that the 
gap is
momentum dependent because $f({\bf k})=\cos k_x\cos k_y$. This means that
the ratio between the gaps at the FS is determined by
$|f({\bf k}_i)/f({\bf k}_j)|$, where $i$ and $j$ can take the values 1, 2, 3 
corresponding to the three bands that determine the FS, i.e., 1 (2) for the 
interior (exterior) hole pocket, and 3 for the electron pockets. 
This creates a constraint on how different these 
gaps can be if $s\pm$ represents the actual pairing symmetry of the pnictides.

\begin{figure}
\subfigure{\includegraphics[width=0.47\textwidth]{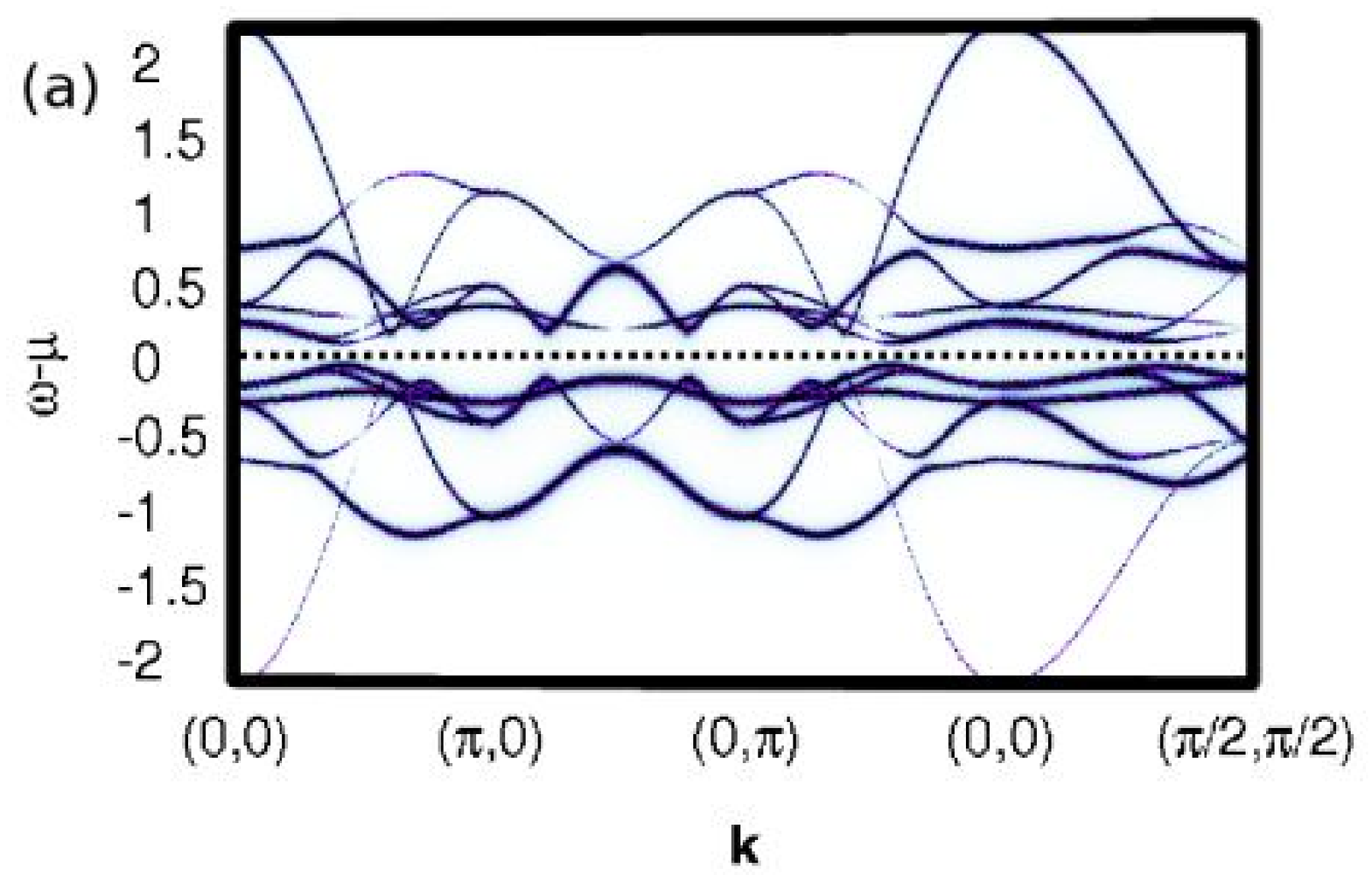}\label{l8l2}}
\subfigure{\includegraphics [width=0.47\textwidth,trim = 10 10 35
 25,clip]{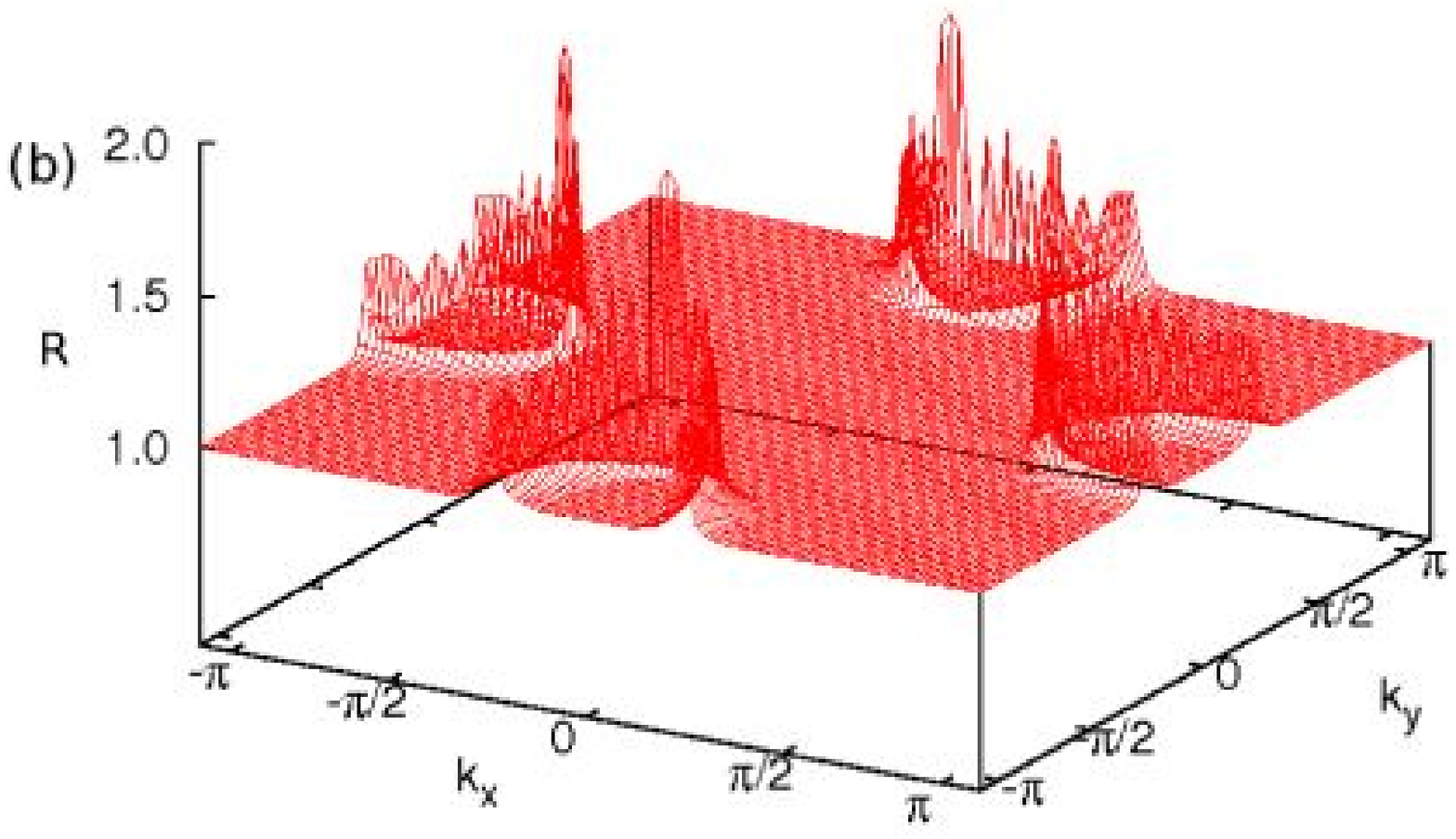}\label{gap}}
\caption{(Color online) (a) The intensity of the points represents the 
values of 
the spectral function $A({\bf{k},\omega})$ for the three-orbital model with 
pairing interaction  $V=0.2$ for the pairing state $s_{IB}$ along the indicated
high symmetry directions in the folded BZ. (b) Ratio $R$ between the gaps for
pairing $s_{IB}$ and pairing $s\pm$ for $V=0.05$ in the unfolded
BZ.
\label{Sib}}
\end{figure}

As previously mentioned, we can define a pairing operator $s_{IB}$ by 
combining operators \#I and \#II with $f({\bf k})=\cos k_x\cos k_y$ so that
the orbital part of the basis is given by $A\lambda_0+B\lambda_8$, where $A$ 
and $B$ are constants. This 
pairing operator is diagonal in the orbital representation, and transforms 
according to $A_{1g}$ thus, it has $S$ symmetry; it also has intra and interband 
terms in the band representation. This is why this operator 
is called $s_{IB}$. 
For a robust range of values of $A$ and $B$, a nodeless gap opens on
all FSs for any finite value of $V$. For example, we can choose the
parameters in such a way that for any given ${\bf k}$, 
the pairings for the
three orbitals have the same sign. The spectral functions
for $V=0.2$ along high symmetry directions in momentum space are shown in 
Fig.~\ref{l8l2} for the special case in which $A=3/2$ and $B=-\sqrt{3}/2$. 
The major difference with the results for the $s\pm$ state 
[see Fig.~\ref{akwv0spm}(b)] is that the interband pairing present in $s_{IB}$ 
opens gaps between the bands away from the FS. This is a feature that should
be observed in ARPES experiments. Also the band spectral functions in both 
cases are very different close to $(0,\pi)$ and $(\pi,0)$. 

We have also investigated how the gaps on the different FSs differ for pairing
$s\pm$ and $s_{IB}$. In Fig.~\ref{gap} we show the ratio $R$ between the two 
gaps in the unfolded BZ for $V=0.05$. It can be seen that
on the hole pockets $R=1$, but an appreciable difference is observed on the 
electron pockets where $R=2$ at the point where the electron pocket is entirely
formed by $xy$, and diminishes as the hybridization of $xy$ with $xz$ or $yz$ 
becomes stronger. The maximum value of $R$ is a function of the values of $A$
and $B$ in the linear combination that defines $s_{IB}$. Thus, while $s\pm$ 
is characterized by gaps with a weak momentum dependence and with similar 
magnitudes on the hole and electron pockets, the $s_{IB}$ state is 
characterized by a different gap on the electron pockets with stronger 
momentum dependence due to the hybridization.

\begin{figure}
\includegraphics[width=0.47\textwidth]{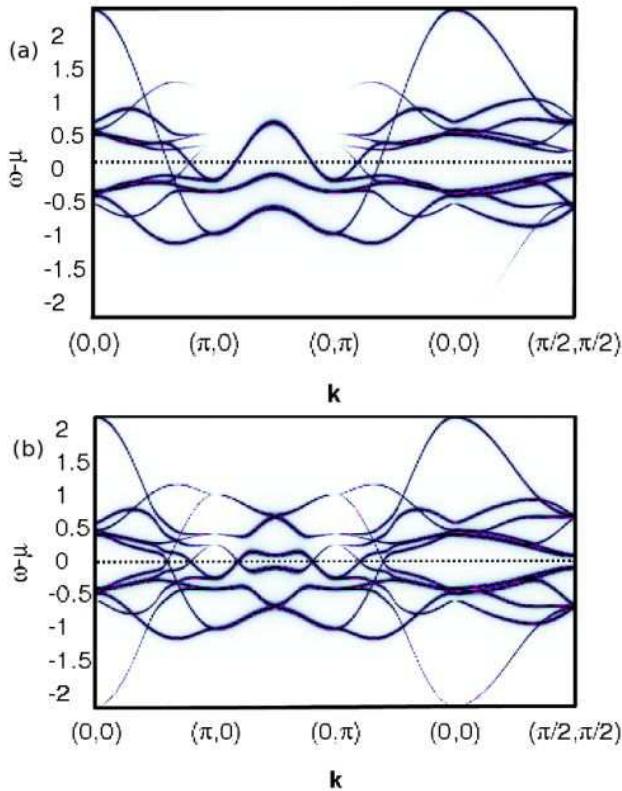}
\caption{(Color online) The intensity of the points represent the values of the
spectral function $A({\bf{k},\omega})$ for the three-orbital model with the pairing
interaction  $V=0.2$, for the pairing operators (a) $B_{2g}$ and 
(b) $B_{2g}^{\rm ext}$ discussed in the text.\label{IVVg}}
\end{figure}

Now we focus on the spectral function for operator $B_{2g}$
presented in Fig.~\ref{IVVg}(a). 
This is the pairing operator that was favored by numerical calculations in
the intermediate $U$ regime of the two-orbital model and it only pairs
electrons in orbital $xz$ with electrons in orbital $yz$.\cite{moreo} 
Although neither this pairing operator nor the Fermi surfaces defining the two hole pockets
involve the $xy$ orbital, the results around the hole pockets differ in the
two- and three-orbital models. In the latter, much lower values of the pairing
attraction $V$ are sufficient to remove the extra nodes found close to the
hole-pocket FS along the $\Gamma$-$X(Y)$ directions in the two-orbital
model. This happens because the bands forming the two hole pockets are now
degenerate at $\Gamma$ and the pockets are consequently at almost
the same momenta of the extended BZ, while they were separated by $(\pi,\pi)$
in the two-band model. As a result, a small interorbital pairing 
can now overcome the
separation between the two FSs and induce a full gap at the hole
pockets.\cite{ib} At the electron pockets a third node, in addition to the 
two of
the two-orbital model, is found  along the $\Gamma$-$X(Y)$ direction, where the
pocket has purely $xy$ character and is, thus, not affected by the operator 
$B_{2g}$. Thus, this pairing operator would show full gaps 
on the hole pocket FS and nodal gaps on the electron pocket FS.

Figure~\ref{IVVg}(b) shows the spectral functions for pairing 
operator $B_{2g}^{\rm ext}$, which is a linear combination of 
$B_{2g}$  with the pairing
operator $V_g$ (with crystal momentum ${\bf Q}$), i.e. next-nearest-neighbor 
interorbital pairing among electrons in all three orbitals is allowed. We find 
that nodes occur only at the
electron pockets. As $V$ increases, nodes at the electron pockets remain only
along the $\Gamma$-$X(Y)$ directions because one of the electron pockets is 
formed by a non-hybridized orbital $xy$ along this direction, and the
relevant pairing interaction is zero for the Fermi momentum. We also
investigated $A({\bf{k},\omega})$ for a similarly extended $B_{1g}$
pairing $B_{1g}^{\rm ext} = (\cos k_x + \cos k_y)\lambda_3 -
a(\cos k_x - \cos k_y)(\lambda_0 -\sqrt{3} \lambda_8)/3 $ (not shown),
which is nodal for extremely small $V$, with nodes in the
$\Gamma$-$M$ direction on the hole pockets, but where the nodes are
already lifted for finite but small $V\gtrsim 0.01$ for many non-zero
values of $a$.

Summarizing, we have found that among the nearest and next-nearest neighbor
pairing operators allowed by symmetry only the $s\pm$ pairing operator is 
purely intraband and produces
nodeless gaps for all values of the pairing attraction $V$. Thus, purely 
intraband pairing interactions occur only if 
electrons in each of the three orbitals are subjected to the same pairing
attraction, i.e. when the identity matrix $\lambda_0$ characterizes the 
orbital portion of the pairing operator. We also found that some linear
combinations of pairing operators \#I and \#II produce nodeless gaps for any 
finite $V$ if $f({\bf k})=\cos k_x \cos k_y$, but interband attraction 
appears in parts of the BZ. Finally,
the interorbital pairing operators
$B_{2g}$ and $B_{2g}^{\rm ext}$ favored by numerical studies 
in a two-orbital model, present
nodeless gaps on the hole pockets but nodes appear on the electron pockets.

\section{Conclusions}\label{sec:conclusions}

In this work, a simple three-orbital Hamiltonian has been constructed
involving the $3d$ orbitals $xz$, $yz$, and $xy$. These orbitals 
have the largest weight at the FS of the 
pnictide LaOFeAs, according to LDA calculations. It was shown that 
it is possible to 
qualitatively reproduce the shape of the LDA-FS by fixing the electron
filling to 4 electrons per Fe. Moreover, two features that have been 
criticized in the two-orbital model have now been corrected: both hole pockets now 
arise from bands degenerate at the $\Gamma$-point, and there is no pocket 
around $M$ in the extended BZ. In addition, the $xy$ character of a small piece
of the electron pockets is now properly reproduced. 

Numerical calculations using a small $2\times 2$
lattice show a tendency to the development of magnetic $(\pi,0)$-$(0,\pi)$ stripes 
when Coulombic interactions are added, result consistent 
with experimental observations.
A mean-field analysis confirms this tendency for physically relevant values 
of $J/U$. As in the case of the 
two-orbital model, an antiferromagnetic metallic phase occurs only at intermediate values of
the Coulomb repulsion. At large $U$, the ground state is magnetic, but it is an insulator 
that is also orbitally ordered. Additionally, a metallic, magnetic and 
orbitally ordered
phase is encountered just before the metal-insulator transition. In
the most interesting regime with a spin-$(\pi,0)$ antiferromagnetic metal without
pronounced orbital order, the bands are similar to the uncorrelated
ones, but their bandwidth is reduced with increasing $U$. The Fermi
surface is also very similar to the
uncorrelated one but, depending on $U$, we find small additional
electron-like pockets near the original hole pockets around $\Gamma$ (small
$U$) or hole-like pockets between the electron- and hole-pockets
(at slightly larger $U$).

The possible pairing operators  that are allowed by the symmetry of the lattice and the 
orbitals have been constructed for pairs made of electrons separated by a distance up to one
diagonal lattice spacing.  If on-site 
pairing is disregarded due to the large Coulomb repulsion, 
it was found that the only purely intraband 
pairing operator that has a full gap on the FS is  
\#I with $f({\bf k})= \cos k_x\cos k_y$  which corresponds to 
the $s\pm$ pairing operator with a momentum dependent OP that has opposite 
signs on the hole and electron FSs. This operator arises from a purely 
intraband pairing attraction equal for each of 
the three bands. 
Note that the  pairing operator \#I is the {\it only} one that leads to purely 
intraband pairing interactions. Since this pairing operator is proportional to
the identity matrix $\lambda_0$ both in the orbital and the band 
representations we found that the ratio $|\Delta_i/\Delta_j|$ between the
gaps in two different FSs can differ only by the ratios 
$|f({\bf k}_i)/f({\bf k}_j)|$; then, any experimental indication of a 
different kind of ratio would indicate some degree of interband 
pairing.\cite{ib} Thus, order parameter ratios predicted by several 
authors\cite{bang,parker,hu,dolgov} with calculations based on purely intraband 
pairing (they allow interband hopping of intraband pairs) are not allowed by 
the symmetry of the lattice and the orbitals. In this regard, our calculations
seem to indicate that unrelated gaps in different FSs can occur only in 
systems in which at least one orbital (or a group of orbitals) is not strongly 
hybridized with the remaining ones.

We found that all the other pairing operators, except for \#I, lead to 
interband pairing attraction in the band representation. In addition, all the 
pairing operators with interband pairing studied here have nodal band 
structures at small $V$ with the exception of pairing operator $s_{IB}$. 
In this case, the gap on the 
electron pockets is expected to have a stronger variation at different points 
in the BZ 
that the gap at the hole pockets. Thus, a strong indication that $s\pm$ is 
the appropriate pairing symmetry would be provided by experiments in the 
pnictides showing a nodeless gap in all FSs, relatively independent
of momentum, and with similar values on all FSs. 

Summarizing, we have shown that the addition of a third orbital corrects the 
shortcomings pointed out in the two-orbital model: the two hole pockets now 
arise
from bands degenerate at the $\Gamma$ point while the electron pockets contain
a small piece with $xy$ character. However, the dependence of the magnetic 
phases with $U$ for the undoped case appears to be similar for three and two  
orbitals except for a magnetic, orbital ordered, metallic
phase that appears in the three-orbital case. In both models it is found that
the only pairing operator allowed by symmetry with next or diagonal 
nearest-neighbor interactions which is purely intraband and produces a 
nodeless gap is the $s\pm$ state. In addition, the only change observed in the 
interorbital $B_{2g}$ pairing state, favored by numerical simulations in 
the two-orbital model, is that, at the mean-field level, the addition of the 
$xy$ orbital renders the gap on the hole pockets nodeless for much smaller 
values of the pairing attraction.

\section{Acknowledgments}
This research was sponsored by the National Science Foundation grant DMR-0706020 
(M.D., A.N., A.M., and E.D.) and the
Division of Materials Science and Engineering, Office of Basic Energy Sciences,
U.S. Department of Energy (A.M. and E.D.).
                                                                                  
\appendix

\section{Mean-Field Equations}\label{app:mf}

In this appendix, we discuss the mean-field approach
used here to study the Hamiltonian given by the kinetic energy
Eq.~(\ref{E.H0k}) and the onsite Coulomb interaction
Eq.~(\ref{eq:Hcoul}). Depending on the ordering vectors ${\bf q}_1$ and 
${\bf q}_2$, listed in Table~\ref{tab:phases}, for magnetic and orbital order, 
the real-space unit cell contains one, two or four sites: One for the
ferro-orbital and ferromagnetic case, four if both ordering vectors are
different from each other and from $(0,0)$, and two in all other
cases. Following Refs.~\onlinecite{Nomura:2000p647,yu}, we only keep the expectation
values of density operators, as given in Eqs.~(\ref{eq:mf_xy})
and~(\ref{eq:mf_xz_yz}) in the Coulomb interaction
Eq.~(\ref{eq:Hcoul}). Together with Eq.~(\ref{E.H0k}), this determines the
mean-field Hamiltonian, see below. We then solve the self-consistency
equations for the six parameters $n_{xy}$, $m_{xy}$, $n$, $m$, $p$,
and $q$ in Eqs.~(\ref{eq:mf_xy}) and~(\ref{eq:mf_xz_yz}) for various
combinations of ordering momenta, see Tab.~\ref{tab:phases}, which
corresponds to minimizing the total energy. This is done for all
considered phases and the one with the lowest energy is taken to be
the stable solution.
Depending on the size of the unit cell, one to four
momenta are coupled by the Coulomb interaction. In
the following, we will provide the Hamiltonians for 
several ordering patterns with different unit cells. In all cases, the
sums run over the whole extended BZ corresponding to the one-iron
unit cell. The calculations were carried out in momentum
space for up to $400\times 400$ ${\bf k}$-points. We did not observe
any pronounced dependence on the number of momenta, except for very
small lattice sizes. 

\subsection{Ferromagnetic and Ferro-Orbital Order: One-Site Unit Cell}

In this case ${\bf q}_1={\bf q}_2=(0,0)$ and

\begin{align}
H_{\textrm{MF}}({\bf k}) &= H_{\textrm{TB}}({\bf k}) 
+U\sum_{{\bf k}, \mu,\sigma} n_{\mu} d^\dagger_{\mathbf{ k},\mu,\sigma} d^{\phantom{\dagger}}_{\mathbf{ k},\mu,\sigma}\nonumber\\
&\quad+ (2U'-J)\sum_{{\bf k}, \mu\neq\nu,\sigma} n_{\nu} d^\dagger_{\mathbf{ k},\mu,\sigma}
d^{\phantom{\dagger}}_{\mathbf{ k},\mu,\sigma}\nonumber\\
&\quad - U \sum_{{\bf k}, \mu,\sigma} \frac{\sigma}{2} m_\mu 
d^\dagger_{\mathbf{ k},\mu,\sigma} d^{\phantom{\dagger}}_{\mathbf{ k},\mu,\sigma}\\
&\quad - J \sum_{{\bf k}, \mu\neq\nu,\sigma} \frac{\sigma}{2} m_\nu 
d^\dagger_{\mathbf{ k},\mu,\sigma} d^{\phantom{\dagger}}_{\mathbf{ k},\mu,\sigma}+NC\;,\nonumber
\end{align}
where $n_{\mu} = n_{xy}$, $m_{\mu} = m_{xy}$ for the $xy$ orbital, and
$n_\mu=n\pm p/2$, $m_\mu=m\pm q$ for $xz$
and $yz$. The sum over ${\bf k}$
runs through the whole BZ, $N$ is the number of lattice sites, and the constant $C$ is
given by
\begin{align}
\label{eq:c_fo_fm}
C&= -U \sum_{\mu} n_{\mu}^2  
+ U/4\sum_{\mu}m_{\mu}^2\\
&\quad
- (2U' -J)\sum_{\mu\neq\nu} n_{\mu}n_{\nu} + J/4 \sum_{\mu\neq\nu} m_{\mu}m_{\nu}\;.\nonumber
\end{align}

\subsection{Antiferromagnetic and Ferro-Orbital: Two-Site Unit Cell}

For AF order with ${\bf q}_1=(\pi,\pi)$, $(0,\pi)$ or $(\pi,0)$, 
the real-space unit cell doubles, and momenta ${\bf k}$
and ${\bf k} + {\bf q_1}$ are  coupled by the interaction.
\begin{align}
H_{\textrm{MF}}({\bf k}) &= H_{\textrm{TB}}({\bf k}) 
+U\sum_{{\bf k}, \mu,\sigma} n_{\mu} d^\dagger_{\mathbf{ k},\mu,\sigma} d^{\phantom{\dagger}}_{\mathbf{ k},\mu,\sigma}\nonumber\\
&\quad+ (2U'-J)\sum_{{\bf k}, \mu\neq\nu,\sigma} n_{\nu} d^\dagger_{\mathbf{ k},\mu,\sigma}
d^{\phantom{\dagger}}_{\mathbf{ k},\mu,\sigma}\nonumber\\
&\quad - U \sum_{{\bf k}, \mu,\sigma} \frac{\sigma}{2} m_\mu 
d^\dagger_{\mathbf{ k}+{\bf q}_1,\mu,\sigma} d^{\phantom{\dagger}}_{\mathbf{ k},\mu,\sigma}\\
&\quad- J \sum_{{\bf k}, \mu\neq\nu,\sigma} \frac{\sigma}{2} m_\nu 
d^\dagger_{\mathbf{ k}+{\bf q}_1,\mu,\sigma} d^{\phantom{\dagger}}_{\mathbf{ k},\mu,\sigma}+NC\;.\nonumber
\end{align}
Again, $n_{\mu} = n_{xy}$, $m_{\mu} = m_{xy}$ for the $xy$ orbital, and
$n_\mu=n\pm p/2$, $m_\mu=m\pm q$ for $xz$ and $yz$; and the same
constant Eq.~(\ref{eq:c_fo_fm}) as above. The case of ferromagnetic
order and alternating orbitals is treated in an analogous manner.

\subsection{Antiferromagnetic and  Alternating Orbital Order with the
  Same Ordering Vector: Two-Site Unit Cell}

In some phases, both the orbital and the magnetic order alternate with
the same ordering vector ${\bf q}={\bf q}_1={\bf q}_2=(\pi,\pi)$, $(0,\pi)$ 
or $(\pi,0)$. In this case, the
Hamiltonian is given by
\begin{align}
H_{\textrm{MF}}({\bf k}) &= H_{\textrm{TB}}({\bf k}) \nonumber\\
&+[(4U'-2J)n + U n_{xy} ]
\sum_{{\bf k},\sigma} d^\dagger_{\mathbf{ k},xy,\sigma}d^{\phantom{\dagger}}_{\mathbf{ k},xy,\sigma}\nonumber\\
&+[Un+(2U'-J)(n_{xy}+n)] 
\sum_{\substack{{\bf k},\sigma\\\mu = xz,yz}} d^\dagger_{\mathbf{
    k},\mu,\sigma} 
d^{\phantom{\dagger}}_{\mathbf{ k},\mu,\sigma}\nonumber\\
&-(U+J)q\sum_{\substack{{\bf k},\sigma\\\mu = xz,yz}}
\frac{\sigma\alpha}{2} d^\dagger_{\mathbf{ k},\mu,\sigma}
 d^{\phantom{\dagger}}_{\mathbf{ k},\mu,\sigma}\\
&- [(U+J) m +Jm_{xy}]\sum_{\substack{{\bf k},\sigma\\\mu = xz,yz}} \frac{\sigma}{2}
d^\dagger_{\mathbf{ k}+{\bf q}_1,\mu,\sigma} d^{\phantom{\dagger}}_{\mathbf{ k},\mu,\sigma}\nonumber\\
&-(Um_{xy} + 2Jm) \sum_{{\bf k}, \sigma} \frac{\sigma}{2}
d^\dagger_{\mathbf{ k}+{\bf q}_1,xy,\sigma} d^{\phantom{\dagger}}_{\mathbf{ k},xy,\sigma}\nonumber\\
&+(U-2U'-J)p\sum_{\substack{{\bf k},\sigma\\\mu = xz,yz}}
\frac{\alpha}{2}
d^\dagger_{\mathbf{ k}+{\bf q}_1,\mu,\sigma}
d^{\phantom{\dagger}}_{\mathbf{ k},\mu,\sigma}+NC.\nonumber
\end{align}
Here, $\alpha = \pm 1$ distinguishes between the $xz$ and $yz$
orbital flavors as $\sigma$ does for the spin. The constant $C$ reads
\begin{align}
C&= -U (n_{xy}^2-m_{xy}^2/4)
-U(n^2 + p^2/2 - m^2/2 - q^2/2)\nonumber\\
&\quad-(8U'-4J)n_{xy}n  - (4U'-2J)(n^2-p^2/4) \nonumber\\
&\quad+Jm_{xy}m + J(m^2 - q^2)/2\;.
\label{eq:c_nmpq}\end{align}

\subsection{Antiferromagnetic and Alternating Orbitals with Different
  Ordering Momenta: Four-Site Unit Cell}

If both orbital occupation and magnetic order alternate with different
ordering momenta, so that ${\bf q}_i=(\pi,\pi)$, $(0,\pi)$ or $(\pi,0)$ with
${\bf q}_1\ne{\bf q}_2$, 
the real-space unit cell contains four sites, and
consequently all four momenta ${\bf k}$, ${\bf k} + {\bf q}_1$, ${\bf
  k} + {\bf q}_2$, ${\bf k} + {\bf q}_1 + {\bf q}_2$ are  coupled, but
apart from this, the Hamiltonian is very similar to the previous case:
\begin{align}
H_{\textrm{MF}}({\bf k}) &= H_{\textrm{TB}}({\bf k}) \\
&+[(4U'-2J)n + U n_{xy} ]
\sum_{{\bf k},\sigma} d^\dagger_{\mathbf{ k},xy,\sigma}d^{\phantom{\dagger}}_{\mathbf{ k},xy,\sigma}\nonumber\\
&+[Un+(2U'-J)(n_{xy}+n)] 
\sum_{\substack{{\bf k},\sigma\\\mu = xz,yz}} d^\dagger_{\mathbf{
    k},\mu,\sigma} 
d^{\phantom{\dagger}}_{\mathbf{ k},\mu,\sigma}\nonumber\\
&- [(U+J) m +Jm_{xy}]\sum_{\substack{{\bf k},\sigma\\\mu = xz,yz}} \frac{\sigma}{2}
d^\dagger_{\mathbf{ k}+{\bf q}_1,\mu,\sigma} d^{\phantom{\dagger}}_{\mathbf{ k},\mu,\sigma}\nonumber\\
&-(Um_{xy} + 2Jm) \sum_{{\bf k}, \sigma} \frac{\sigma}{2}
d^\dagger_{\mathbf{ k}+{\bf q}_1,xy,\sigma} d^{\phantom{\dagger}}_{\mathbf{ k},xy,\sigma}\nonumber\\
&+(U-2U'-J)p\sum_{\substack{{\bf k},\sigma\\\mu = xz,yz}}
\frac{\alpha}{2}
d^\dagger_{\mathbf{ k}+{\bf q}_2,\mu,\sigma}
d^{\phantom{\dagger}}_{\mathbf{ k},\mu,\sigma}\nonumber\\
&-(U-J)q\sum_{\substack{{\bf k},\sigma\\\mu = xz,yz}}
\frac{\sigma\alpha}{2}
d^\dagger_{\mathbf{ k}+{\bf q}_1+{\bf q}_2,\mu,\sigma}
d^{\phantom{\dagger}}_{\mathbf{ k},\mu,\sigma}+NC.\nonumber
\end{align}
The constant $C$ is still
given by equation (\ref{eq:c_nmpq}).

\section{$\lambda_i$ matrices}
\label{app:lambda}

The $\lambda_i$ matrices used in the text are presented here:
\begin{align*}
  \lambda_0=
  \left(\begin{array}{ccc}
      1 & 0               &   0\\
      0 & 1               &   0\\
      0 & 0               &   1
    \end{array} \right),
  &\quad
  \lambda_1=
  \left(\begin{array}{ccc}
      0 & 1               &   0\\
      1 & 0               &   0\\
      0 & 0               &   0
    \end{array} \right),
\end{align*}

\begin{align*}
  \lambda_2=
  \left(\begin{array}{ccc}
      0 & -i               &   0\\
      i & 0               &   0\\
      0 & 0               &   0
    \end{array} \right),&\quad
  \lambda_3=
  \left(\begin{array}{ccc}
      1 & 0               &   0\\
      0 & -1               &   0\\
      0 & 0               &   0
    \end{array} \right),
\end{align*}

\begin{align*}
  \lambda_4=
  \left(\begin{array}{ccc}
      0 & 0               &   1\\
      0 & 0               &   0\\
      1 & 0               &   0
    \end{array} \right),&\quad
  \lambda_5=
  \left(\begin{array}{ccc}
      0 & 0               &   -i\\
      0 & 0               &   0\\
      i & 0               &   0
    \end{array} \right),
\end{align*}

\begin{align*}
  \lambda_6=
  \left(\begin{array}{ccc}
      0 & 0               &   0\\
      0 & 0               &   1\\
      0 & 1               &   0
    \end{array} \right),&\quad
  \lambda_7=
  \left(\begin{array}{ccc}
      0 & 0               &   0\\
      0 & 0               &   -i\\
      0 & i               &   0
    \end{array} \right),
\end{align*}

\begin{equation*}
  \lambda_8=\frac{1}{\sqrt{3}}
  \left(\begin{array}{ccc}
      1 & 0               &   0\\
      0 & 1               &   0\\
      0 & 0               &   -2
    \end{array} \right).
\end{equation*}


\begin{thebibliography}{10}
\bibitem{note:signt8}
Choosing $t_8 = +t_7/2$ would also give us the desired results: The purely $xy$
  character is then found along the $X$-$M$ and $Y$-$M$ directions, which are
  equivalent to $\Gamma$-$X$ and $\Gamma$-$Y$ in the reduced BZ.
\bibitem{note_t7}
Another way to obtain $yz$ character between the $X$ and $M$ points of the
  extended BZ would be to choose $t_3$ and $t_4$ very large and $t_6$ much
  smaller, so that the $xy$ band would remain below the Fermi surface and
  $xz$/$yz$ rise above it. However, the degeneracy of $xz$ and $yz$ at
  $M=(\pi,\pi)$ would then force an additional hole pocket around $M$
  similar to the FS in Ref.~\onlinecite{plee}.
\bibitem{note_GK}
Note that the large number of inter- as well as intra-orbital hopping processes
  between NN and NNN sites renders the usual Goodenough-Kanamori rules
  inapplicable.
\bibitem{hop} Note that the ``interband'' superconductivity proposed by
some authors\cite{parker,bang,hu,dolgov} is still represented by an intraband 
(i.e., diagonal) pairing matrix but with 
$\Delta_1=\Delta_1(\Delta_2,\Delta_3)$, $\Delta_2=\Delta_2(\Delta_1,\Delta_3)$,
and $\Delta_3=\Delta_3(\Delta_1,\Delta_2)$ .
\end{thebibliography}
\end{document}